\documentclass[twocolumn]{emulateapj}
\usepackage{apjfonts}
\usepackage{graphicx}
\usepackage{epstopdf}
\newcommand{\be}{\begin{equation}}
\newcommand{\ee}{\end{equation} }
\newcommand{\ba}{\begin{eqnarray}}
\newcommand{\ea}{\end{eqnarray}}

\newcommand{\nn}{\mbox{} \nonumber \\ \mbox{} }

\newcommand{\Te}{\widetilde T_e}

\shorttitle{Hot Electromagnetic Outflows. III.}
\shortauthors{Thompson \& Gill}	
\slugcomment{Submitted to the Astrophysical Journal}
\begin{document}
\title{Hot Electromagnetic Outflows. III.  \\ Displaced Fireball in a Strong Magnetic Field}
\author{Christopher Thompson and Ramandeep Gill}
\affil{Canadian Institute for Theoretical Astrophysics, 60 St. George St., Toronto, ON M5S 3H8, Canada.}


\begin{abstract}
The evolution of a dilute electron-positron fireball is calculated in the regime of strong magnetization and very high compactness
($\ell \sim 10^3-10^8$).  Heating is applied at a low effective temperature ($< 25$ keV), and the fireball is allowed to expand,
so that the formation of a black-body spectral distribution is inhibited by pair annihilation.  The diffusion equation 
for Compton scattering is coupled to a single-temperature pair gas and an exact (trans-relativistic) cyclo-synchrotron photon source.  
We find that the photon spectrum develops a quasi-thermal peak, with a power-law slope below it that is characteristic of gamma-ray 
bursts ($F_\omega \sim $ const).  The formation of a thermal high-frequency spectrum is checked using the full kinetic equations at $\ell \sim
10^3$.  These results have several implications for the central engine of GRBs, and the mechanism
of energy transport.   1. Baryon rest mass carries less than $\sim 10^{-5}$ of the energy flux at jet breakout at
$\lesssim 10^{12}$ cm from the engine, with most carried by the magnetic field.  2. This degree of baryon purity points to the 
presence of an event horizon in the engine, and neutrons play a negligible role in the prompt emission mechanism.
3. X-ray flashes are emitted by outflows carrying enough baryons that the photosphere is pair-depleted, which we show results in faster
thermalization.   4. The relation between observed peak frequency and burst luminosity is bounded below
by the observed Amati et al. relation if jet Lorentz factor $\sim ({\rm opening~angle})^{-1}$ at breakout.  
5. Stellar models are used to demonstrate an inconsistency between the highest observed GRB energies, and a hydrodynamic nozzle:  
magnetic collimation is required.  6. The magnetized pair gas is dilute enough that high-frequency Alfv\'en waves may 
become charge starved.  Finally, we suggest that limitations on magnetic reconnection from plasma collisionality have been overestimated.
\end{abstract}


\section{Introduction}

The spectrum of a gamma-ray burst (GRB) peaks at energies around the electron rest mass,
and below this peak it is usually much flatter than the Rayleigh-Jeans
tail of a black body.  (See \citealt{piran04}, \citealt{meszaros06}, and \citealt{kouveliotou12} for reviews.)
Very bright bursts peaking in the X-ray band, or at lower photon energies,
appear to be absent.  We show that these fundamental properties of GRBs have, collectively,
a simple and cogent explanation:  the outflowing material is strongly magnetized, rich
in electron-positron pairs, and depleted in ions.  It is heated before, and during, breakout from
a confining medium such as a stellar envelope, or a cloud of neutron-rich debris.  The 
Lorentz factor during breakout is modest, $\Gamma \sim ({\rm opening~angle})^{-1}$, and
the outflow is compact enough that the pairs follow a nearly thermal energy 
distribution described by a single temperature.  No additional species of particles, such 
as neutrons or pions, are invoked or required.  

The observed floor to the spectral peak frequency, first identified by \cite{amati02}
for long GRBs, is recovered if the outflow carries a total energy that is comparable to the binding 
energy of a massive stellar CO core.  We show that the largest isotropic-equivalent luminosities measured in GRBs 
are inconsistent with a hydrodynamical nozzle forming in a collapsing stellar envelope, and point to 
magnetic collimation.  Outflows which are strongly magnetized, but have photospheres dominated 
by the electron-ion component, are shown to have softer spectra.  We identify them with X-ray flashes.  

The pairs largely annihilate following breakout, if heating temporarily subsides.  This allows the
outflow to be accelerated outward by a combination of the Lorentz force and radiation pressure.  
The compact thermalization phase must, however, be followed by additional heating
after the outflow has attained $\Gamma \sim 10^2-10^3$.  This second phase is considered briefly at the end of this paper.
What powers the continuing dissipation, and the production of the relativistic particles that emit
the hard tail of the spectrum, depends on finer details of the outflow and is therefore more conjectural.  
The leading candidates are the reconnection of a reversing magnetic field, or the interaction of the fast,
magnetized material with a slower baryonic shell ahead of it.  We view the calculations presented here as compelling
enough that some popular dissipation channels, such as hydrodynamic collisions between baryonic shells,
or inelastic collisions between dilute flows of neutrons and charged ions, can now be disfavored.

The strong baryon purity of the outflow points to the presence of an event horizon in the central engine.
The extreme radiant energy of most GRBs is inconsistent with any known type of stellar magnetic flare,
even with those most extreme flares of the Soft Gamma Repeaters (SGRs).  Nonetheless, we find a genuine commonality 
in the {\it physical properties} of the outflows that give rise to GRBs and magnetar flares:
they are simultaneously photon-rich and strongly magnetized, and, during a critical phase in the emission process,
only mildly relativistic.  They differ in overall energy scale, and the degree of rotationally forced collimation.

The plan of this paper is as follows.  In Section \ref{s:review} we place our work in the context of the
voluminous literature on Comptonization in thermal baryonic plasmas, and previous approaches to
the prompt emission of GRBs.  Section \ref{s:pplas} describes our approach to calculating the photon spectrum,
and pair density and temperature.  The results of the numerical calculations are shown in Section 
\ref{s:spectrum}.  These are compared with a semi-analytic scaling solution for the plasma temperature
and scattering depth in Section \ref{s:scaling}.   The temperature and spectrum of a magnetized outflow with a 
baryon-dominated photosphere is considered in Section \ref{s:baryon}.  The effects of bulk Comptonization
during jet breakout on the emergent spectrum are shown in Section \ref{s:multscatt}.  Finally, we draw
together our results on the magnetized pair plasma with a global model of a Poynting-dominated jet in
Section \ref{s:poyntingjet}, showing in Section \ref{s:amati} how together they provide a simple motivation 
of the Amati et al. boundary.  Section \ref{s:summary} summarizes our results.  Appendix \ref{s:appA} reviews 
the various channels for soft-photon emission, and \ref{s:appB} presents details of our calculation of $e^\pm$ 
pair creation and annihilation.  Throughout this paper, we use the notation $X = X_n\times 10^n$ to denote quantity $X$ in units of $10^n$.

\section{Previous Approaches to Hot Comptonizing Plasmas and the Spectra of GRBs}\label{s:review}

A physical explanation for the $\sim m_ec^2$ peak of GRBs must simultaneously account for the 
relatively soft shape of the spectrum below the peak, which is mostly inconsistent
with a black body \citep{goldstein12,goldstein13}.  For this last reason alone, a simple fireball model 
\citep{goodman86,shemi90} is inadequate.  Distributed energy release in baryonic 
fireballs allows for a wider range of low-frequency spectra (e.g. \citealt{vurm13}),
but fine-tuning of the location of the dissipation, and the introduction of more complicated particle
distributions, are required. 

Matter and radiation interact in a GRB outflow over many decades in radius.  Most 
theoretical attempts to understand the spectrum have i) assumed that the low- and high-frequency
parts of the spectrum form in the same region, and ii) that this process is localized to
a particular radial zone.  Because the {\it high-frequency} part of the spectrum requires
high Lorentz factors, attempts were made to reproduce simultaneously the spectral peak and low-frequency
spectrum at high $\Gamma$ (e.g. \citealt{peer04,giannios05}).  This approach immediately
rules out a thermalization process, such as is investigated here, because the observed spectral
peak must sit well above $\sim 1$ MeV.  Another difficulty lies in finding a robust mechanism
for localizing the dissipation in radius, given the wide range of possibilities.  

The approach taken here, following \cite{thompson06} and \cite{vurm13}, is to divide the problem into 
two parts:  
\vskip .05in\noindent
1. The low-frequency spectrum of GRBs, up to and including the peak, is assumed to originate in
a separate zone from the high-frequency tail.
\vskip .05in\noindent
2. The Lorentz factor in the inner thermalization zone is much smaller than
that in the hard tail.  This inner zone is associated with jet breakout from a confining
medium, and is significantly displaced from the central engine \citep{tmr07,lmmb13}.
Confining material is present at intermediate radii:
either a Wolf-Rayet envelope \citep{woosley93,paczynski98,macfadyen99} or 
a neutron-rich wind \citep{duncan86,eichler89,dessart09}.  
\vskip .05in\noindent
3. The hard tail to the spectrum originates further out in the outflow, due to continuing
magnetic dissipation.  Magnetized outflows could be driven by rapid rotation of a compact star \citep{dt92, usov92, usov94,
thompson94, meszaros97, lyutikov03, mckinney06}, or possibly by magnetic flaring in an accretion disk \citep{npp92}.  
A jet emitted by the collapsed remnant of a massive neutron star would ultimately be powered by such accretion.

\subsection{Pair Creation in Magnetized Outflows}

\cite{guilbert83} and \cite{svensson87} have considered steady-state electron-positron pair cascades triggered by the injection
of relativistic particles into a compact, soft photon source, noting that this will lead to the accumulation
of a considerable optical depth in cold pairs.  

Pair creation in relativistic outflows has also been studied for some
time \citep{cavallo78,goodman86,paczynski90,shemi90,krolik91,grimsrud98}.  But when {\it baryonic} kinetic energy dominates
the outflow luminosity, thermally created pairs are present in negligible concentration except
very close to the engine, even in the presence of delayed dissipation (e.g. \citealt{beloborodov13}).  Pairs can
be regenerated by bulk heating near the photosphere of a turbulent MHD outflow \citep{thompson94, thompson97}, during collisions 
between dilute baryonic shells \citep{ghisellini99}, or by non-thermal particle acceleration at shocks \citep{meszaros00}.  

On the other hand, if the baryon concentration in the outflow is pushed to very low values --
that is, if it is magnetically dominated -- then thermally created pairs can dominate the scattering 
opacity over many decades of radius.  Our focus here is on the region inside the photosphere, as influenced
by radially distributed heating.  We do not consider non-local pair creation effects which would dominate 
outside the photosphere \citep{tm00,beloborodov02}.

Previous work by \cite{usov92} focused on pair-creation near the engine by a unipolar inductor mechanism, although in
practice this would be dominated by other pair creation channels such as neutrino collisions (e.g. \citealt{eichler89,zalamea11})
or damping of hydromagnetic turbulence \citep{tb98}.  \cite{usov94} and \cite{meszaros97} considered a pair gas that is advected passively from
the engine out to the photosphere of a magnetized wind or jet, assuming the same radial Lorentz factor profile as a thermal fireball 
inside the photosphere.

The closest treatment to ours
is by \cite{thompson97}, who studied the equilibrium of continuously heated, thermal pair plasmas in strong magnetic fields
but did not make a detailed assessment of thermal cyclo-synchrotron emission.  In the context of magnetar flares,
\cite{td01} considered thermal pair creation in super-QED magnetic fields, where other photon creation processes contribute.

Outflows with comparable energy flux in toroidal magnetic field and thermal radiation have been investigated by
\cite{thompson94,thompson06}, \cite{meszaros97}, \cite{drenkhahn02}, \cite{giannios06}, \cite{giannios07}, and \cite{russo13a,russo13b}.
The direct involvement of thermal radiation in the prompt emission from Poynting-dominated outflows has, by contrast,
been discounted by \cite{usov94}, \cite{lyutikov00}, \cite{lyutikov03}, and \cite{zhang11}.  These authors proposed 
instead that this component would decouple from the outflow (forming e.g. a soft precursor) and
that residual pairs trapped in the magnetic field would act as seeds for synchrotron emission at larger distances from
the engine.

\subsection{Multiple Compton Scattering}

Even though the theory of multiple Compton scattering (Comptonization)
in dense baryonic plasmas has a long history (with much of the fundamental work done 
by the Soviet school: \citealt{pozdnyakov83}), the analogous problem in highly compact, thermal
pair plasmas has received remarkably little attention.  In part, that may be because pairs
in the primeval fireball are only present in a state of enormous optical depth.  

Here we calculate in some detail the response of a dilute, and strongly magnetized, pair gas to steady heating.
The radiation compactness is still very high
($10^3 - 10^8$), but the equivalent black-body temperature is low enough ($\ll 25$ keV)
that pairs are much less numerous than photons.  We evolve the Kompaneets equation coupled to a detailed
calculation of pair creation and annihilation, and an exact evaluation of
cyclo-synchrotron emission.

Separately we allow for a small fraction ($\lesssim 10^{-2}$) of the plasma
energy to be injected in relativistic particles, which have a small direct effect on the photon spectrum 
below $\sim m_ec^2$, but can spawn a higher density of cold pairs than expected in equilibrium with a
thermal photon gas.
As a check of our treatment of pair creation and annihilation, we also evolve the full kinetic 
equations for photons and pairs in a more dilute plasma with a compactness $\ell \sim 10^3$.   

During the approach to black-body equilibrium in a Comptonizing plasma, one generally 
finds an intermediate, flat component of the spectrum ($F_\omega \sim$ const), 
which connects to a distinct Wien peak.  \cite{ghisellini99} noted that this 
intermediate portion of the spectrum might correspond to the low-frequency spectral 
slopes of GRBs (see also \citealt{thompson98}).  However, they focused on plasmas of relatively
low compactness ($\ell \sim 10^2$), with a goal of explaining both the low- and
high-frequency components of GRB spectra, and did not consider strong magnetization or
the effects of expansion (both which we find to be crucial to the low-frequency slope).
\cite{peer04} and \cite{vurm13} showed that a flat low-frequency spectrum 
can arise from distributed heating in baryonic outflows with secondary pair creation -- but only for a much
narrower range of compactness than must be experienced by GRBs, and inconsistent
with the high compactness expected at jet breakout.
The conditions in which a Wien peak {\it fails} to emerge from a compact pair plasma (it is usually
absent from GRB spectra) are addressed quantitatively for the first time here.

In the dilute pair plasma considered here, thermalization is limited by 
a relatively low pair density, and by a finite source of soft photons.  We
show that the end of heating is followed by rapid pair annihilation and
only modest spectral cooling.  The rest-frame spectral peak is, therefore,
buffered to a value $\sim 0.1\,m_ec^2$ over a wide range of compactness.   
Further flattening of the spectrum is shown
to occur as the photons flow through a magnetized jet past its
breakout point:  here the scattering depth drops precipitously and the jet experiences
a strong outward Lorentz force combined with pressure from the collimating radiation field
\citep{russo13a,russo13b}.

\subsection{Other Emission Models}

Considerable attention has already been given to the emergent synchrotron-self-Compton spectrum
in relativistic pair plasmas with a modest compactness $\ell \lesssim 100$:  initially in
the context of accretion disk coronae \citep{lightman87}, and then for GRB outflows
(\citealt{peer04}; \citealt{stern04} adopt essentially the same approach but exclude pair creation).
The main goal in these works was to reproduce all the main components of the 
spectrum within a dissipation zone of limited (but uncertain) size.  In the case of GRBs, the
Lorentz factor must be high in the high-frequency emission zone, so it was
also assumed to be high in the zone that determines the final spectral peak.  

Sometimes a separate 
black body component has been introduced (e.g. \citealt{peer06}), representing an adiabatically evolved echo 
of a fireball phase closer to the engine.   Incomplete thermalization inside the scattering photosphere of
a baryon-dominated outflow naturally leads to a distinct Wien peak in the spectrum
\citep{beloborodov13}, but this hardly represents the low-frequency part of a typical GRB.  
It is possible to combine non-thermal particle acceleration and synchrotron emission at moderate
scattering depth in a baryon-dominated fireball to produce a GRB-like spectrum \citep{peer06,vurm11,vurm13},
but this solution appears sensitive to the placement of the dissipation zone, and different
choices are shown to give quite different results.

Continuous heating in a relativistic outflow, which has some motivation in the magnetized
case \citep{thompson94,spruit01}, has been shown to produce promising high-frequency spectral
slopes \citep{giannios06}.  But if the photon seeds are restricted to a black body and
continuing photon creation is turned off, then the low-frequency spectrum
does not deviate much from a Planckian \citep{giannios12} unless the photons have
undergone strong adiabatic softening before being reheated \citep{thompson98}.  

Other approaches to a flat low-frequency spectrum have been considered, including hard-spectrum synchrotron cooling 
particles \citep{bykov96,uhm13} or black-body emitting jets with sharp angular gradients \citep{lundman13}.
Finally, we note that dissipation due to $n$-ion collisions, which is a possible source of 
non-thermal pairs \citep{beloborodov10,vurm11}, is negligible during jet breakout at low $\Gamma$, and especially
if the electron-ion component is subdominant to thermal pairs.


\section{Nearly Thermal Pair Plasma in a \\ Strong Magnetic Field}\label{s:pplas}

A pair plasma differs in an important respect from baryonic plasmas:  as the temperature
drops below $\sim 0.1 m_ec^2$, the pairs annihilate.  This has a strong buffering
effect on the rate of Compton scattering, and the upward flux of photons in frequency space.
Complete thermalization -- the formation of a black-body spectral distribution -- is pushed
to a much higher compactness ($> 10^8$) than would be the case in a baryonic plasma.  

The temperature adjusts so that a nearly constant flux of photons is maintained up to a frequency 
just below the peak.  This corresponds to a low-frequency spectrum $U_\omega \sim$ constant; 
we find that an isolated Wien peak ($U_\omega \sim \omega^3$ below the peak) is absent if
the pairs are created only on the thermal Boltzmann tail.  

Higher optical depths can develop
if a modest fraction of the dissipation is in relativistic particles, which we show does tend
to harden the low-frequency spectrum.   In this way, measurements of GRB spectra offer constraints
on the intermittency of the heating process.

The thermal and magnetic energy densities are conveniently parameterized in terms of the compactness,
\be\label{eq:comp}
\ell_{\rm th} \equiv {\sigma_T U_{\rm th} ct\over m_ec^2};\quad\quad
\ell_B \equiv {\sigma_T (B^2/8\pi) ct\over m_ec^2},
\ee
where as usual $\sigma_T$ is the Thomson cross section, $m_e$ the electron rest mass, and $c$ the
speed of light.  A high compactness suppresses the temperature of the pairs and allows their 
rapid thermalization, even though the scattering depth does not exceed $d\tau_T/d\ln(t) = 
n_e \sigma_T ct \sim 10-100$.  

By contrast, calculations of more dilute relativistic plasmas,
such as Blazar jets, are complicated by uncertainty in the input spectrum of relativistic particles, and
the mechanism by which they are accelerated.  These uncertainties partly disappear in the problem examined here,
because relativistic particles cool much too rapidly to contribute significantly to the Comptonization process,
and because hard photons lose energy by recoil.  

Seed relativistic particles Compton cool on a timescale
\be
t_C(\gamma) \sim {t\over \ell_{\rm th}\gamma}.
\ee
Supposing that a fraction $f_{\rm rel}$ of the radiation energy is supplied by these
particles (with the remainder by gradual heating of thermal particles), the time-averaged energy density in
relativistic particles is
\be
U_{\rm rel} \sim f_{\rm rel}{t_C\over t}U_{\rm th},
\ee
and the time-averaged compactness is
\be
\ell_{\rm rel} = {\sigma_T U_{\rm rel} ct\over m_ec^2} \sim {f_{\rm rel}\over\gamma} \ll 1 \ll \ell_{\rm th}.
\ee
The equilibrium Compton parameter is
\be\label{eq:yCrel}
y_{\rm C,rel} \sim \gamma^2\sigma_T {U_{\rm rel}\over\gamma m_ec^2}ct \sim f_{\rm rel},
\ee
which is tiny compared with the Compton parameter of the thermal plasma.
As a result, the emergent spectrum is determined almost entirely by a competition between soft-photon 
emission and multiple Compton scattering by thermal particles.  

We therefore focus on thermal cyclo-synchrotron emission and absorption.   The mean energy of the pairs is in a range, 
$\sim (0.05-0.2) m_ec^2$, where an exact calculation of the emission spectrum is required.  The details are 
reviewed in Appendix \ref{s:appA}, and the result shown in Figure \ref{fig:cyclotron}.  This emission channel 
dominates if the magnetic energy density exceeds the thermal energy density.  Other soft photon sources 
(bremsstrahlung and double Compton) are included for completeness.

\begin{figure}[h]
\epsscale{1.15}
\plotone{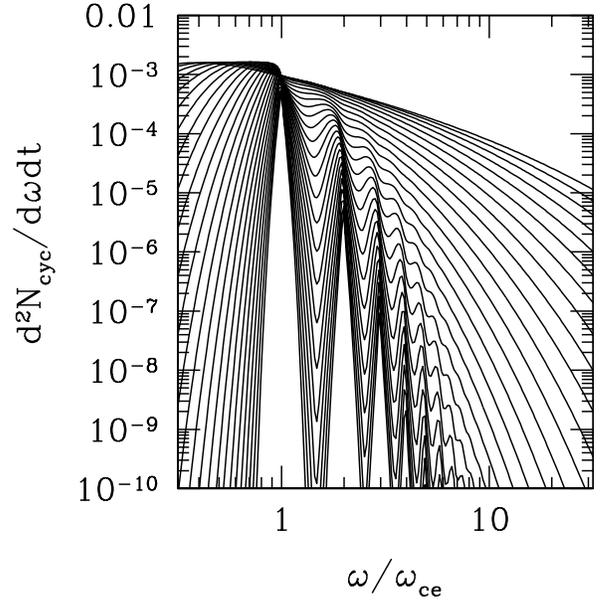}
\caption{Classical cyclo-synchrotron emission from electrons (positrons) with a relativistic Boltzmann distribution
(\ref{eq:relBol}) and an isotropic distribution of pitch angles.  Average rate per particle.  Temperature varies in increments 
$\Delta{\rm log}_{10}(T_e/m_ec^2) = 0.1$ from $T_e = 10^{-2.5} m_ec^2 = 1.6$ keV up to $T = m_ec^2 = 511$ keV (bottom to top).  
Note that, classically, $d^2N_{\rm cyc}/d\ln\omega dt$ depends on $B$ only via the ratio of wave frequency $\omega$
to cyclotron frequency $\omega_{\rm ce}$.  Integrating $\hbar\omega\,d^2N_{\rm cyc}/d\omega dt$ over $\omega$ reproduces 
the synchrotron power.}
\label{fig:cyclotron}
\vskip .2in
\end{figure}

\subsection{Plasma Dynamics}\label{s:dynam}

In addition to these regulating effects, we find that expansion, combined with continued 
heating {\it and} photon creation, has a significant regulating effect on the low-frequency
spectrum.  A sharp thermal peak formed in an initial
thermalization event at very high compactness is noticeably reduced if the plasma
expands and there is a continuing upward flux of soft photons.

We choose a simple heating model representing a conical jet expanding at a constant Lorentz factor.
Then the proper energy density scales with radius $r$ and bulk-frame time coordinate\footnote{We
work in the co-moving frame in Sections \ref{s:pplas}-\ref{s:multscatt}, and
label co-moving flow parameters with a $^\prime$ only to avoid ambiguity.}  
 $t \sim r/\Gamma c$ as $B^2/8\pi \sim t^{-\delta}$, where $\delta = 2$.
Heat is injected at a rate $t dU_{\rm th}/dt = \delta\cdot U_{\rm th,0}(t/t_0)^{-\delta}$ per unit volume, where $U_{\rm th,0}$
is the initial thermal energy density (in photons, rest energy of pairs, and thermal energy of all 
material particles).  Parameterizing 
\be
U_{\rm th} = f_{\rm th} {B^2\over 8\pi},
\ee
we explore the regime of moderately strong magnetization, $f_{\rm th} = 10^{-1}-1$.

In the Kompaneets calculations, we must start a 
simulation with a finite energy density so that the seed electrons do not overheat
(become relativistic). Then as time evolves
\ba\label{eq:uthvst}
U_{\rm th}(t) &=& (t/t_0)^{-4\delta/3} U_{\rm th,0} + t^{-4\delta/3} \int^t_{t_0} d\tilde t\, \tilde t^{4\delta/3} {dU\over d\tilde t}\nn
&=& \left[3 - 2(t/t_0)^{-\delta/3}\right] (t/t_0)^{-\delta}U_{\rm th,0}.
\ea
In a given simulation, the fraction of the final thermal energy that results from distributed
heating is\label{eq:fdis}
\be
f_{\rm dis} \simeq 1 - {1\over 3}(t/t_0)^{-\delta/3}.
\ee

The simplest, and perhaps most generic, form of heating involves the damping of hydromagnetic turbulence.
Small-scale irregularities in the flow could be triggered by ideal or resistive MHD instabilities; or by 
an interaction of the magnetized jet material with denser baryonic material.  While almost all work on 
magnetic reconnection focuses on non-ideal effects near current sheets, it should be emphasized that 
reconnection has simpler effects, by changing the topology of the magnetic field and converting magnetic 
energy to bulk kinetic energy.  In the presence of a dense photon gas and at moderate scattering depths, 
Compton drag effectively damps the differential motion of all particle species with respect to the 
mean flow \citep{thompson94}.

Nonetheless, Alfv\'en waves can easily cascade to higher wavenumbers before damping by Compton drag.
In a strongly magnetized plasma, they anisotropically heat the electrons and positrons:  either because
the wave current becomes charge starved \citep{tb98}; or, if the particle density is high enough, by Landau damping on the motion
of $e^\pm$ parallel to the magnetic field (\citealt{thompson06}; see
\citealt{qg99} for related work on baryonic plasmas).  In Sections \ref{s:rec} and  \ref{s:cs} we consider 
the implications of our thermal plasma solution for reconnection and charge-starvation effects.  

\subsection{Formation of Quasi-Thermal Peak, $T_e, T_\gamma \ll m_ec^2$}\label{s:highcomp}

We evolve the photon spectrum in the diffusion approximation, including stimulated and recoil
effects.  The Kompaneets equation for the photon occupation number $N(\omega)$ is
\ba\label{eq:komp}
&&{\partial N\over \partial t} + \dot\omega{\partial N\over\partial\omega} \;=\; 
{\partial N\over\partial t}\biggr|_{\rm cyc} + {\partial N\over\partial t}\biggr|_{\rm ff} + 
   {\partial N\over\partial t}\biggr|_{\rm dC} + {\partial N\over\partial t}\biggr|_{\rm drift}\nn
  &+& n_e\sigma_T c{1\over \omega^2}{\partial\over\partial\omega}\left\{\omega^3
\left[{T_e\over m_ec^2}{\partial N\over\partial\ln\omega} + {\hbar\omega\over m_ec^2}N(1+N)\right]\right\}.\nn
\ea
Here $n_e \equiv n_{e^+} + n_{e-}$ is the proper density of scattering particles,
and the random particle motion is described by a single temperature.  
As is usual, stimulated scattering makes a net contribution 
only to the recoil term, but cancels from the Doppler upscattering term.  

The source terms include cyclo-synchrotron emission and absorption,
and (non-magnetic) free-free and double-Compton emission and absorption; 
they are reviewed in Appendix \ref{s:appA}. 
We solve (\ref{eq:komp}) in flux-conservative form, meaning that the variable evolved is 
${\cal N}(\omega) \equiv \omega^2 N(\omega)$.  The equation is solved by the method of lines, 
with a second-order differencing in frequency and fourth-order Runge-Kutta evolution in time.  

Both static and expanding plasmas are considered.  In the expanding case, it is essential to consider
{\it large} expansion factor $\gtrsim 10-10^2$.  For example, a jet propagating through the inner core 
of a Wolf-Rayet star may encounter resistance over such a range of radius, and continue to interact 
with entrained stellar material even beyond its photosphere (see Section \ref{s:poyntingjet}).   
Adiabatic expansion in the outflow rest frame corresponds to a dilution $n_e \propto t^{-\delta}$, 
with $\delta = 2$.   Then $\dot\omega = -(\delta/3)\omega/t$ on the left-hand side of (\ref{eq:komp}).
The calculation is stopped if the outflow optical depth $d\tau_T/d\ln(t) < 1$ 
(which generally happens only after heating is turned off).  

For completeness, we also include the relative drift
between photons and pairs, which develops at low optical depth in a spreading jet
\citep{russo13a,russo13b}.  This has the effect of raising the frequency of the photons
as measured in the frame of the particles, and is implemented with the term
\be\label{eq:ndrift}
{\partial N\over \partial t}\biggr|_{\rm drift} = -{c\over d\tau_T/d\ln(t)}\omega{\partial N\over\partial\omega}.
\ee

Multiplying equation (\ref{eq:komp}) by $\hbar\omega$ and integrating over the photon phase
space, one finds, as usual
\ba\label{eq:dugamdt}
{dU_\gamma\over dt}\biggr|_C &=& {1\over m_ec^2}\left[4T_e -
{\langle(\hbar\omega)^2\rangle + \langle(\hbar\omega)^2 N\rangle\over\langle\hbar\omega\rangle}\right]\sigma_T n_e c U_\gamma\nn
&=& 4 {T_e-T_C\over m_ec^2}\sigma_T n_e c U_\gamma.
\ea
Here $T_C$ is the Compton temperature.  

\subsection{Validity of a Single Temperature}

In what follows, we restrict the particle distribution to relativistic Boltzmann,
\be\label{eq:relBol}
{dn_e\over d\gamma} = {n_e\over \Te K_2(1/\Te)} \beta\gamma^2 e^{-\gamma/\Te};\quad\quad \Te\equiv
{T_e\over m_ec^2},
\ee
with temperature $\Te \lesssim 0.2$, and an isotropic distribution of pitch angle.

In this situation, the timescales for heating and cooling of the pairs are both very short and in near balance,
with cooling being primarily by Compton scattering of the thermal photon field:
\ba\label{eq:thc}
{t_{\rm heat}\over t} &\sim& {3n_eT_e/2 \over f_{\rm th} B^2/8\pi} = {3\Te d\tau_T/d\ln(t)\over 2f_{\rm th}\ell_B};\nn
{t_{\rm cool}\over t} &\sim& {3m_e c\over 4 \sigma_T U_\gamma t} = {3\over 4\ell_{\rm th}}.
\ea
A thermal distribution presupposes the exchange of energy between the charged particles on a shorter timescale.    
In the absence of such a process, the pair energy distribution will peak around 
\be\label{eq:eeq}
E_{e,\rm eq} = {3m_ec^2\over 4d\tau_T/d\ln(t)}.
\ee
A monoenergetic distribution is approached in the idealized case of uniform heating and cooling.

An important point is highlighted by equation (\ref{eq:eeq}):  the bulk of the pair population remains sub-relativistic 
during a heating episode, $m_ec^2 > T_e > T_C$,  only if the plasma starts off at a large scattering depth -- even if 
the outflow is still very compact.  As we discuss in Section
\ref{s:summary}, this provides a distinction between an early heating phase (before jet breakout) when the low-frequency part
of the GRB spectrum is formed, and a secondary phase (after breakout) that produces the high-energy tail.

Coulomb scattering is relatively slow in this context, due to the low particle density:  
\be\label{eq:tcoul}
{t_{\rm coul}\over t} \sim {\Te^{3/2}\over d\tau_T/d\ln(t)}.
\ee
In a continuously heated pair plasma with $f_{\rm th} \sim 0.1$, we find $d\tau_T/d\ln(t) \sim 10$-$10^2$ and $\Te \sim 0.05$-$0.1$ 
(Section \ref{s:spectrum}), in which case $t_{\rm coul}/t_{\rm heat} \sim \ell_{\rm th} \Te^{3/2} 
[d\tau_T/d\ln(t)]^{-2} \sim 10^{-5} \ell_{\rm th}$.  The effectiveness of Coulomb scattering at thermalizing
the pairs varies over the range of $\ell_{\rm th}$ considered, $\sim 10^2 - 10^8$.

The rapid emission and re-absorption of cyclotron photons will 
thermalize the pairs on a shorter timescale than (\ref{eq:tcoul}) (e.g. \citealt{ghisellini99}).  
In spite of this, it is possible in some circumstances to maintain a significant temperature anisotropy,
because the magnetic pressure greatly exceeds the thermal pressure of the pairs.  For example, an Alfv\'enic 
cascade mainly excites the motion of the pairs parallel to the magnetic field \citep{thompson06}.   

Before a heating episode,
both the perpendicular and parallel temperatures $T_\perp, T_\parallel$ quickly adjust to the 
photon color temperature $T_c$ near the cyclotron fundamental.  Given a photon energy density $U_\omega
= T_c \omega^2/\pi^2 c^3$, the pairs feel a drag force
\be
m_e {dV_\parallel\over dt} \sim {V_\parallel\over c} \int d\omega \sigma_{\rm res}(\omega) U_\omega,
\ee
where 
\be
\sigma_{\rm res} \sim {2\pi^2 e\over B}\omega \delta\left(\omega - \omega_{\rm ce}\right)
\ee
is the absorption cross section at the fundamental \citep{canuto71}.

Then the time for the parallel and perpendicular temperature to adjust to $T_c$ is
\be
{t_{\rm cyc}\over t} \sim {V_\parallel\over dV_\parallel/dt} = {1\over 6\ell_B} \left({T_c\over m_ec^2}\right)^{-1},
\ee
where $\ell_B$ is the magnetic compactness (\ref{eq:comp}).  This is shorter than the Compton timescale 
(\ref{eq:thc}), $t_{\rm cyc} \lesssim t_{\rm cool}$, if $f_{\rm th} \lesssim (9/2)(T_c/m_ec^2)$.  That is the case
in our calculations with $f_{\rm th} \sim 0.1$, which yield $T_c \sim 0.05\,m_ec^2$.

A qualitatively different conclusion is reached in a later phase of the magnetized fireball,
when the high-frequency spectral tail must be generated.  As the magnetic field and the entrained
pairs are accelerated away from the engine by a combination of photon pressure and the Lorentz force,
the thermal peak drops by a factor $\sim 10-10^2$ in the co-moving frame, to $T_c \sim 10^{-3} m_ec^2$. 
Then $f_{\rm th}$ rises as the magnetic field dissipates.  If $T_\perp$ is comparable to $T_c$ at the
beginning of the heating episode, then it remains much smaller than $T_\parallel$.   Perpendicular
heating is mainly by non-resonant Compton scattering, with interesting consequences for the angular
pattern of the scattered radiation \citep{thompson06}.


We describe the volumetric heating of the pairs via
\be\label{eq:duedth}
{dU_e\over dt}\biggr|_{\rm heat} = {f_{\rm th}\over t_{\rm tot}}{B^2\over 8\pi}
= \ell_{\rm th}{m_ec^2\over \sigma_T c t_{\rm tot}^2}.
\ee
The compensating change in energy by Compton scattering is the negative of (\ref{eq:dugamdt}).  
We find that Compton equilibrium is only approximately 
maintained during heating:  $(T_e-T_C)/m_ec^2 \sim \tau_T^{-1}$.  There is a rapid approach to equilibrium
after the heating turns off, due to the very high compactness.  

Even though the photon field cannot be defined by a single temperature if its low-frequency
spectrum is flat, the high-frequency spectrum does maintain a thermal form at high compactness:
$dU_\gamma/d\omega \propto \omega^n e^{-\hbar\omega/T_\gamma}$, with $n \simeq 3$ and
$T_\gamma$ close to $T_e$.  For the purposes of constructing
simple analytic models of the expanding pair plasma, we will sometimes use 
\ba\label{eq:specgrb}
  {1\over\hbar} {dU_\gamma\over d\omega} &=& K \left({\omega_t\over c}\right)^3 e^{-\hbar\omega/T_e} \quad (\omega < \omega_t);\nn
&=& K\left({\omega\over c}\right)^3 e^{-\hbar\omega/T_e} \quad (\omega > \omega_t),
\ea
which matches smoothly at $\omega_t = 3T_e/\hbar$ with coefficient
\be
K = 0.083 {U_\gamma \over a_{\rm SB}T_e^4}.
\ee

\subsection{Pair Creation and Annihilation}\label{s:paireq}

The density of pairs evolves according to annihilation and creation by photon collisions,
$e^ + e^- \leftrightarrow \gamma + \gamma$.  In a warm plasma, $\Te \lesssim 0.1$, the annihilation
cross section can be approximated by $\langle \sigma_{\rm ann} |{\bf v}_{e^+}-{\bf v}_{e-}|\rangle 
\simeq {3\over 8}\sigma_Tc$, so that
\be\label{eq:dnedt}
{dn_e\over dt}\biggr|_{\rm ann} \simeq {3\over 4}\cdot n_{e^-}n_{e^+}\sigma_T c.
\ee
The calculation of the rate of pair creation, given by equation (\ref{eq:dnedtgg}), involves 
convolutions over the photon distribution function, and is reviewed in Appendix \ref{s:appB}.
In some calculations we include an additional source of cold pairs, derived from non-thermal 
relativistic particles, through a parameterized term (\ref{eq:dnedtnth}) that is described
in Section \ref{s:nth}.  In all,
\be\label{eq:dnedttot}
{dn_e\over dt} = {dn_e\over dt}\biggr|_{\rm ann} + {dn_e\over dt}\biggr|_{\gamma\gamma} + 
{dn_e\over dt}\biggr|_{\rm nth}.
\ee

If both pairs and photons follow thermal distributions with the same temperature $T$, then
their densities have a simple relation.  The chemical potentials are $\mu_\gamma = 
\mu_{e^+} = \mu_{e^-} = \mu$, all vanishing in a black-body gas.  Further restricting to
$T \ll m_ec^2$, we have
\be
n_e = n_{e^+} + n_{e^-} = 2g_e\left({m_e T\over 2\pi\hbar^2}\right)^{3/2}e^{(\mu - m_ec^2)/T}
\ee
and
\be
{1\over \hbar}{dU_\gamma\over d\omega}\biggr|_{\hbar\omega = m_ec^2} = 
{g_\gamma\over 2\pi^2}\left({m_e c\over\hbar}\right)^3e^{(\mu-m_ec^2)/T},
\ee
where $g_e = g_\gamma = 2$ is the number of spin degrees of freedom.   Hence
\be\label{eq:ne_eq}
n_{e,{\rm eq}} = (2\pi)^{1/2}\left({T\over m_ec^2}\right)^{3/2}
    {1\over\hbar} {dU_\gamma\over d\omega}\biggr|_{\hbar\omega = m_ec^2}.
\ee
Written in this way, the result is insensitive to the numerically determined photon
temperature, and to the shape of the spectrum below the peak.  

\begin{figure}[h]
\epsscale{0.9}
\plotone{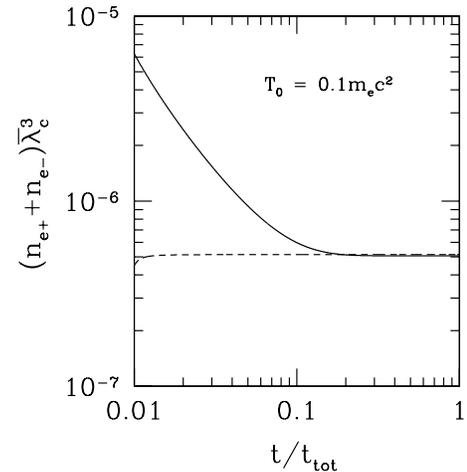}
\caption{Relaxation of pair density to equilibrium value (\ref{eq:ne_eq}) 
in a black-body gas of initial temperature $T_0 = 0.1m_ec^2$,
and an initial excess of pairs.  Normalization:  (reduced Compton wavelength)$^{-3}$.}
\label{fig:pairtest}
\end{figure}

A test of the pair-creation algorithm is provided by a black-body gas 
interacting with a thermal pair gas that has an initial density different from the
equilibrium value (\ref{eq:ne_eq}): see Figure \ref{fig:pairtest}.

\begin{figure}[h]
\epsscale{1}
\plotone{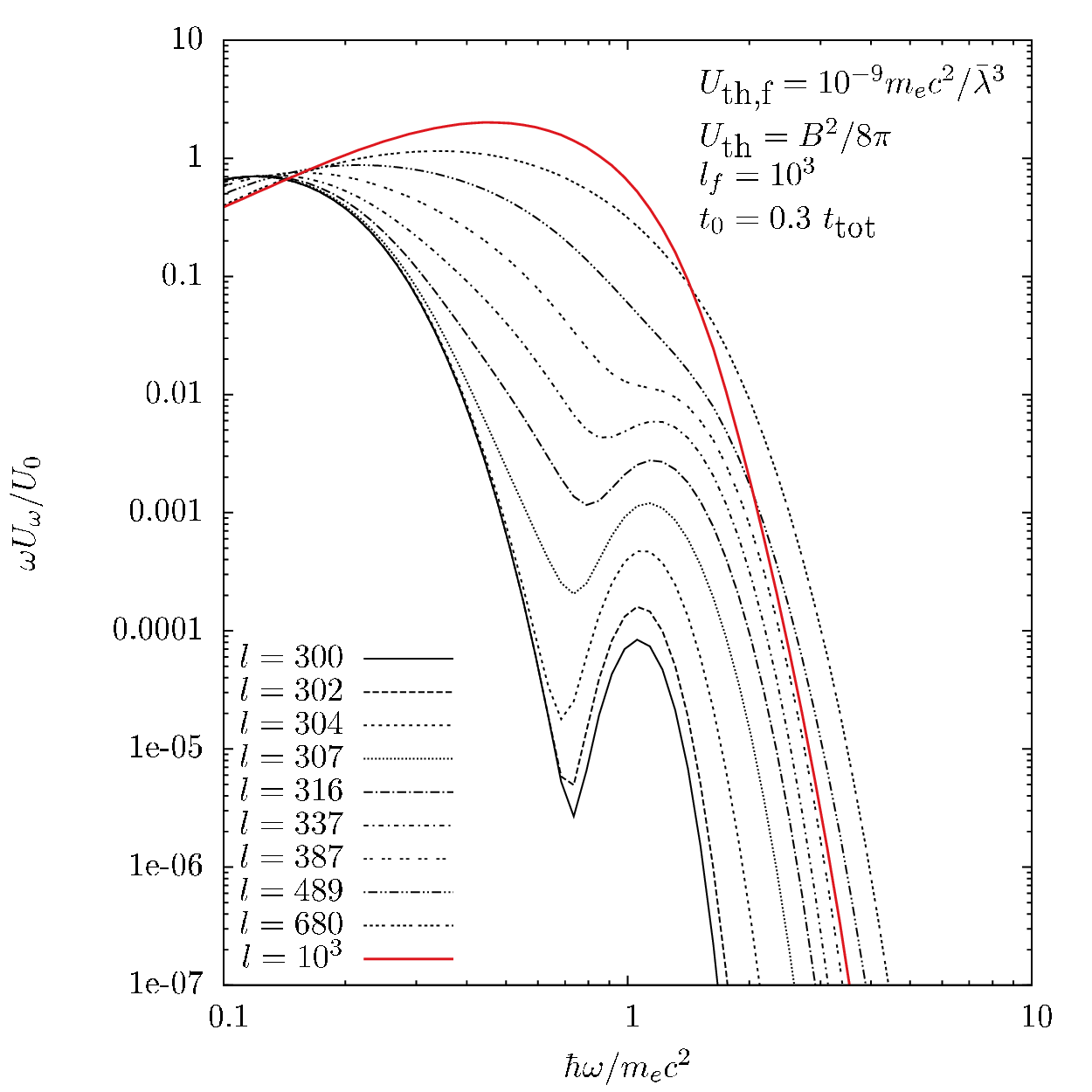}
\caption{Development of the photon spectrum at final compactness $10^3$, using a full
relativistic, kinetic treatment of Compton scattering and pair creation and annihilation.
Steady heating, no expansion, with an initial excess of pairs leading to a prominent
annihilation line during the early evolution.  The line becomes insignificant 
at $\ell_{\rm th} \gtrsim 10^3$.}
\label{fig:kinetic}
\vskip .1in
\end{figure}

\subsection{Supplemental Non-thermal Particle Injection}\label{s:nth}

Relativistic pairs injected with a small power $f_{\rm rel} U_{\rm th}/t$
contribute a small Compton parameter (\ref{eq:yCrel}) compared with the thermal pairs,
but can have a larger influence
on the scattering depth.  Given a thermal photon peak energy $E_{\rm pk}$,
pairs injected with a Lorentz factor $\gamma \gtrsim (m_ec^2/E_{\rm pk})^{1/2} \sim 2-3$ 
cool mainly by emitting hard photons of energy $> m_ec^2$.  

The thermal pair population could briefly be raised above this energy threshold.
For example, it is possible that a significant fraction of the magnetic field energy
density is transferred to the pairs (by a simple cascade process) on a timescale $t_{\rm heat} < t$
and within a small volume $\lesssim (c t_{\rm heat})^3$.  Then each particle is heated 
and cooled at an average rate
\be\label{eq:heat}
{dE_e\over dt} = {B^2\over 8\pi n_e t_{\rm heat}} - {4\over 3}(\gamma^2-1)\sigma_T c U_\gamma.
\ee

Synchrotron cooling can be neglected here, because i) $\gamma$ is low enough
that the synchrotron emission is self-absorbed; ii) $f_{\rm th} = 8\pi U_\gamma/B^2$ is perhaps 
as small as $\sim 0.1$, but not much smaller; and iii) the simplest heating mechanism, a cascade
of Alfv\'en waves, creates a strongly anisotropic particle distribution with particle motion primarily
along the magnetic field \citep{thompson06}.  (This anisotropic distribution is insensitive to
cyclotron and firehose instabilities, given the extremely small value of the plasma $\beta = 8\pi n_e kT_e/B^2$.)

At a high radiation compactness, the particles reach an equilibrium Lorentz factor
\be\label{eq:gamsqeq}
\gamma^2-1 = {3\over 4f_{\rm th} d\tau_T/d\ln(t)} \left({t_{\rm heat}\over t}\right)^{-1}.
\ee
Given $f_{\rm th} \sim 0.1$ and $d\tau_T/d\ln(t) \sim 30$, we see that $t_{\rm heat}$
must be shorter than $\sim 0.03\, t$ for $\gamma^2$ to exceed $m_ec^2/E_{\rm pk} \sim 10$.  This
may be uncomfortably short for a Kelvin-Helmholtz instability driven by velocity shear on 
a lengthscale $\sim ct \sim r/\Gamma$, but not for impulsive bursts of magnetic reconnection.  

It is straightforward to incorporate this additional source of pairs into the Kompaneets calculation 
through an additional source term in equation (\ref{eq:dnedttot}).  The injection of pair rest energy,
after averaging over the plasma volume, is described by a single parameter,
\be\label{eq:dnedtnth}
m_ec^2 {dn_e\over dt}\biggr|_{\rm nth}  = f_{\rm nth} {dU_e\over dt}\biggr|_{\rm heat}\quad
(f_{\rm nth} < f_{\rm rel}),
\ee
which depends on the energy spectrum and luminosity of the non-thermal particles.   After rapidly
cooling off the thermal photons, these pairs annihilate at the rate (\ref{eq:dnedt}), leaving
a net optical depth $d\tau_T^{\rm nth}/d\ln(t) \sim (16f_{\rm nth} \ell_{\rm th}/3)^{1/2}$.
This dominates the optical depth of the thermally created pairs if
\be
f_{\rm nth} > {3\over 16 \ell_{\rm th}} \left[{d\tau_T^{\rm th}\over d\ln(t)}\right]^2.
\ee
A larger Compton parameter can now be maintained, leading to a more strongly peaked photon spectrum
and a lower $E_{\rm pk}$.

\subsection{Validity of the Kompaneets Equation at High Energies}

The Kompaneets equation obviously cannot be used to evolve the $\sim 511$ keV annihilation
feature in a cold pair gas.  However, we are considering temperatures low enough that 
$n_e m_ec^2$ is a tiny fraction (typically much less than a percent) of $U_\gamma$, and the
Thomson scattering depth is moderately large.  A demonstration that the annihilation line is weak
at high $\ell_{\rm th}$ is provided by a full kinetic calculation of the photon and
electron/positron distributions.  (This code will be described in a separate
publication.)  The development of the spectrum in a non-expanding box up to a final
compactness $\ell_{\rm th} = 10^3$ is shown in Figure \ref{fig:kinetic}.  The annihilation
feature indeed becomes negligible; our calculations typically focus on yet higher $\ell_{\rm th}$.

We must also consider whether the calculated
photon distribution is accurately described by the solution to the Kompaneets
equation (\ref{eq:komp}) near the pair-creation threshold.  As long as the spectrum has a well-defined
thermal peak, the solution to (\ref{eq:komp}) is the same as the thermal equilibrium solution,
with the high-frequency expansion $dU_\gamma/d\omega \sim e^{-\hbar\omega/T}$.  The solution is
therefore valid even though the approximation of Thomson scattering breaks down at $\hbar\omega \sim m_ec^2$.  

A photon temperature variable is easily extracted from a distribution 
of the form $dU_\gamma/d\omega = K \omega^\beta e^{-\hbar\omega/T_\gamma}$.  Then
$\langle\hbar\omega\rangle = \beta T_\gamma$, $\langle(\hbar\omega)^2\rangle = 
\beta(\beta + 1)T_\gamma^2$, and inverting gives
\be\label{eq:Tgapproxa}
T_\gamma = {\langle(\hbar\omega)^2\rangle\over\langle\hbar\omega\rangle} - 
     \langle \hbar\omega\rangle.
\ee
A simple check is provided by a Wien distribution, $\beta =3$, for which
$\langle \hbar\omega\rangle = 3T$, $\langle (\hbar\omega)^2\rangle = 12T^2$.  
In a GRB-like spectrum (\ref{eq:specgrb}) with $\beta \simeq 0$, one has instead
\be\label{eq:Tgapprox}
T_\gamma \simeq {\langle(\hbar\omega)^2\rangle\over\langle\hbar\omega\rangle}.
\ee

\subsection{Temperature Evolution of the Pairs}

The pairs exchange energy both with thermal photons, and (a much smaller number of) annihilation 
photons of a somewhat higher frequency.  We write
\ba\label{eq:dTedt}
{dT_e\over dt} &\;=\;& {dT_e/d\langle K_e\rangle\over n_e}\biggl({dU_e\over dt}\biggr|_{\rm heat} -
{dU_\gamma\over dt}\biggr|_C + {dU_\gamma\over dt}\biggr|_{\rm ann}\biggr)\nn
&& - {\delta\over 3t}{dT_e\over d\ln \langle p_e\rangle},
\ea
where the first two terms on the right-hand side are given by equations (\ref{eq:dugamdt}) and 
(\ref{eq:duedth}), and $\langle p_e\rangle$, $\langle K_e\rangle$ are the mean thermal momenta 
and kinetic energies.  We have also added to (\ref{eq:dugamdt}) a contribution from
the net rate of change of photon energy due to pair creation and annihilation:
\be
{dU_\gamma\over dt}\biggr|_{\rm ann} = m_ec^2
\left({dn_e\over dt}\biggr|_{\rm ann} - {dn_e\over dt}\biggr|_{\gamma\gamma}\right).
\ee
The final term in (\ref{eq:dTedt}) represents adiabatic cooling.  We now comment on the signs
within the annihilation term.

In this situation, annihilation photons lose energy 
primarily by the Compton recoil off the colder electrons.    Therefore any energy put
into annihilation photons goes quickly into the kinetic energy of the pairs.  Although
energetic $e^\pm$ so created will return part of their energy to the photon field
by Compton scattering before equilibrating with the thermal pair population by Coulomb
scattering, the description of the pair distribution by a single temperature provides
a self-consistent way of accounting for the rest energy of pairs created and destroyed.  
To be consistent, we must account for the creation of $e^\pm$ by high-energy photon collisions
by subtracting their rest energy from the thermal energy of the existing particles.

\begin{figure} 
\epsscale{0.9}
\plotone{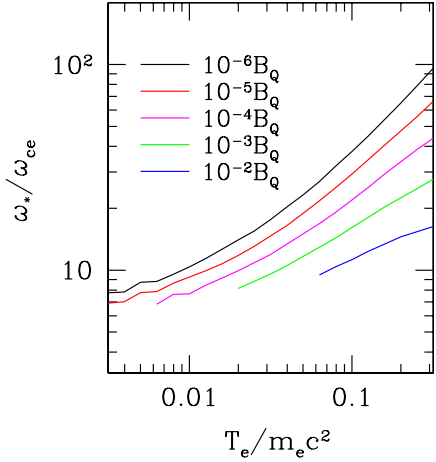}
\caption{Break frequency $\omega_* = m_*\omega_{\rm ce}$ separating the low-frequency Rayleigh-Jeans tail from the intermediate,
flat component of the spectrum.  Colors correspond to different magnetic energy densities,
normalized to the QED magnetic field $B_Q = 4.4\times 10^{13}$ G. Only temperatures exceeding $(B^2/8\pi a_{\rm SB})^{1/4}$ are plotted.
The irregularities at low $T_e$ reflect the harmonic structure of the emissivity.}
\label{fig:nstar}
\vskip .2in
\end{figure}

\section{Results of Compton Evolution}\label{s:spectrum}

We now present the results of a numerical solution of equations (\ref{eq:komp}), (\ref{eq:dnedttot}),
and (\ref{eq:dTedt}), starting with some semi-analytic considerations.

\begin{figure*}
\centerline{
\includegraphics*[width=0.5\hsize]{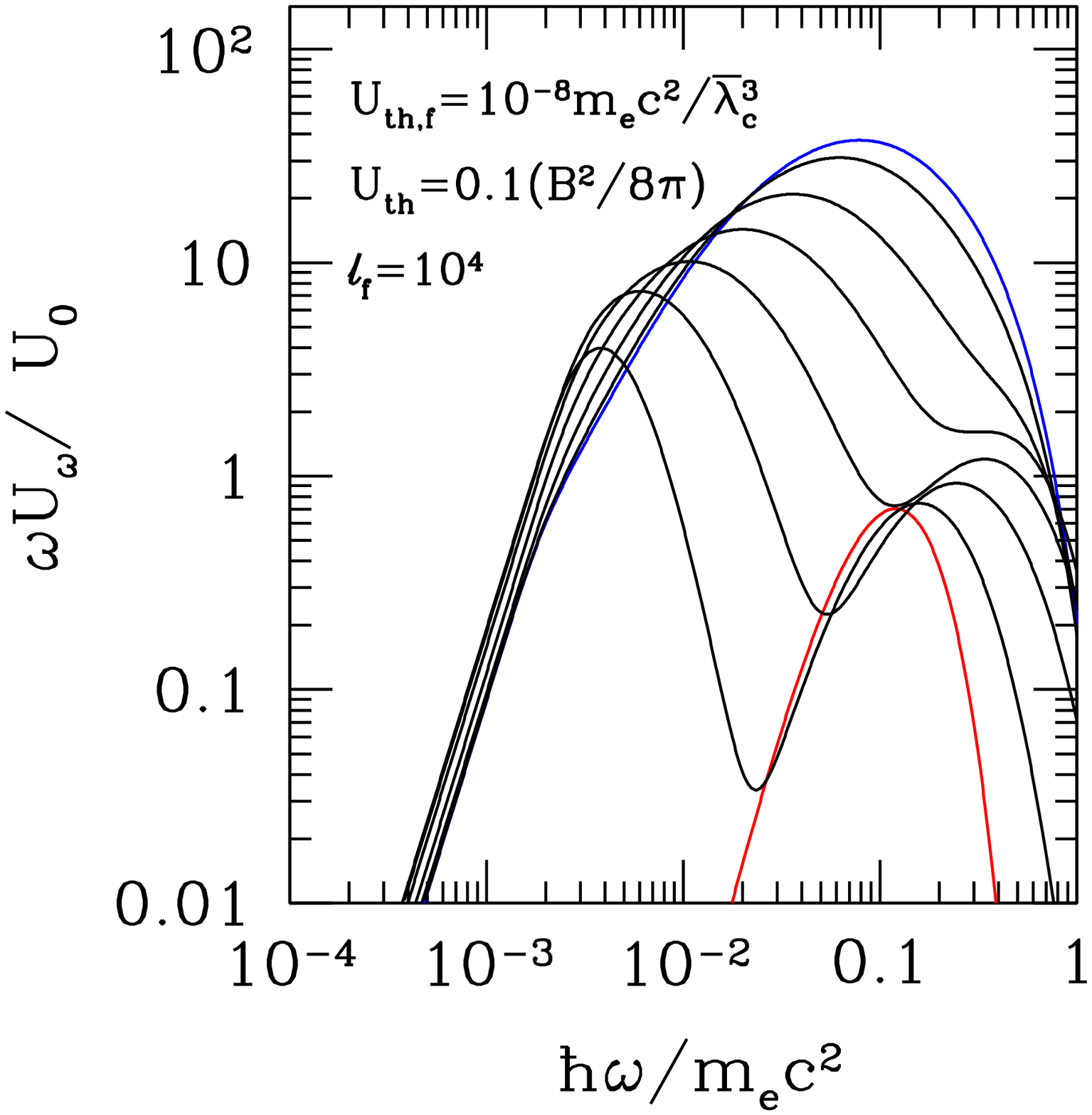}
\includegraphics*[width=0.5\hsize]{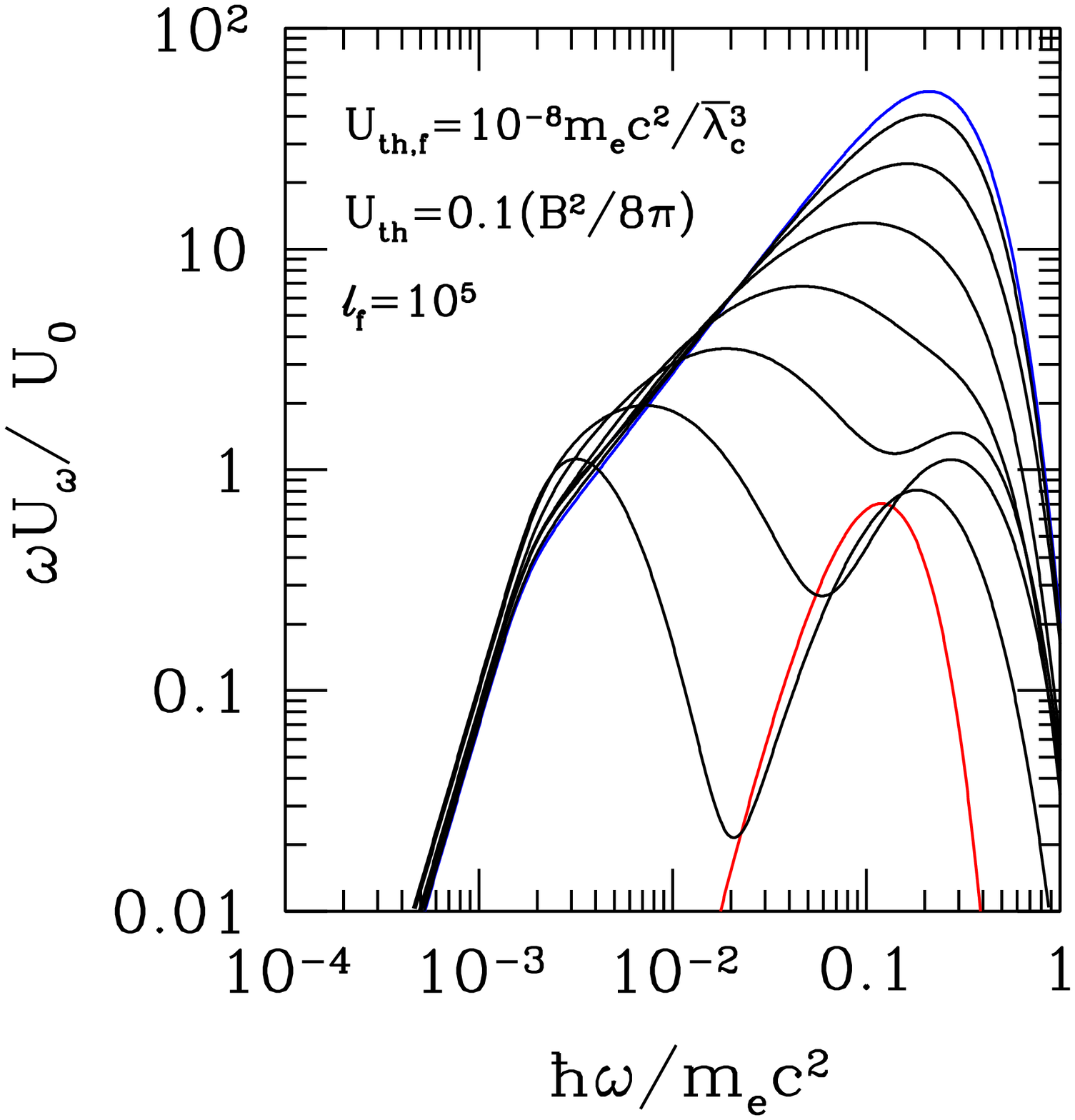}
}
\vskip .5in
\centerline{
\includegraphics*[width=0.5\hsize]{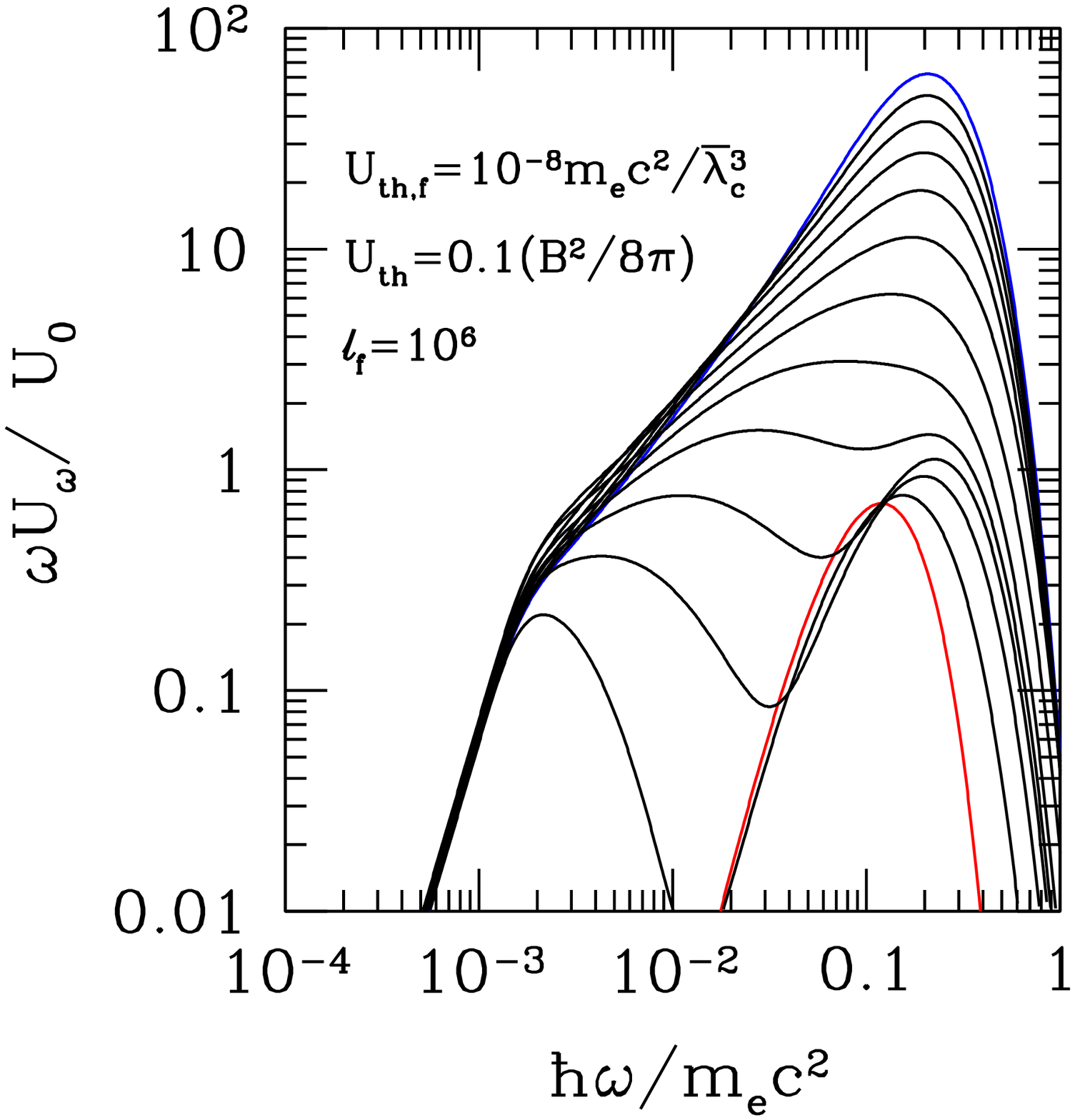}
\includegraphics*[width=0.5\hsize]{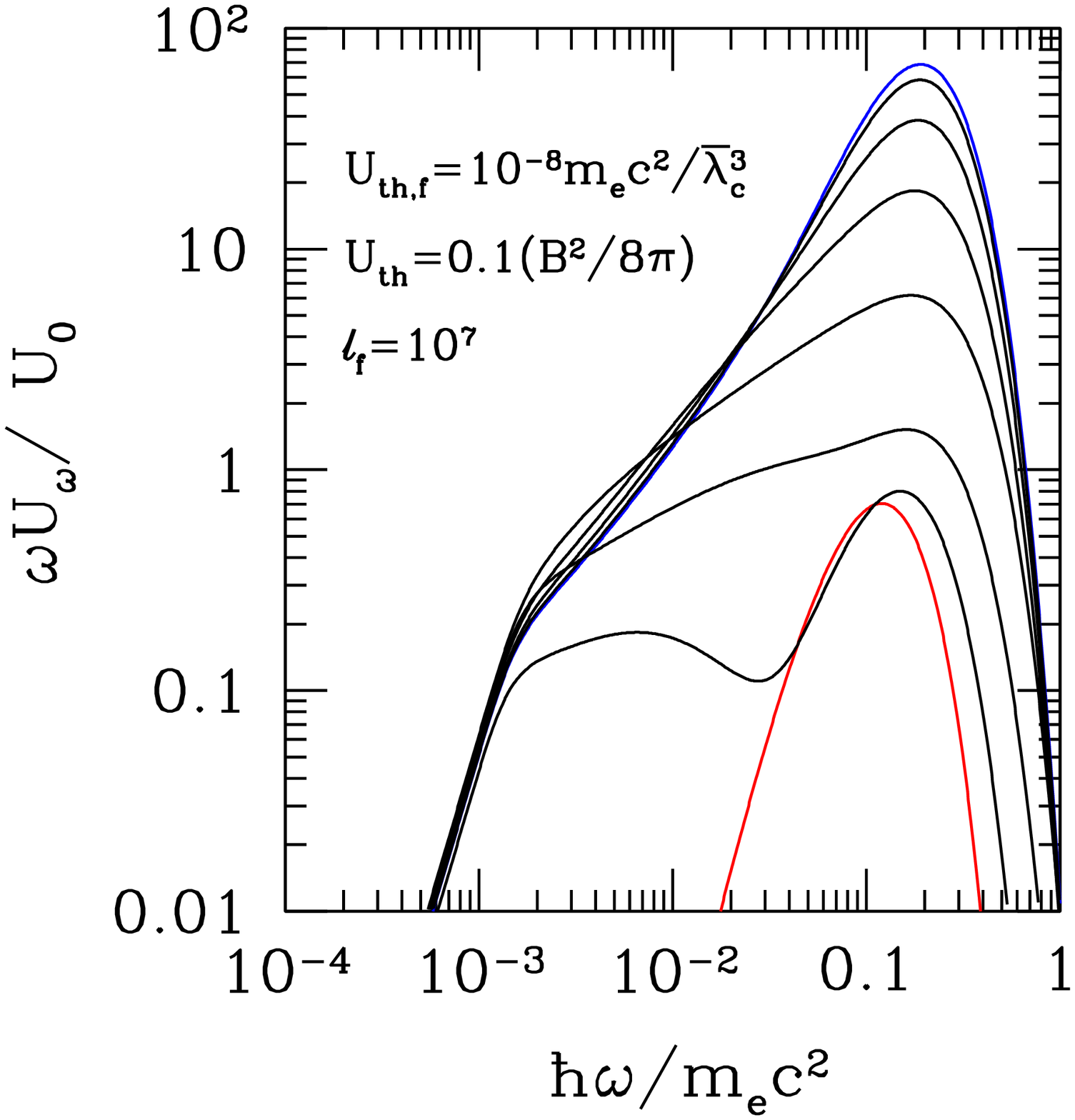}
}
\caption{Development of the photon spectrum in a fixed volume, in increments
$\Delta y_C = 0.25$, 0.4, 0.4, 1 (top left to bottom right) of the Compton parameter (\ref{eq:yC}).  Final energy density $U_{\rm th,f} = 
10^{-8}\,m_ec^2/\bar{\lambda}_c^3$, and final compactness 
$\ell_{\rm f} = \sigma_T U_{\rm th,f} ct_{\rm tot}/m_ec^2 = 10^4-10^7$ (top left to bottom right).  Blue line:
final spectrum.  Red: seed spectrum (Wien gas with temperature $T_0 = 0.03\,m_ec^2$ and
energy density $U_{\rm th,0} = 10^{-2}U_{\rm th,f}$).  Initial pair density $n_{e,0} = 10^{-1}U_{\rm th,0}/m_ec^2$.}
\label{fig:noexpand}
\end{figure*}
\begin{figure*}
\centerline{
\includegraphics*[width=0.5\hsize]{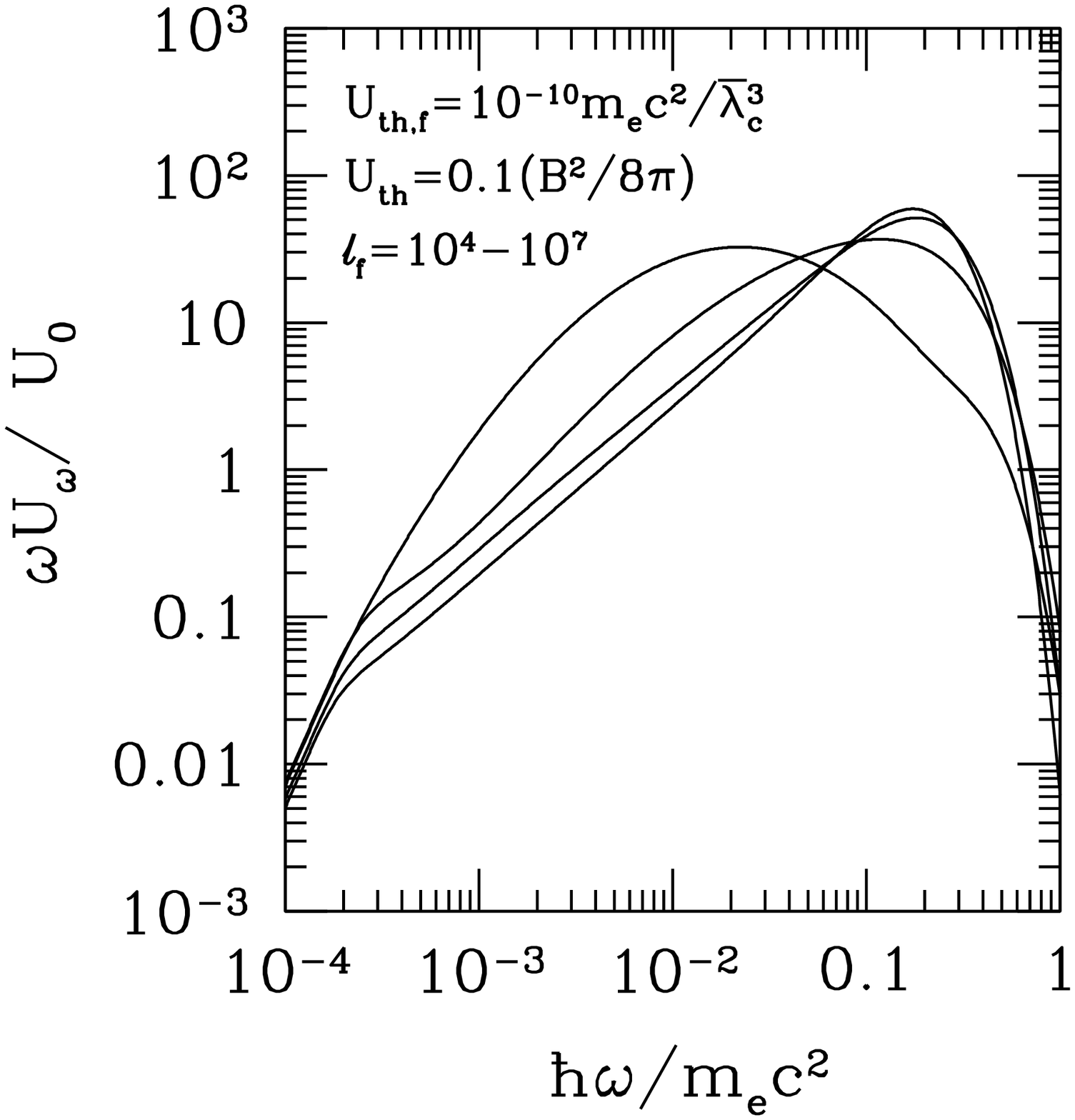}
\includegraphics*[width=0.5\hsize]{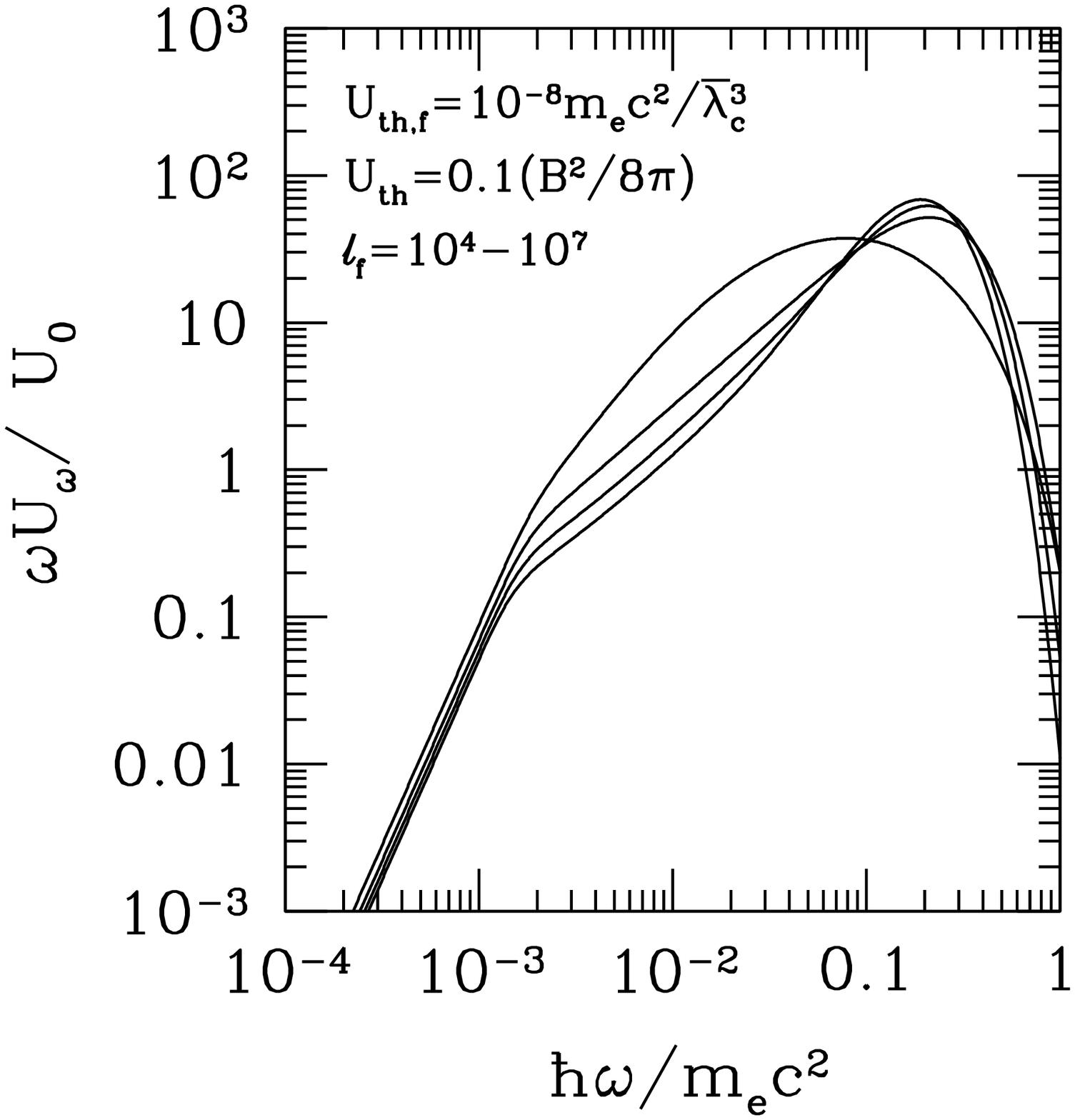}
}
\vskip .6in
\centerline{
\includegraphics*[width=0.5\hsize]{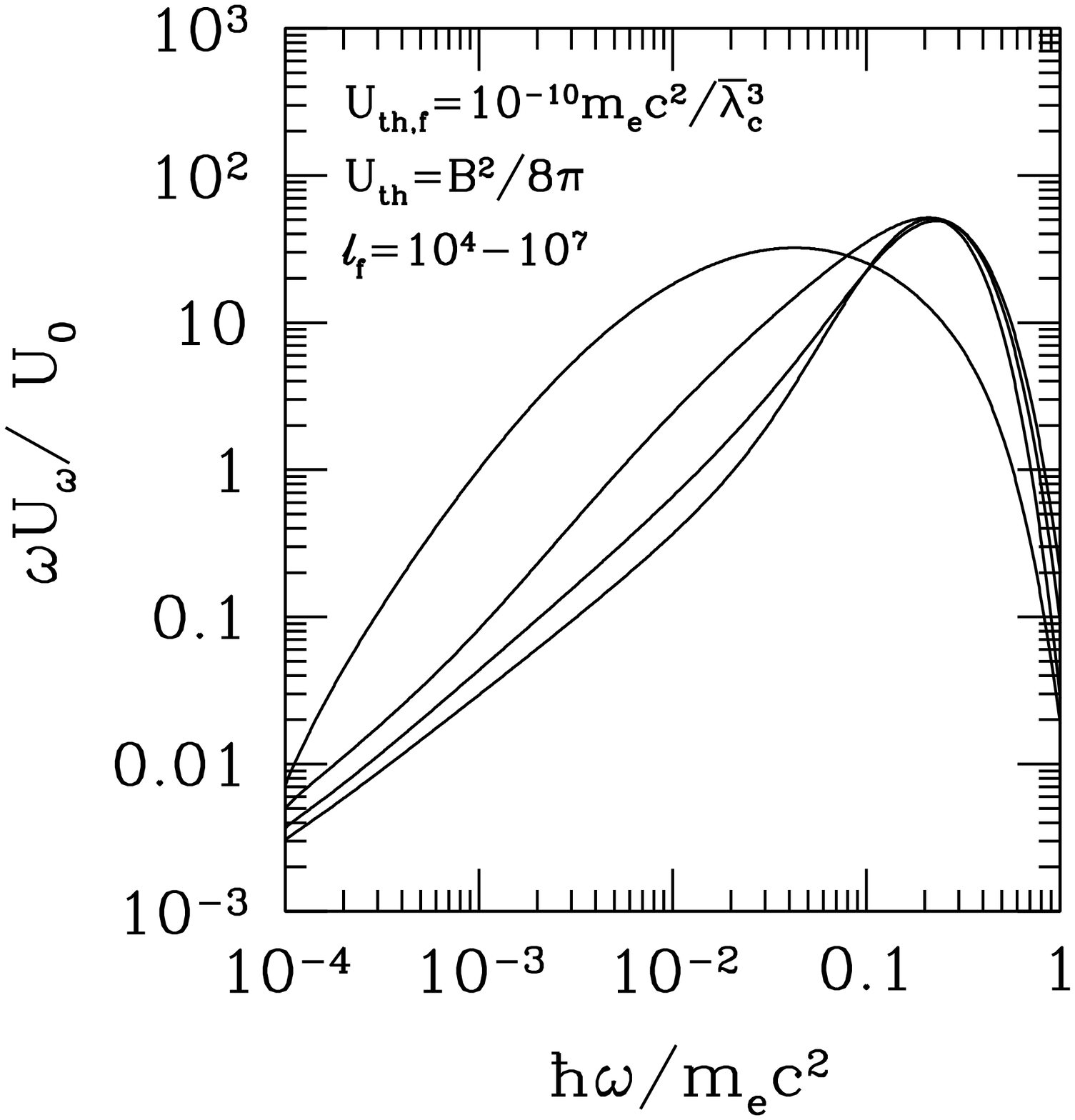}
\includegraphics*[width=0.5\hsize]{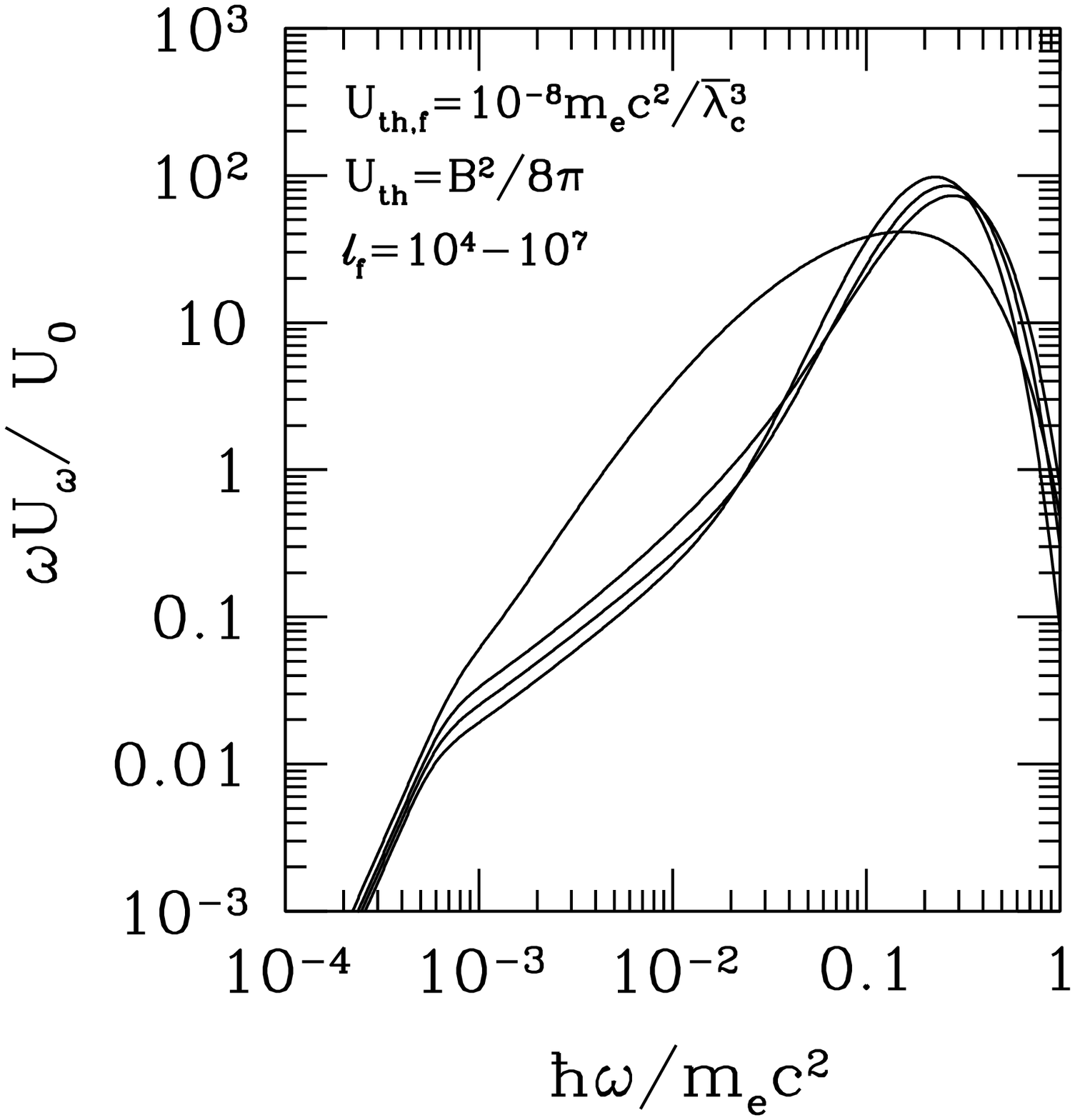}
}
\caption{Final photon spectrum, fixed volume, as a function of final compactness.  
Final energy density $U_{\rm th,f} = 10^{-10}$, $10^{-8}\,m_ec^2/\bar{\lambda}_c^3$ (left, right),
and magnetization $8\pi U_{\rm th}/B^2 = 0.1, 1$ (top, bottom).  The flat part
of the spectrum connects more directly to the thermal peak as the magnetization is raised - see equation (\ref{eq:epsthcr}).
}
\label{fig:noexpand2}
\end{figure*}

\begin{figure*}
\centerline{
\includegraphics*[width=0.5\hsize]{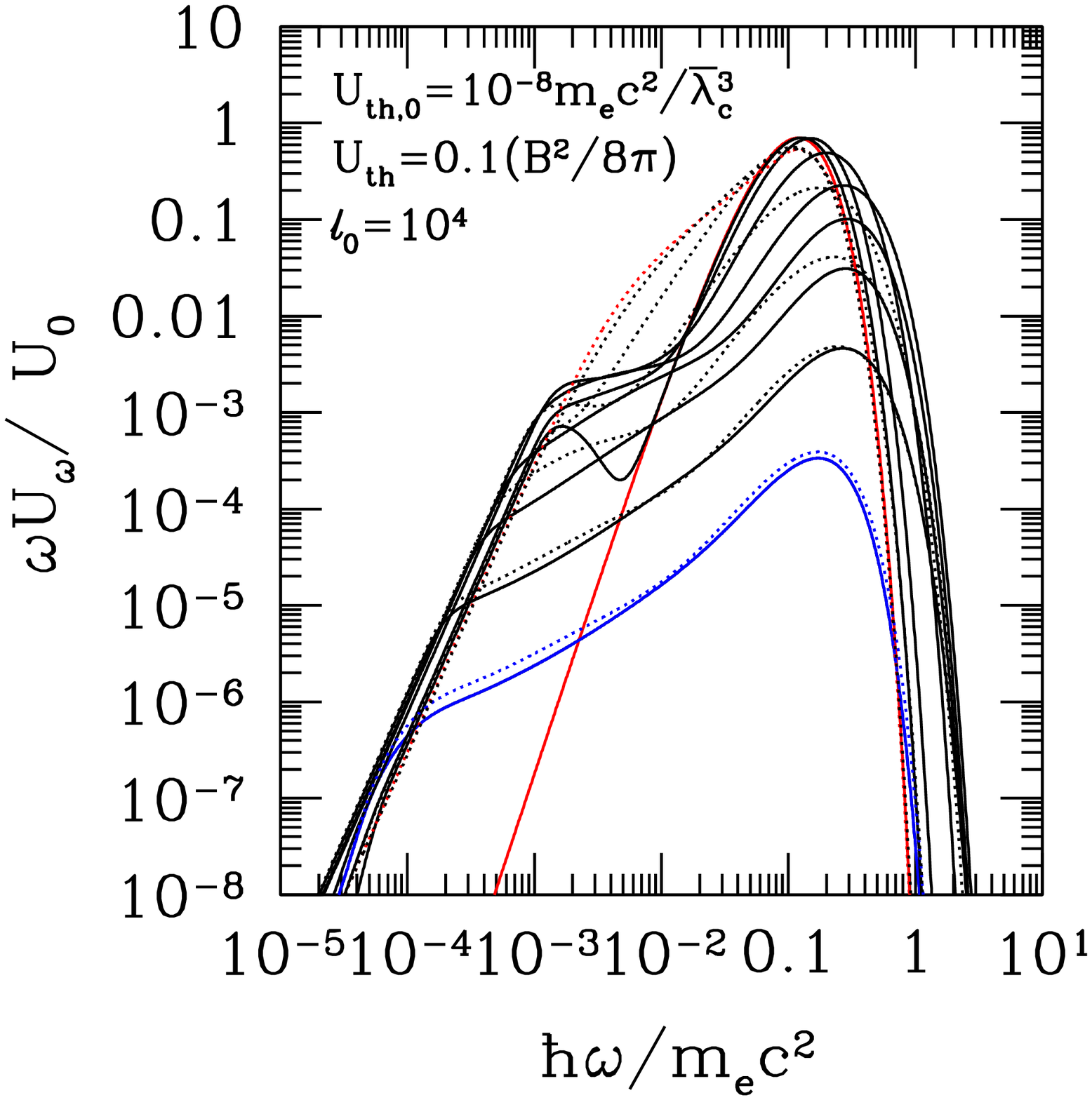}
\includegraphics*[width=0.5\hsize]{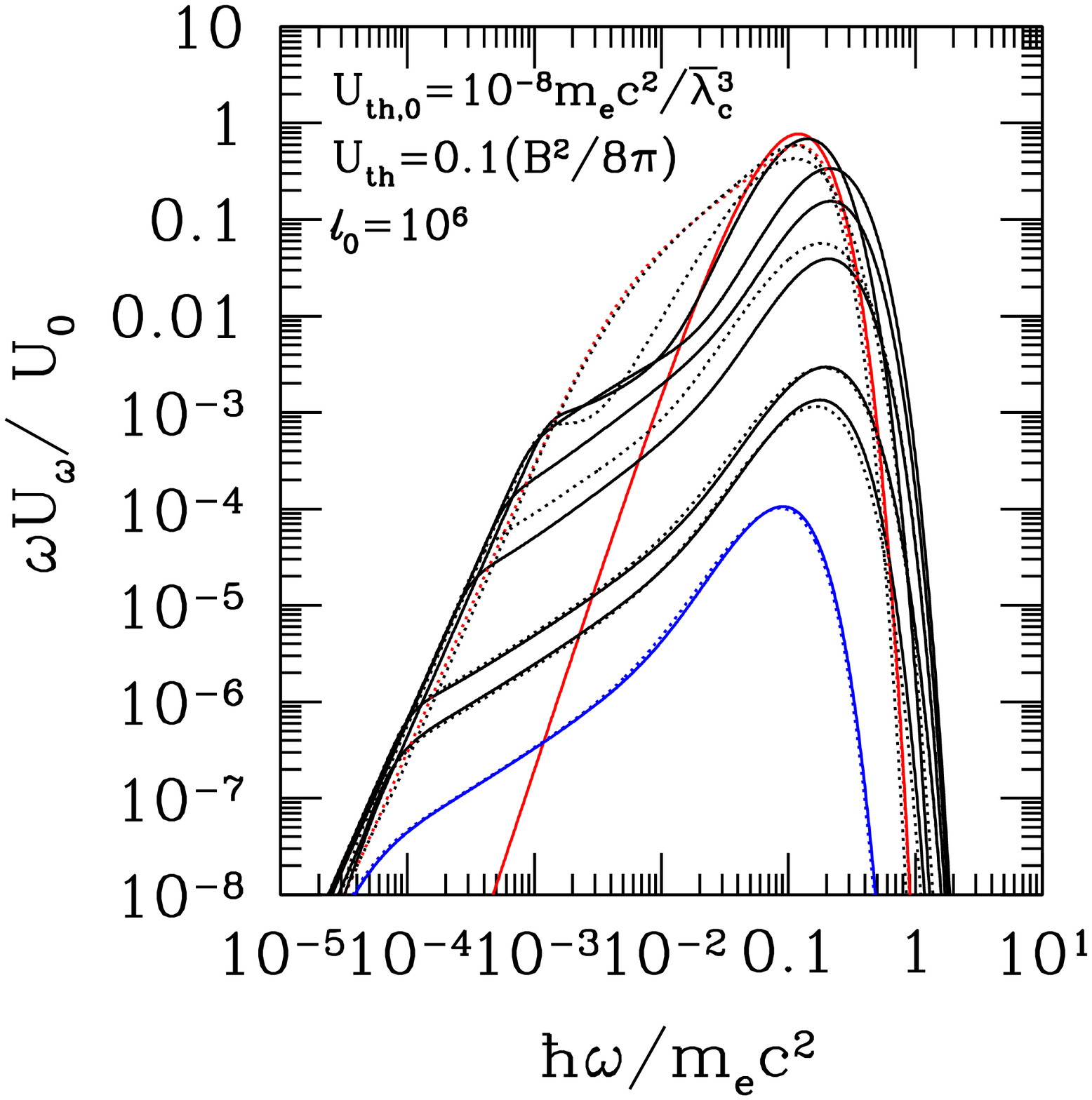}
}
\caption{Development of the photon spectrum in an expanding volume with initial thermal 
energy density $10^{-8}\,m_ec^2/\bar{\lambda}_c^3$, net expansion factor 
$t_{\rm tot}/t_0 = 10^2$, and heating turned off at $10^{-0.5}\,t_{\rm tot}$.  After
expansion by a decade or so, the spectrum is insensitive to the seed spectrum
(solid red line: Wien;  dotted red line: GRB-like (\ref{eq:specgrb}) with low frequency Rayleigh-Jeans
cutoff; both $T_0 = 0.03\, m_ec^2$).  Initial thermal compactness $10^4$ (left) and $10^6$ (right).}
\label{fig:expand}
\vskip .1in
\end{figure*}
\begin{figure*}
\centerline{
\includegraphics*[width=0.5\hsize]{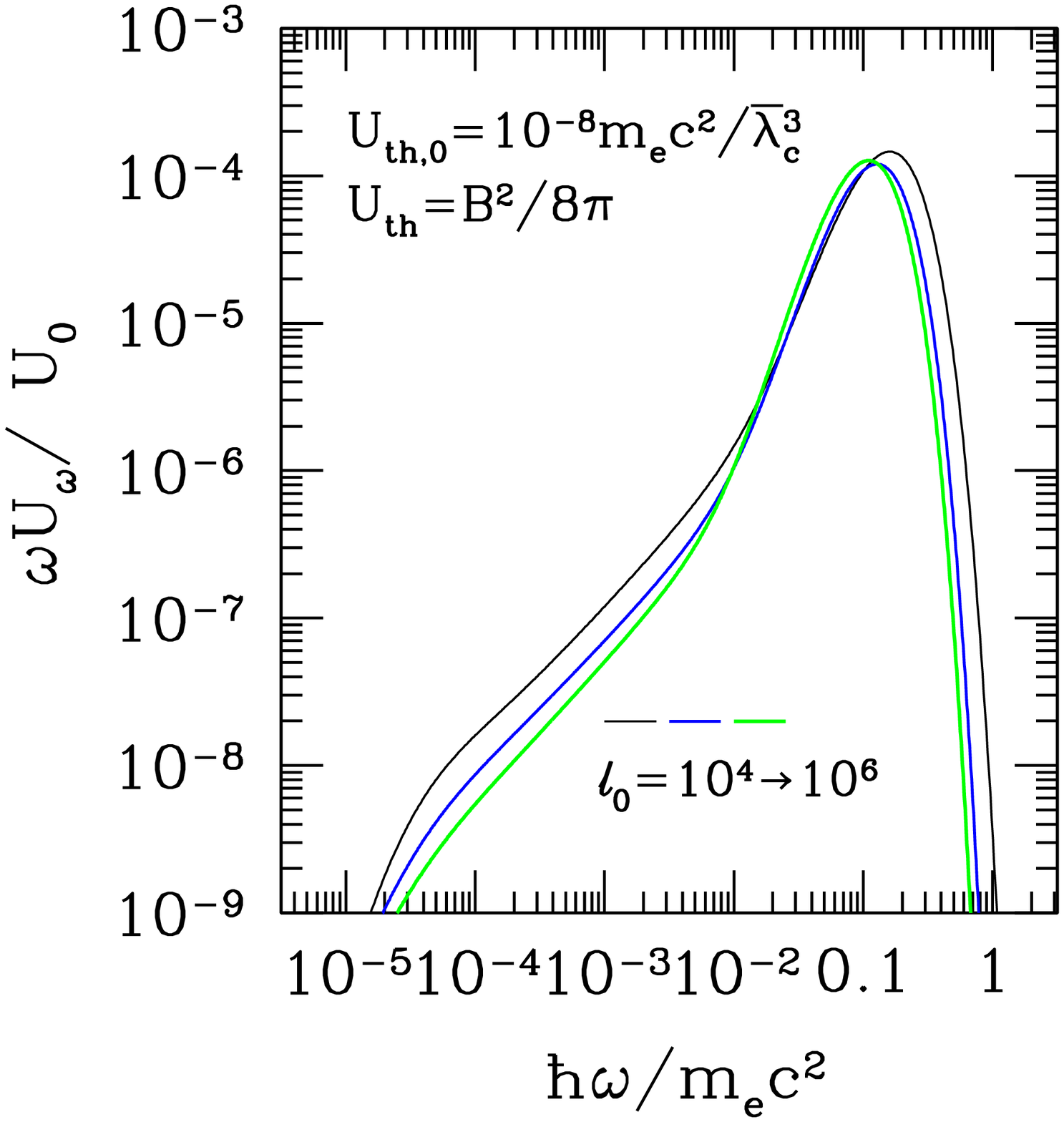}
\includegraphics*[width=0.5\hsize]{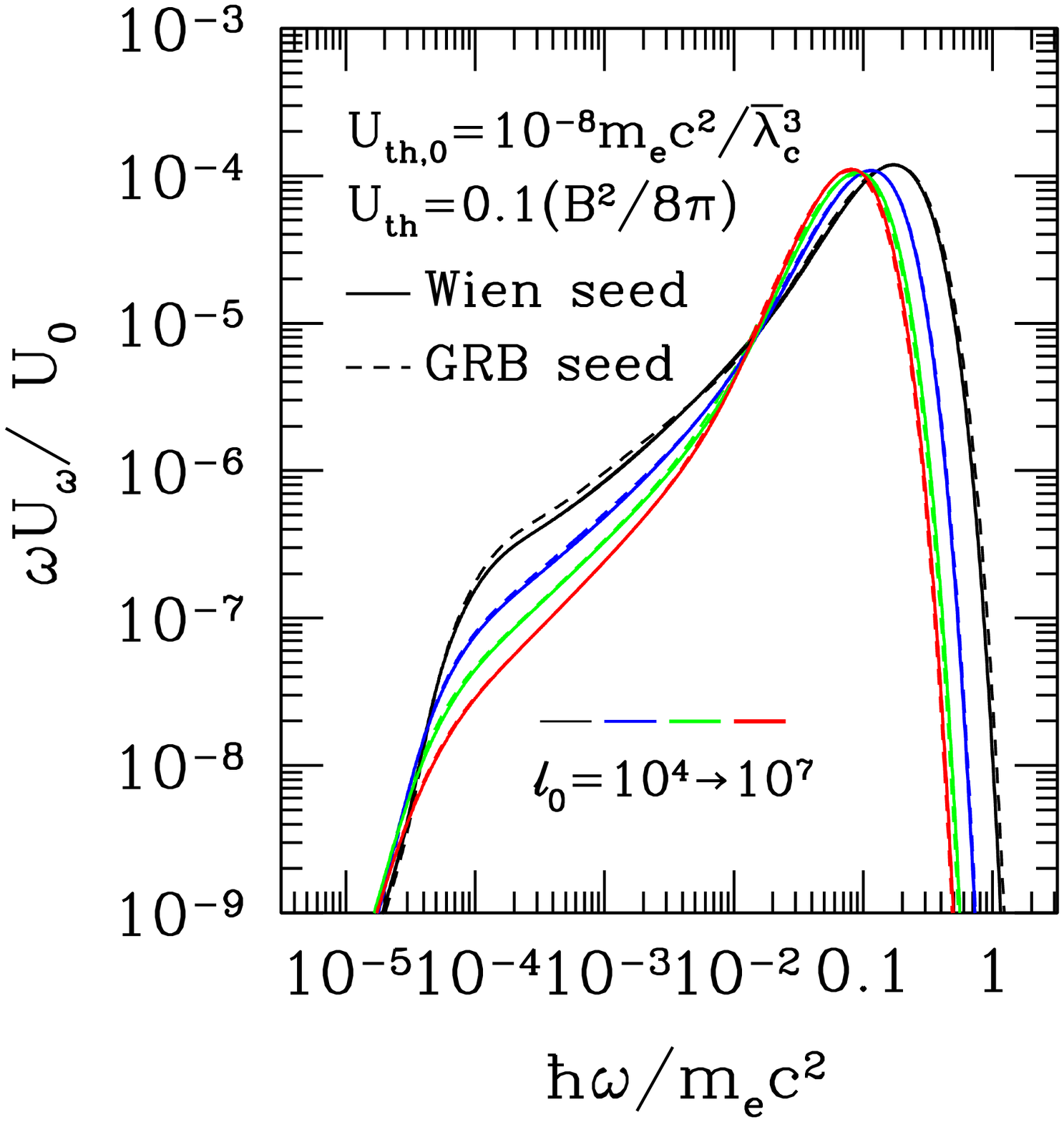}
}
\caption{Effect of magnetization and compactness on the final photon spectrum 
in an expanding, magnetized pair plasma.  Initial thermal energy density $10^{-8}\,m_ec^2/\bar{\lambda}_c^3$,
net expansion factor $t_{\rm tot}/t_0 = 10^2$, and heating turned off at $10^{-0.5}t_{\rm tot}$.  Here we
have compensated the effect of adiabatic expansion on the spectral amplitude in models which reach 
$\tau_T = 1$ at $t < t_{\rm tot}$, so as to afford a direct comparison.}
\label{fig:expandfin}
\vskip .1in
\end{figure*}

\subsection{Slope and Normalization of the Low-frequency Spectrum in a Strongly Magnetized Plasma}\label{s:slope}

A flat component of the photon spectrum ($U_\omega \sim$ const) generally appears at intermediate
frequencies in a dilute gas approaching thermodynamic equilibrium \citep{pozdnyakov83}.  Here
we consider how this component connects with a thermal peak (near which Doppler upscattering
by thermal $e^\pm$ motions is balanced by recoil energy loss).  We show that a distinct Wien
peak does not form in a strongly magnetized, thermal pair plasma.  The normalization of the
flat spectral component, relative to the peak, depends directly on the ratio of thermal to 
magnetic energy densities $f_{\rm th}$.

At lowest frequencies, the spectrum is black body, and breaks to $U_\omega \sim$ const at a frequency
$\omega_*$ and harmonic $m_* = \omega_*/\omega_{\rm ce}$ where the Compton upscattering rate
\be\label{eq:break}
{1\over \omega}{d\omega\over dt}\biggr|_{\rm C} \sim 4\Te n_e\sigma_T c = \alpha_{\rm cyc}(\omega)c.
\ee
Here 
\be\label{eq:alphacyc}
\alpha_{\rm cyc}(\omega) = {\hbar\omega\over 4\pi B_\omega} {d^2n_{\rm cyc}\over d\omega dt}
= {\hbar\omega\over 4\pi B_\omega} n_e  {d^2N_{\rm cyc}\over d\omega dt}
\ee
is the absorption coefficient, $d^2N_{\rm cyc}/d\omega dt$ is the rate of emission of cyclo-synchrotron
photons by a single $e^\pm$ (Figure \ref{fig:cyclotron}), and the Planck function 
$B_\omega \simeq  g_\gamma(2\pi)^{-3} \omega^2 T_e/c^2$ when $\hbar\omega \ll T_e$.
For completeness, we correct for the enhancement in upscattering by trans-relativistic $e^\pm$, for which
\be\label{eq:omegadot}
{1\over\omega}{d\omega\over dt}\biggr|_C \sim {4\over 3}\langle\gamma^2-1\rangle \sigma_T n_e c
= 4f_{\rm rel}\Te\sigma_Tn_ec
\ee
at low frequencies.  If the pairs follow a Boltzmann distribution (\ref{eq:relBol}), 
the correction factor is
\be
f_{\rm rel} \equiv \Te + \langle\gamma\rangle; \quad\quad \langle\gamma\rangle = 3\Te - {K_1(1/\Te)\over K_2(1/\Te)}.
\ee

These equations give the implicit relation for $m_*(T_e)$,
\be\label{eq:nstar}
{1\over m_*(T_e)f_{\rm rel}(T_e)\Te^2}{d^2N_{\rm cyc}\over d\omega dt}\biggr|_{m_*(T_e)} = {32\alpha_{\rm em}^2\over 3\pi}\left({B\over B_Q}\right).
\ee
Here $B_Q = m_ec^3/e\hbar = 4.4\times 10^{13}$ G is a convenient normalization of the magnetic field.  
(Near breakout of a GRB jet one may expect $B \sim 10^{-4}\,B_Q$.)  The left-hand side of this equation 
is a function only of $T_e$, showing that $m_*(T_e)$ depends weakly on $B$. The
solution, for the exact cyclo-synchrotron emissivity, is shown in Figure \ref{fig:nstar}.


The photon spectrum is $U_\omega = \omega_*^2 T_e/\pi^2 c^3$ in the frequency range $\omega_* < \omega < E_{\rm pk}/\hbar$.
Connecting this with a spectrum of the form (\ref{eq:specgrb}), up to a peak energy $E_{\rm pk} \simeq 3T_e$, gives
\be\label{eq:epsthcr}
f_{\rm th} = {24\alpha_{\rm em}m_*^2\over \pi} \Te^2 = 0.056\left({T_e\over 0.05~m_ec^2}\right)^2\left({m_*\over 20}\right)^2.
\ee 
We may interpret this equation as follows:  the low-frequency cyclotron bath does not supply enough
photons to form a localized thermal peak (with a low-frequency slope $d\ln(F_\omega)/d\ln\omega > 0$)
as long as the fraction of the magnetic energy density converted to thermal radiation is smaller
than (\ref{eq:epsthcr}).

\subsection{Constant Heating in a Static Medium}

We consider plasmas with a range of final energy density, $U_{\rm th} = 
10^{-11}-10^{-8} m_ec^2/\bar{\lambda}_c^3$, where $\bar{\lambda}_c$ is the reduced Compton
wavelength.  The development of the spectrum is shown in Figure \ref{fig:noexpand}.
The dependence of the output spectrum on compactness is shown in Figure \ref{fig:noexpand2}.
This spectrum is quite flat below the peak in a range of final compactness $\ell_f \sim 10^5-10^6$,
and is still much flatter than Wien when $\ell_f = 10^7$.

\subsection{Continuous Heating in an Expanding Medium}\label{s:expand}

We now consider expanding plasma with the thermal profile (\ref{eq:uthvst}).  We stop heating at a fixed time 
$t = 0.3t_{\rm tot}$, after which the plasma suffers adiabatic losses, the temperature drops, and pairs rapidly 
annihilate.  As the scattering depth approaches unity, the differential drift between photons and pairs 
begins to counterbalance adiabatic cooling.  The imprint of bulk Compton scattering on the photons propagating through the photosphere
of an accelerating jet is calculated separately in Section \ref{s:multscatt}.

Figure \ref{fig:expand} shows the development of the spectrum starting from a seed Wien peak.  For an expansion
factor $t_{\rm tot}/t_0 = 10^2$, equation (\ref{eq:fdis}) implies that $\sim 98.5\,\%$ of the final energy density
results from distributed heating.  Indeed, the spectrum reaches its final form, modulo changes in amplitude
and the position of the Rayleigh-Jeans tail, well before the expansion is complete.   

\begin{figure}
\epsscale{0.9}
\plotone{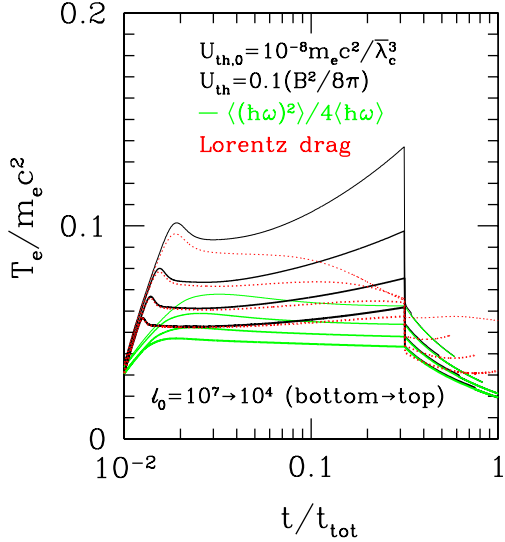}
\caption{Pair temperature corresponding to Figure \ref{fig:expandfin}.
$T_e$ rapidly adjusts downward to Compton equilibrium after heating turns off at $10^{-0.5}\,t_{\rm tot}$.
Dotted red curves:  term (\ref{eq:ndrift}) included in Kompaneets equation, representing
differential acceleration of the magnetic field and entrained $e^\pm$ pairs across the photon field
near the scattering photosphere.  Solid black curves: drag term turned off.  Green curves:  Compton
temperature corresponding to black curves.
}
\label{fig:Te}
\end{figure}
\begin{figure}
\epsscale{0.9}
\plotone{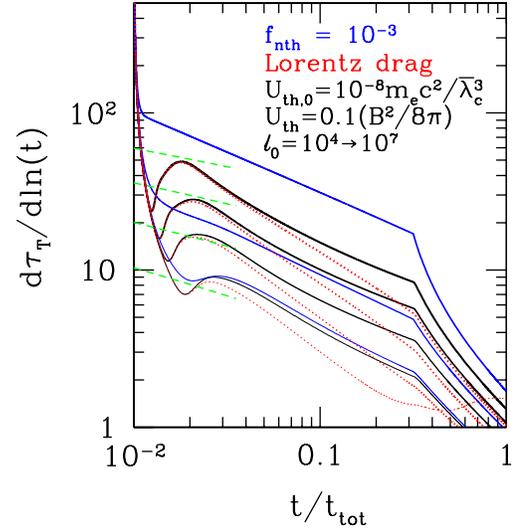}
\caption{Optical depth to Thomson scattering, $d\tau_T/d\ln(t) = n_e(t)\sigma_T ct$.  Corresponding temperature evolution in 
Figure \ref{fig:Te}.  Dotted red curves include term (\ref{eq:ndrift}) in the Kompaneets equation, solid black curves
do not.  Solid blue curves: addition non-thermal pair source term (\ref{eq:dnedtnth}) in equation (\ref{eq:dnedttot}).  
Dashed green curves:  semi-analytic model of soft photon creation, Compton upscattering, and pair creation (Section \ref{s:scaling}).
}
\label{fig:tauT}
\vskip .1in
\end{figure}
\begin{figure}
\epsscale{0.9}
\plotone{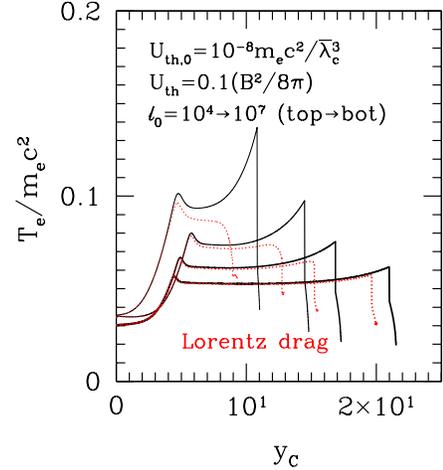}
\caption{Compton parameter corresponding to Figures \ref{fig:Te}, \ref{fig:tauT}.}
\label{fig:yC}
\vskip .2in
\end{figure}
\begin{figure}
\epsscale{0.9}
\plotone{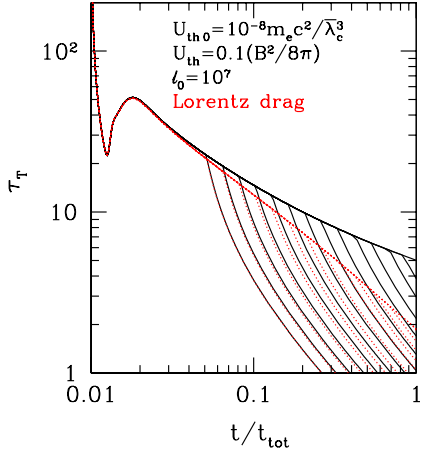}
\caption{Effect of varying the cutoff time of particle heating.}
\label{fig:tauT2}
\end{figure}

The dependence 
of the final spectrum on compactness and bulk-frame magnetization ($\sim f_{\rm th}^{-1}$) is shown 
in Figure \ref{fig:expandfin}.  In agreement with the analytic argument advanced in Section \ref{s:slope}, reducing 
the thermalization efficiency $f_{\rm th}$ flattens the connection between the $F_\omega \sim$ constant
component of the spectrum, and the thermal peak.

Details of the evolution of $T_e$ and $\tau_T$ are shown in Figures \ref{fig:Te} and \ref{fig:tauT}.  The 
cutoff in heating leads to a rapid drop in electron temperature and annihilation of pairs.   The spectral peak 
frequency experiences a more gradual change.  Once $\tau_T$ drops below $\sim 5$, the continuing reduction in 
scattering depth is driven mainly by expansion ($n_e \sim t^{-2}$ corresponding to $\tau_T \sim
t^{-1}$).  This behavior is replicated for a range of cutoff times in Figure \ref{fig:tauT2}.  Finally,
the growth of the Compton parameter (Figure \ref{fig:yC}),
\be\label{eq:yC}
y_C(t) = 4\int^t_{t_0} c\,d\tilde t\,{T_e(\tilde t)\over m_ec^2}\,
\sigma_T n_e(\tilde t),
\ee
saturates after heating turns off.

While heating is applied, the electron and Compton temperatures remain significantly different for the 
highest compactness considered, $\ell_{\rm th} \sim 10^4-10^7$, but rapidly converge after heating is turned off.
(See Section \ref{s:highcomp} for more discussion of this point.)   

\begin{figure} 
\epsscale{0.9}
\plotone{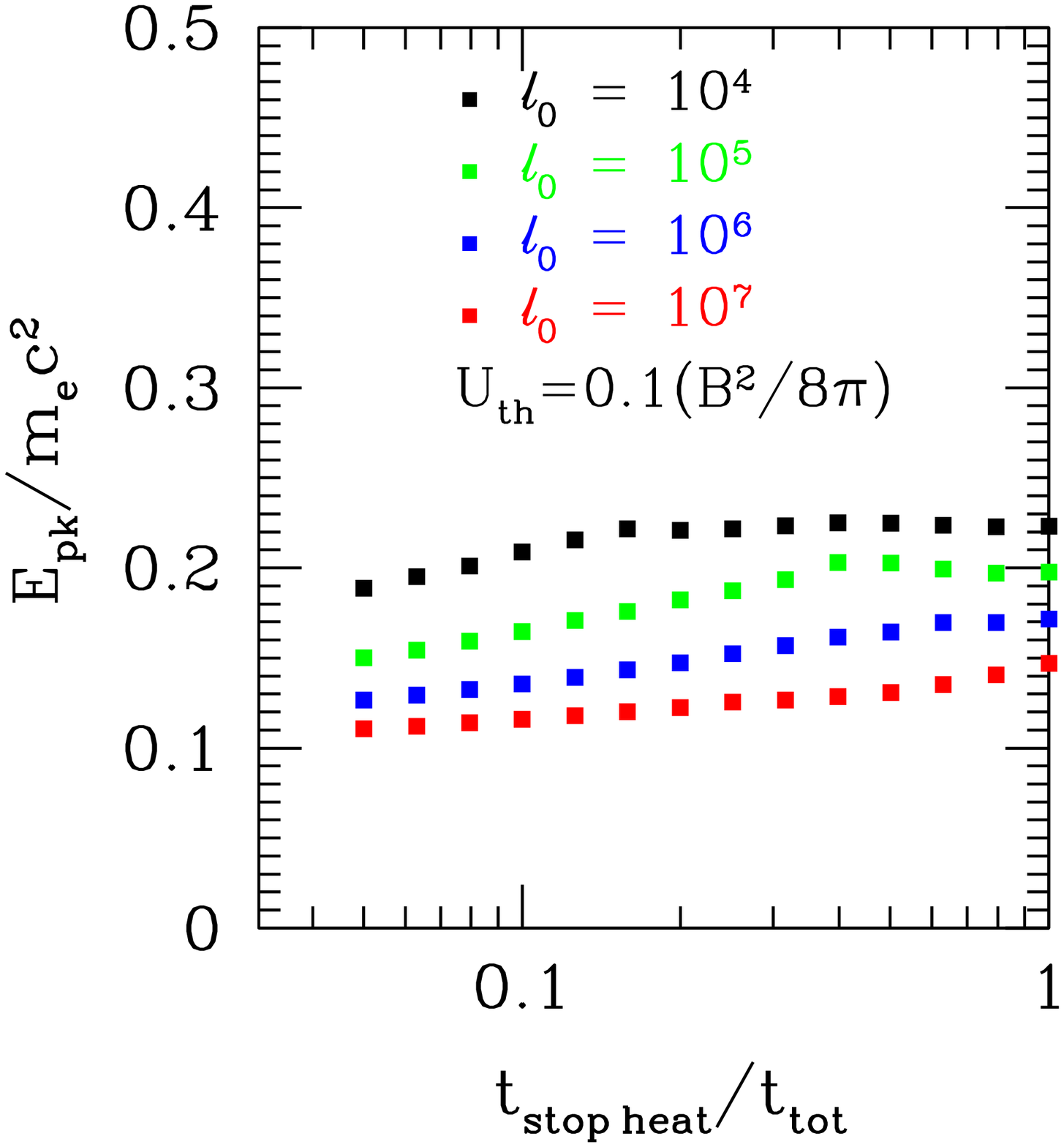}
\plotone{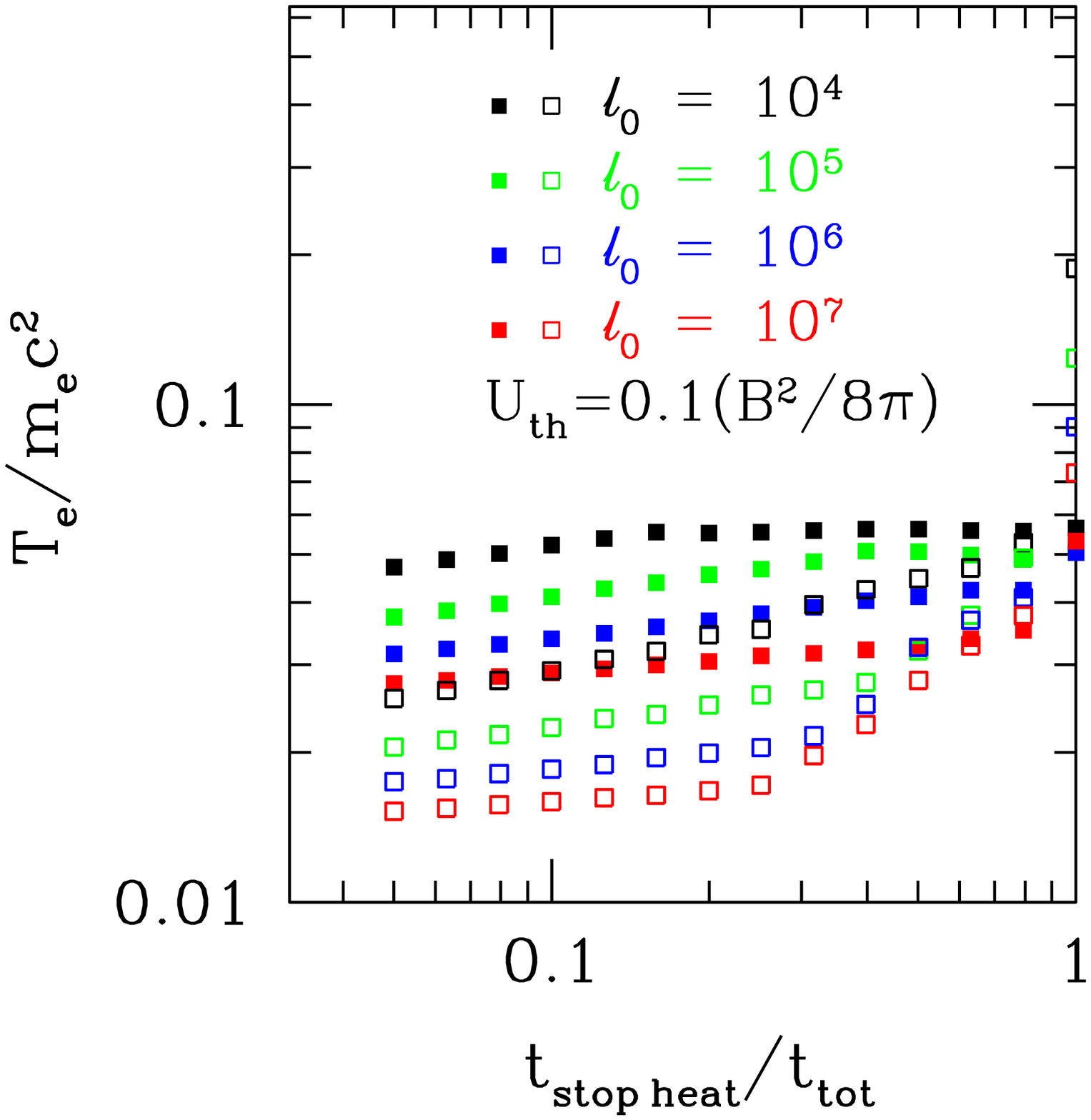}
\caption{Spectral peak energy (top panel) and final pair temperature (bottom panel) 
as a function of initial thermal compactness, for $U_{\rm th,0} = 10^{-8} m_ec^2/\bar{\lambda}_c^3$ 
and expansion  factor $10^2$.  Open points:  adiabatic evolution.  Solid points:  the outflow experiences modest 
spectral heating, that partly compensates adiabatic cooling, during the final transition to transparency, 
as represented by the term (\ref{eq:ndrift}) in the Kompaneets equation.
}
\label{fig:fintemp}
\vskip .1in
\end{figure}

The output values of $T_e$ and $E_{\rm pk} \simeq \langle(\hbar\omega)^2\rangle/\langle \hbar\omega\rangle$ are shown
in Figure \ref{fig:fintemp} as a function of compactness, and the time at which heating is turned off.  As long
as the gas passes through a brief adiabatic, expansionary phase, we find that the output peak energy clusters 
around $E_{\rm pk} \sim 0.1\,m_ec^2$ for final compactness $\gtrsim 10^4-10^5$, and
extends upward to $\sim 0.2\,m_ec^2$ for final compactness $\sim 10^2$.

\subsection{Effect of Non-thermal Pair Creation}

Here we consider the effect of the injection of non-thermal particles on the output spectrum, as parameterized by the yield
(\ref{eq:dnedtnth}) of cold pairs that supplement the thermal particle density.  To keep the calculation self-consistent,
we only consider a high radiation compactness, ranging from $\ell_0 = 10^6$ down to $\ell_f = 10^4$ in the examples given. 
The energy injected directly in cold pairs extends from $f_{\rm nth} = 10^{-5}$ up to $10^{-2}$ of the 
heat that is deposited gradually in the thermal pairs.  

\begin{figure} 
\epsscale{0.9}
\plotone{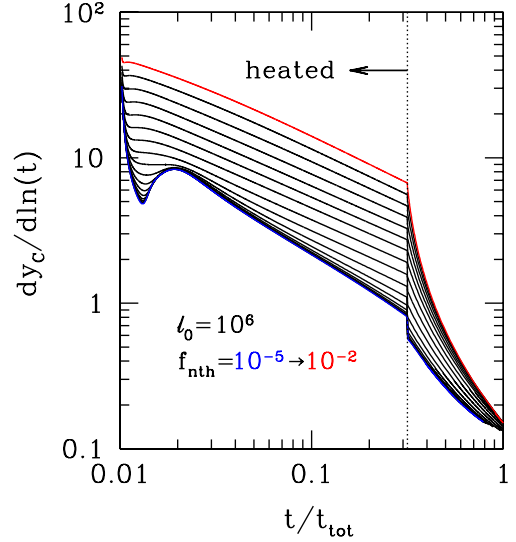}
\caption{Compton parameter of an expanding, magnetized plasma, in which cold pairs are injected directly 
carrying a fraction $f_{\rm nth} =10^{-5}$-$10^{-2}$ of the thermal energy density (equation (\ref{eq:dnedtnth})), 
representing the end product of a cascade from relativistic energies.  Initial thermal
energy density $U_{\rm th,0} = 10^{-8} m_ec^2/\bar{\lambda}_c^3$ and initial (final) radiation compactness 
$\ell_0 = 10^6$ ($\ell_f = 10^4$).  The cumulative Compton parameter in cold pairs is several times larger than the 
plotted value, and is $\gtrsim 10^4$ times greater than that supplied directly by the cascading charges (equation (\ref{eq:yCrel})).
}
\label{fig:yCnth}
\vskip .1in
\end{figure}
\begin{figure} 
\epsscale{0.9}
\plotone{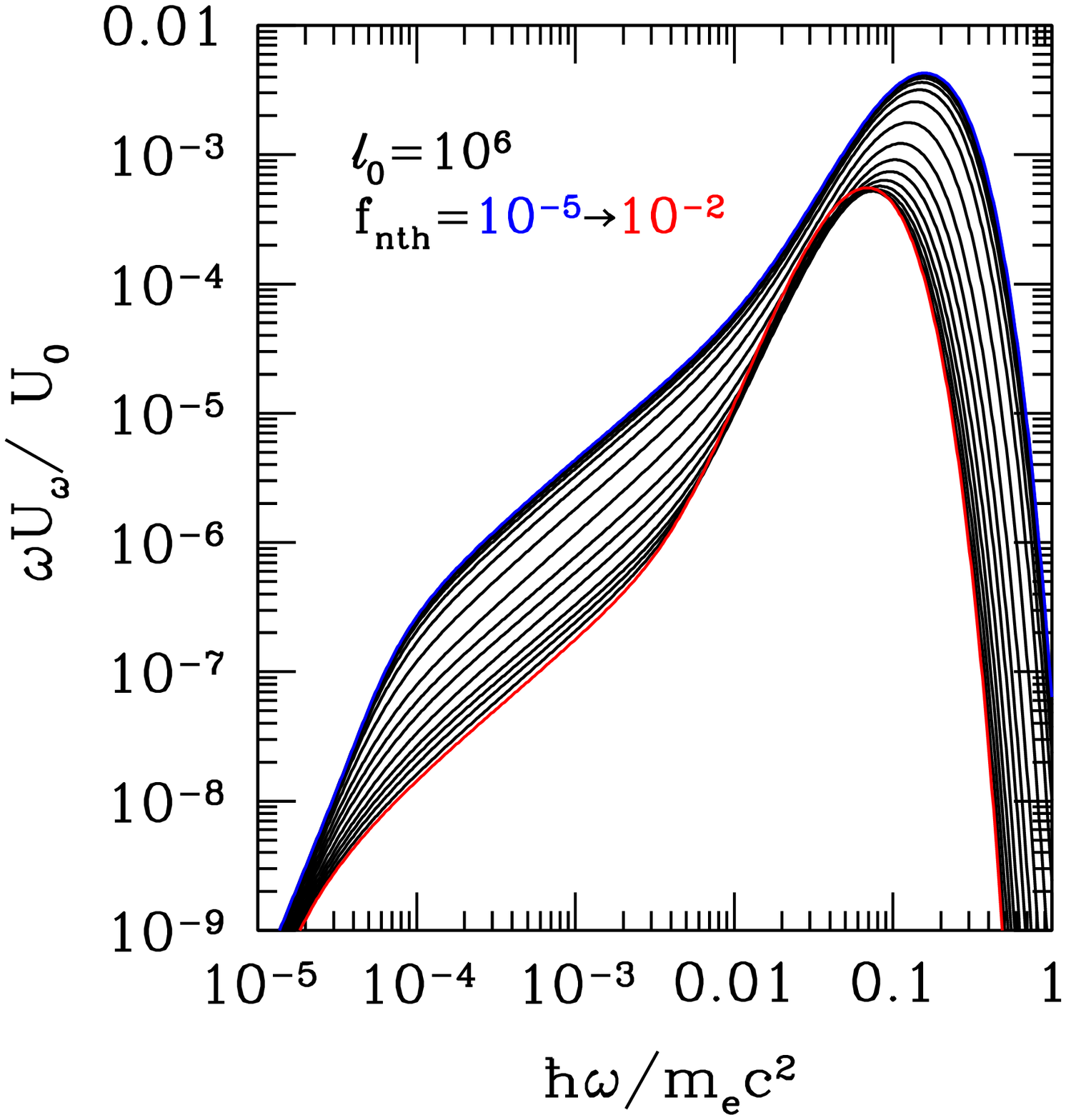}
\plotone{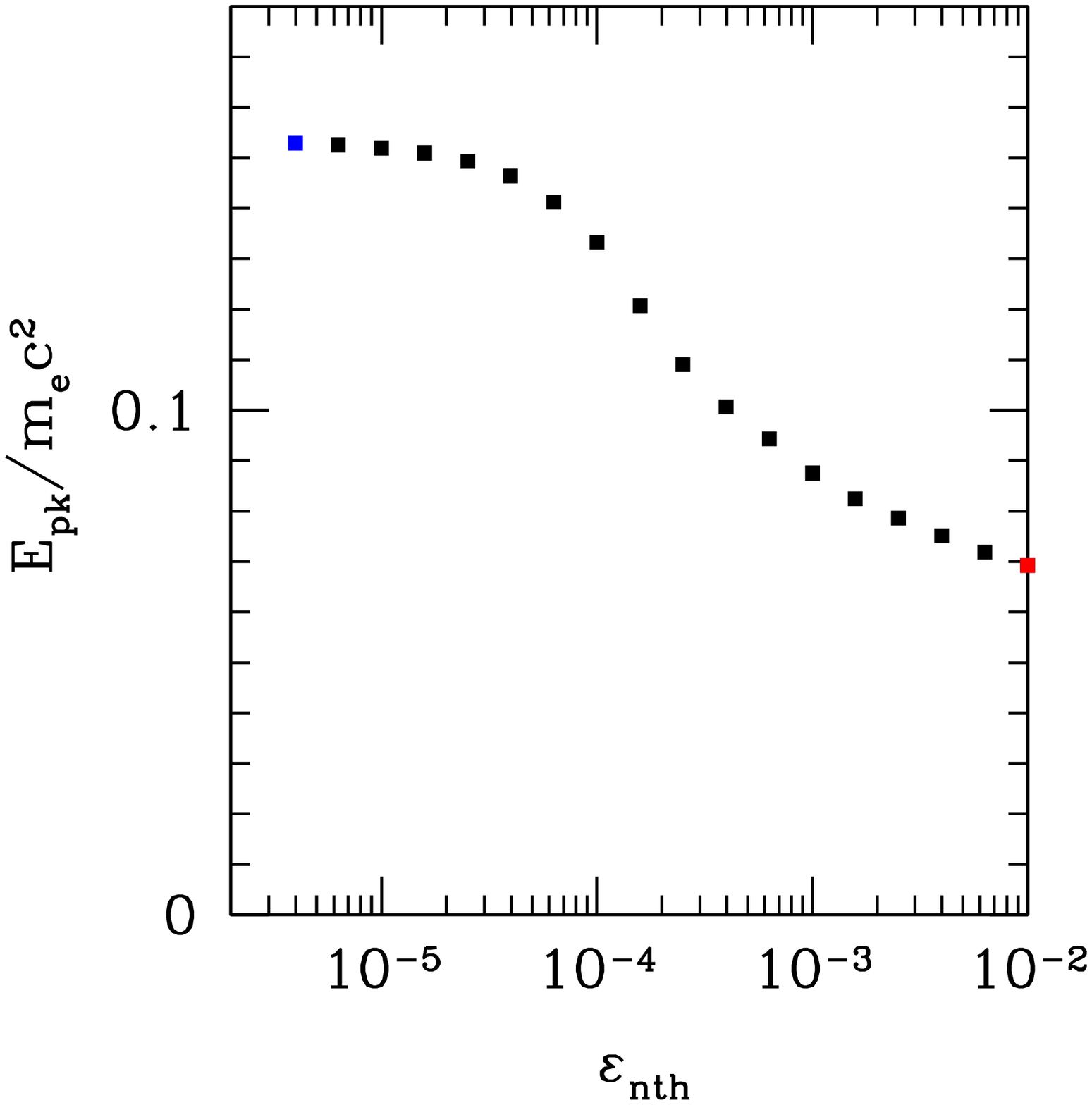}
\plotone{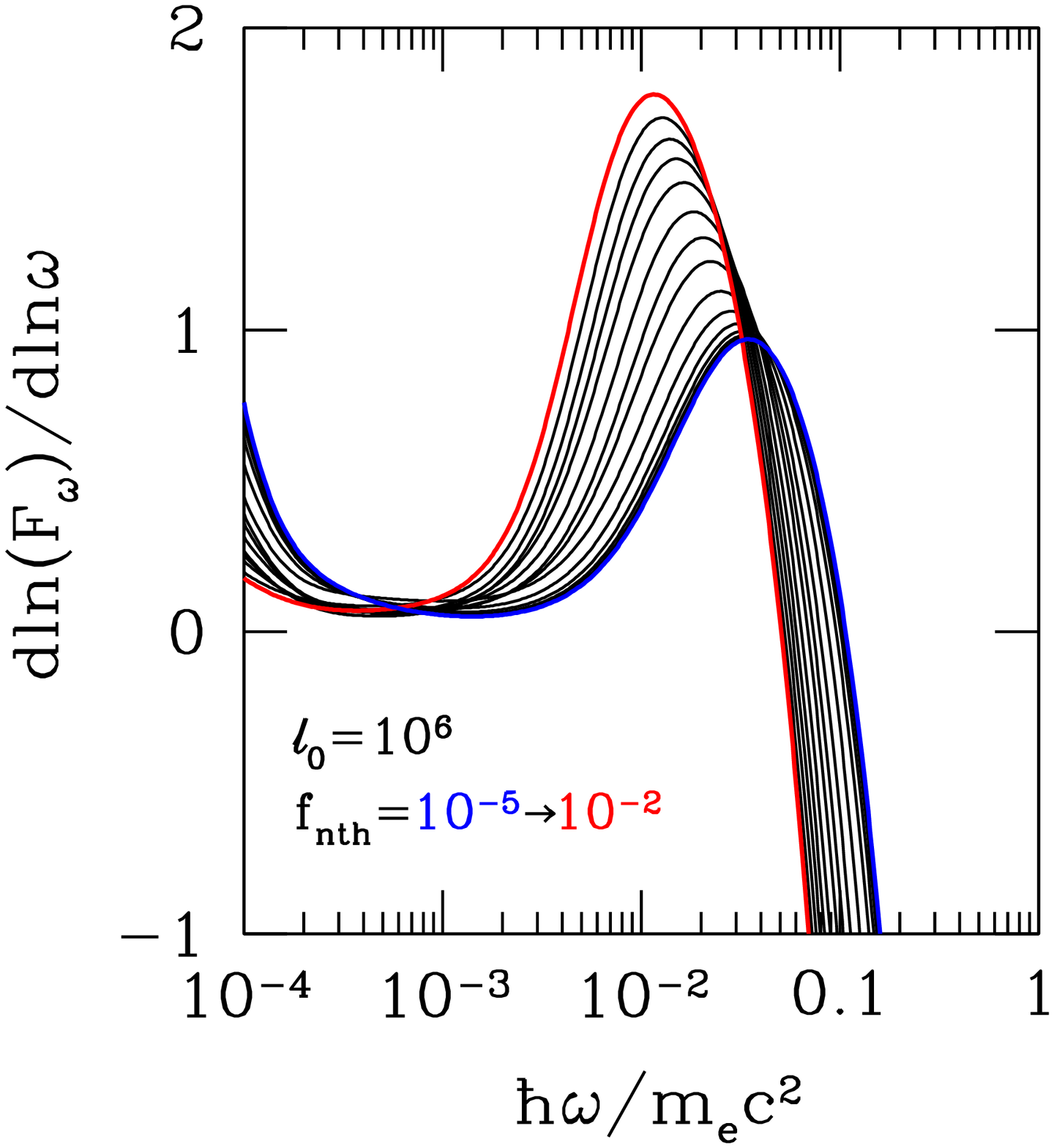}
\caption{Output spectrum, spectral peak energy, and spectral slope in the simulations of Figure \ref{fig:yCnth}.  
As the optical depth in cold pairs increases, due to the increased efficiency of non-thermal particle injection,
the spectrum hardens and forms a more concentrated peak.  The peak frequency is pushed lower, due to the
increased rate of soft photon creation and upscattering.  Processing of this spectrum through the photosphere
of an accelerating MHD jet reduces the peak spectral index by $\sim 0.5$ (Section \ref{s:multscatt}).
}
\label{fig:specnth}
\vskip .1in
\end{figure}

The Compton parameter of the cold pair gas rises significantly at the larger values of the non-thermal energy fraction:
see Figure \ref{fig:yCnth}.  As a result, the spectrum is more strongly peaked (Figure \ref{fig:specnth}), with
a harder spectrum right below the peak.  While heating is ongoing, the peak of the spectrum is pushed to a lower 
frequency due to the increased efficiency of soft photon creation and upscattering. We preserve the temporal heating profile
that was applied previously to purely thermal plasmas.  After heating turns off, at $t = 10^{-0.5} t_{\rm tot}$, 
the pairs rapidly annihilate and the optical depth through them converges to a common value.  

We conclude that moderate rates of non-thermal heating ($f_{\rm nth} \sim 10^{-3}$-$10^{-2}$) result in
a harder low-frequency spectrum than that usually measured in GRBs.   

\section{Scaling Solution for Optical Depth and Temperature in an Expanding Medium}\label{s:scaling}

A useful check of the numerical results described in Section \ref{s:spectrum} is provided by a simple
scaling model.  This applies to the initial transient phase during which the plasma reaches a scaling
behavior, and most of the co-moving photon number is accumulated.  Thereafter, according to equation 
(\ref{eq:uthvst}), the (differential) rate of photon creation drops off.  

The state of an expanding, and continuously heated, thermal pair plasma is conveniently 
described by $T_e$ and the optical depth $d\tau_T/d\ln(t) \equiv \sigma_T n_e ct$.   We consider expansion 
at constant Lorentz factor, with magnetic energy density $\sim t^{-2}$, bulk-frame volume $\sim t^2$, 
and constant ratio of injected thermal to magnetic energy, corresponding to $\delta = 2$
in equation (\ref{eq:uthvst}).   Generalization to other expansion profiles is straightforward.

\begin{figure} 
\epsscale{0.9}
\plotone{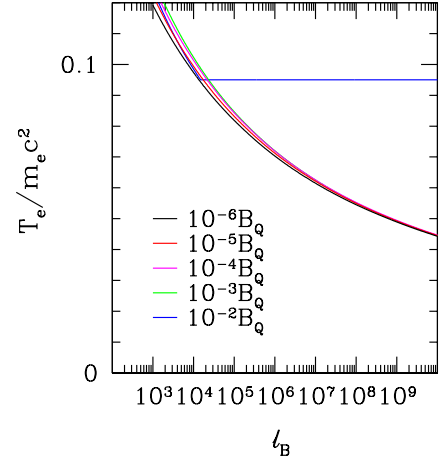}
\caption{Pair temperature in a continuously heated and expanding 
outflow, as a function of the magnetic compactness $\ell_B$, according to the scaling solution
derived in Section \ref{s:scaling}.  Thermal (mainly photon) energy 
density $U_{\rm th} = 0.1(B^2/8\pi)$.  Temperature is regulated by a competition between
two effects:  Compton upscattering of thermal cyclo-synchrotron photons, against 
the exponential dependence of pair depth on temperature.}
\label{fig:temp_equil}
\vskip .1in
\end{figure}
\begin{figure}
\epsscale{0.9} 
\plotone{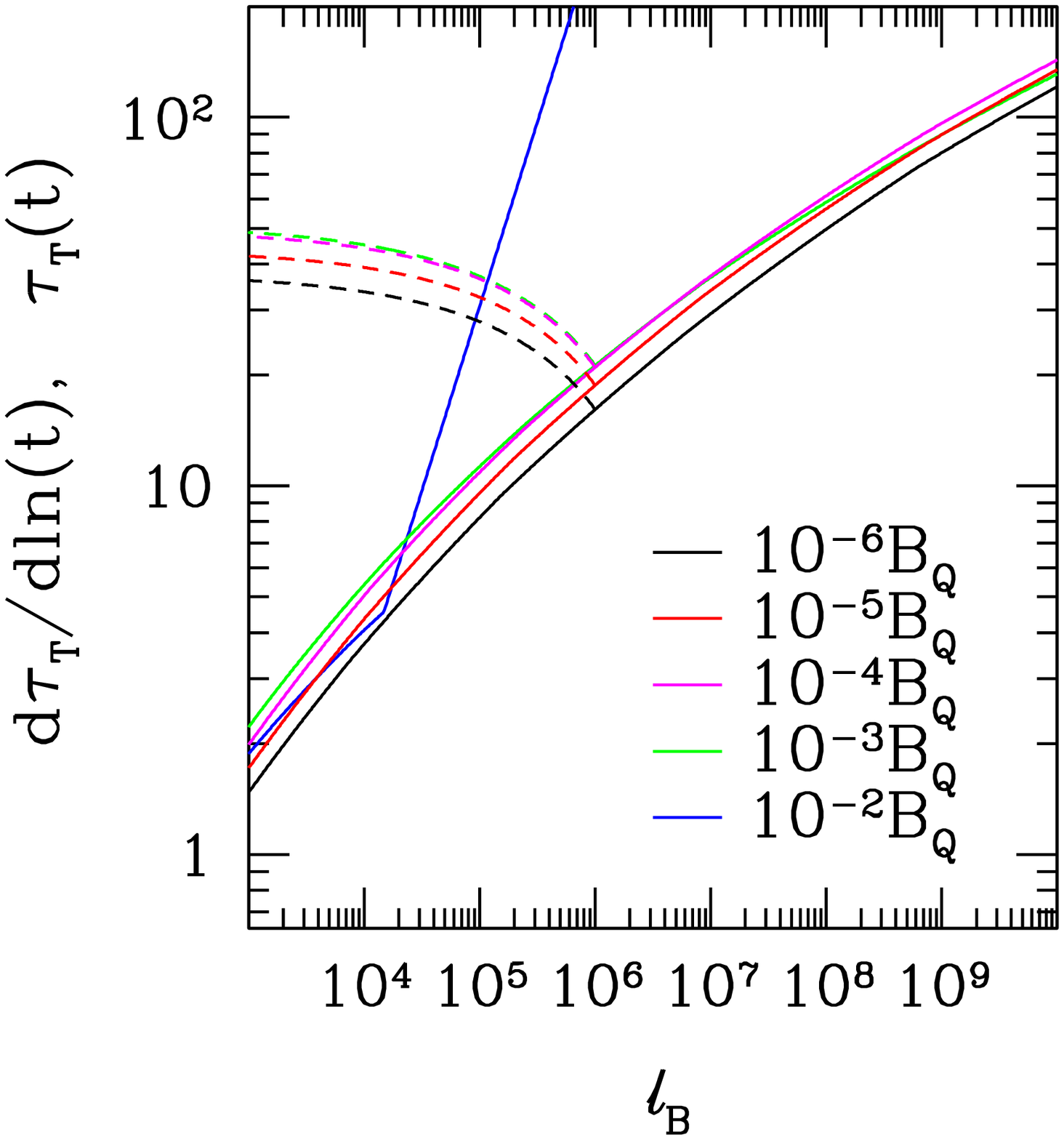}
\plotone{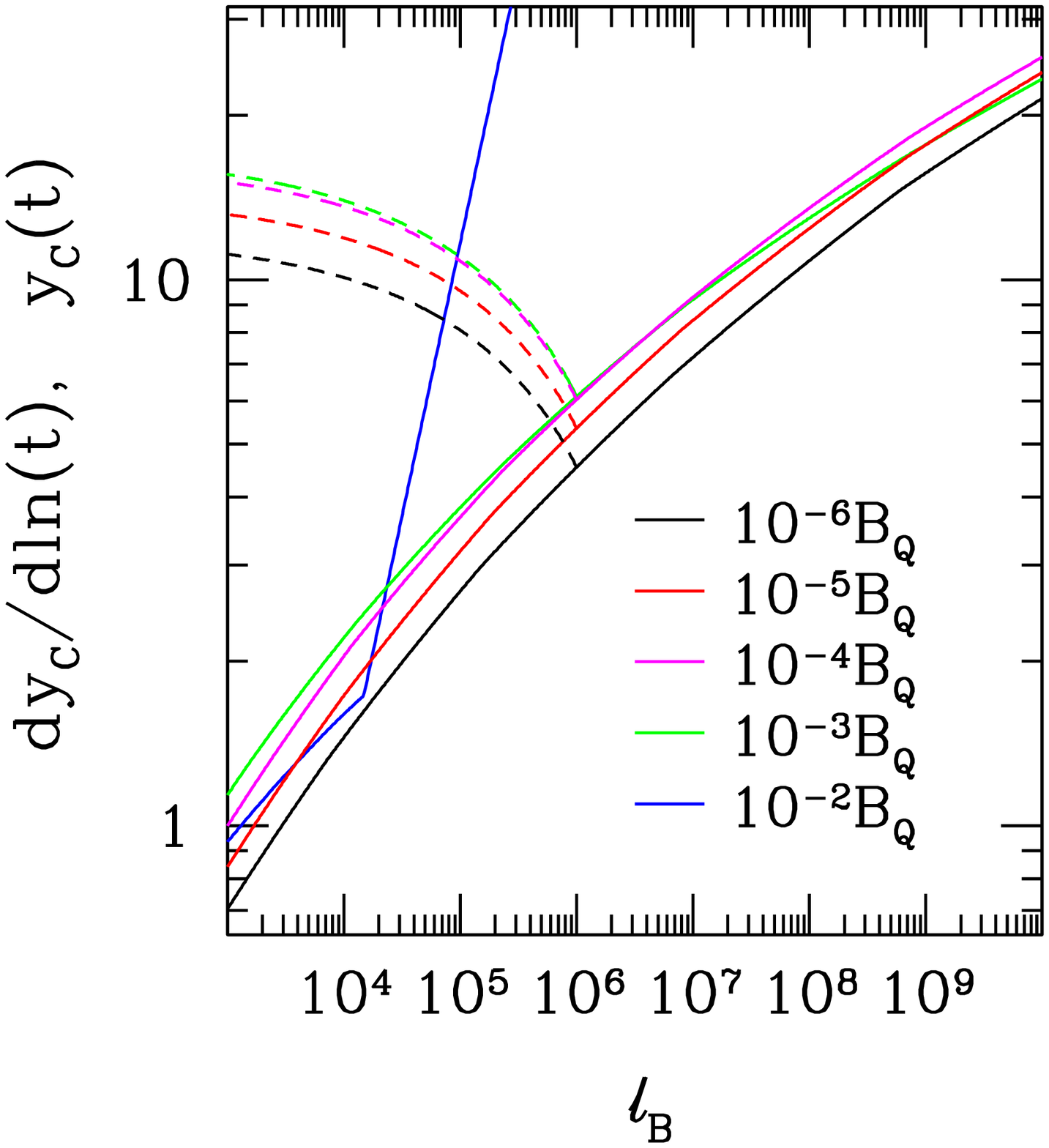}
\caption{Thomson scattering depth and Compton parameter, per logarithm of expansion
time $t \propto \ell_B^{-1}$, in the same system described in Figures \ref{fig:nstar}
and \ref{fig:temp_equil}.  The cumulative $\tau_T$ and $y_C$, integrated forward
from time corresponding to $\ell_B = 10^6$, is shown in the dashed curves.}
\label{fig:yC_equil}
\end{figure}
\begin{figure}
\epsscale{0.9} 
\plotone{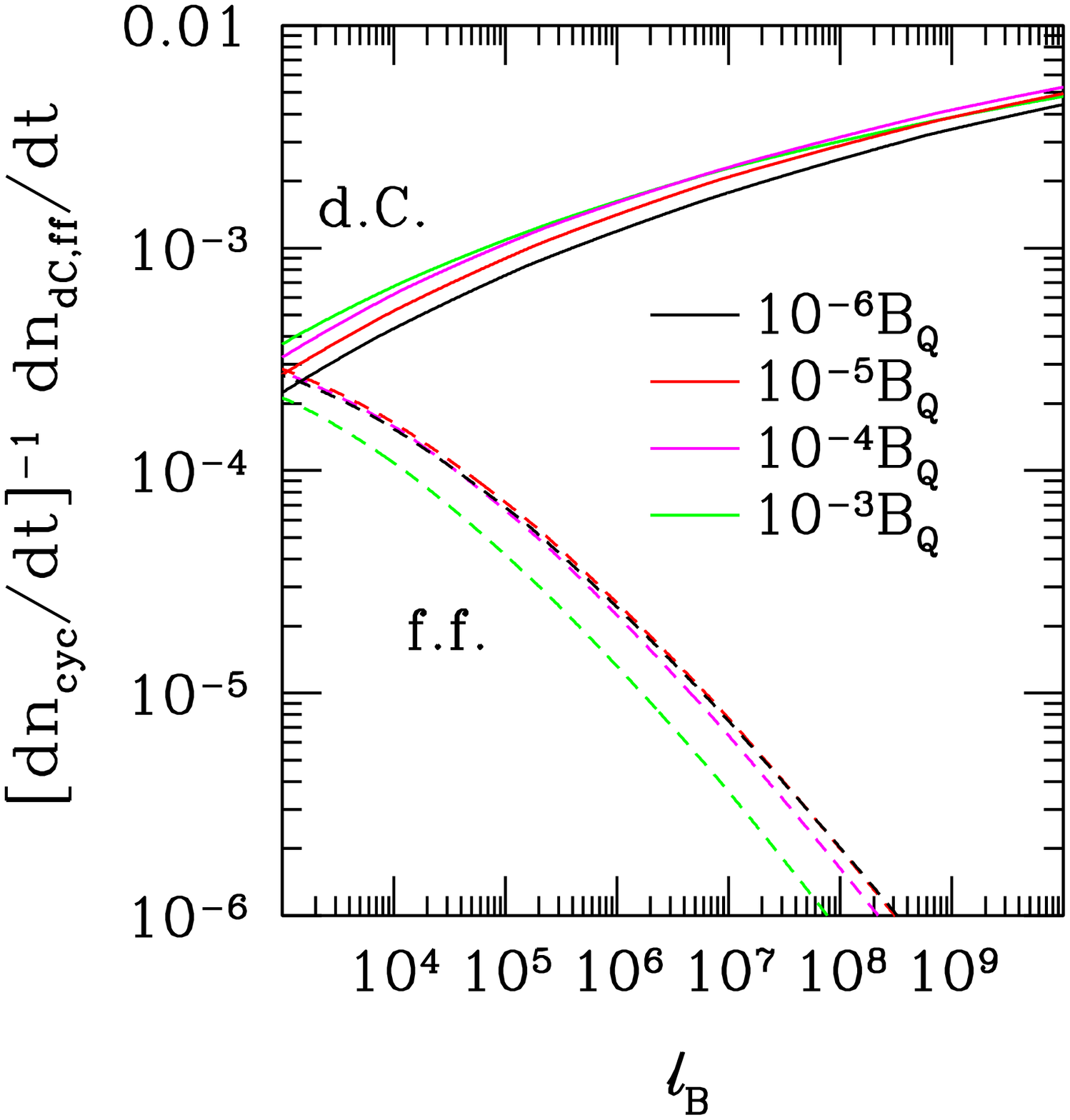}
\caption{Relative contributions of thermal cyclo-synchrotron, free-free,
and double-Compton emission to the creation of soft photons, in the same
system described in Figures \ref{fig:nstar}, \ref{fig:temp_equil}, and \ref{fig:yC_equil}.}
\label{fig:coolchan_equil}
\vskip .1in
\end{figure}

We found that $T_e$ and $\tau_T$ vary slowly and in 
opposing ways, so that $dy_C/d\ln(t)$ is nearly constant.  The net Compton 
parameter accumulates logarithmically with time.  In the flat portion of the spectrum, 
the left- and right-hand sides of (\ref{eq:komp}) both approximately vanish.  Then one can write
\be
{1\over t^\delta}{\partial\over\partial t}\left(t^\delta {\partial n_\gamma\over\partial\omega}\right)
= -{\partial F_\gamma\over\partial\omega} \simeq 0, 
\ee
where $\partial n_\gamma/\partial\omega = N \omega^2/\pi^2 c^3$ and
\be 
F_\gamma = n_e \sigma_T c\left({T_e\over m_ec^2}\right)\left({\omega_*\over c}\right)^3{3N\over \pi^2}
  \propto n_e \omega_*^2 \propto t^{-3}
\ee
is the rate at which photons flow toward the spectral peak, per unit volume.   In the approximation
that all the photons are near the peak, these equations integrate to $n_\gamma(t) = F_\gamma(t) t$.  

The break frequency
$\omega_* = m_*(T_e) \omega_{\rm ce}$ bounding the low-frequency black-body tail is given in Figure 
\ref{fig:nstar}.  
Further setting $\langle \hbar\omega\rangle n_\gamma = f_{\rm th} B^2/8\pi$, and
writing $\langle \hbar\omega \rangle = f_\omega T_e$, one obtains a relation 
between temperature and optical depth,
\be\label{eq:phcrbal}
{d\tau_T(T_e)\over d\ln(t)} m_*^2(T_e) f_{\rm rel}(T_e) \Te^3 = 
{\pi\over 24\alpha_{\rm em}}\,{f_{\rm th}\over f_{\omega,\rm C}}.
\ee
As long as pair creation and annihilation are nearly in equilibrium, equation (\ref{eq:ne_eq})
implies,
\be\label{eq:phcrbal2}
{d\tau_T(T_e)\over d\ln(t)} = \left({\pi\over 2}\right)^{1/2} {\ell_{\rm th}\over f_{\omega,\tau}}\Te^{-5/2}e^{-1/\Te},
\ee
giving
\be\label{eq:analytic}
m_*^2(T_e)f_{\rm rel}(T_e)\Te^{1/2} e^{-1/\Te} = {9.6\over \ell_B}
\left({f_{\omega,\tau}\over f_{\omega,\rm C}}\right).
\ee

The coefficients $f_\omega$ in equations (\ref{eq:phcrbal}) and (\ref{eq:phcrbal2}) are not
entirely equivalent.  The number density of photons at the pair-creation threshold is determined essentially
by $T_e$ and $U_\gamma$.  The photon energy density receives only a modest supplement from
the flat-spectrum band:  one finds $f_{\omega,\tau} \simeq 3.9$ for a spectrum of the form (\ref{eq:specgrb}),
as compared with $3$ for a Wien spectrum.  

On the other hand, equation (\ref{eq:phcrbal}) is derived in the approximation that all the photons
are upscattered to a thermal peak, whereas in fact the flat-spectrum band contributes logarithmically to the total.  
We find that taking $f_{\omega,C} = 3/\ln(E_{\rm pk}/m_*\hbar\omega_{\rm ce})$ allows an accurate fit of this 
semi-analytic model to the transient peak, for a compactness varying between $\sim 10^5$ and $\sim 10^7$.
See the dashed green curves in Figure \ref{fig:tauT}.
These values of $f_{\omega,\tau}, f_{\omega,C}$ are used to construct the plots here and in Section \ref{s:baryon}.

The electron temperature is shown in Figure \ref{fig:temp_equil}, and the optical depth and Compton parameter
in Figure \ref{fig:yC_equil}.  The differential optical depth $d\tau_T/d\ln(t)$ varies only slowly with compactness.  
The cumulative $\tau_T$ and $y_C$ are significantly larger:  see the dashed lines in Figure \ref{fig:yC_equil}.
Photon creation is dominated by the cyclo-synchrotron channel, except at very low compactness (Figure \ref{fig:coolchan_equil}).

\section{Strongly Magnetized Outflow with Baryon-dominated Photosphere}\label{s:baryon}

The presence of baryons in the outflow imposes a lower bound on the scattering depth.
This has two effects:  first, there is strong adiabatic cooling of any thermal photon gas 
unless the baryon loading is fine-tuned to a critical value \citep{shemi90} or, alternatively, unless the
outflow is heated continuously out to its photosphere \citep{thompson94,spruit01,giannios06}.  Second, 
a large Compton parameter develops at large $\tau_T$, pushing the photons closer
to a Wien peak, with a harder low-frequency spectrum.  

As an example, consider 
an expanding plasma with the same heating profile as we have studied previously: a
cutoff in heating is followed by adiabatic expansion.  But now we introduce baryons into the outflow
and allow the expansion to continue well beyond our previous cutoff time $t_{\rm tot}$, so that
the integration is stopped only when $d\tau_T/d\ln(t) < 1$.  For heavily baryon-dominated
outflows, the plasma must expand by an additional factor $\sim 30$.  

\begin{figure} 
\epsscale{0.9} 
\plotone{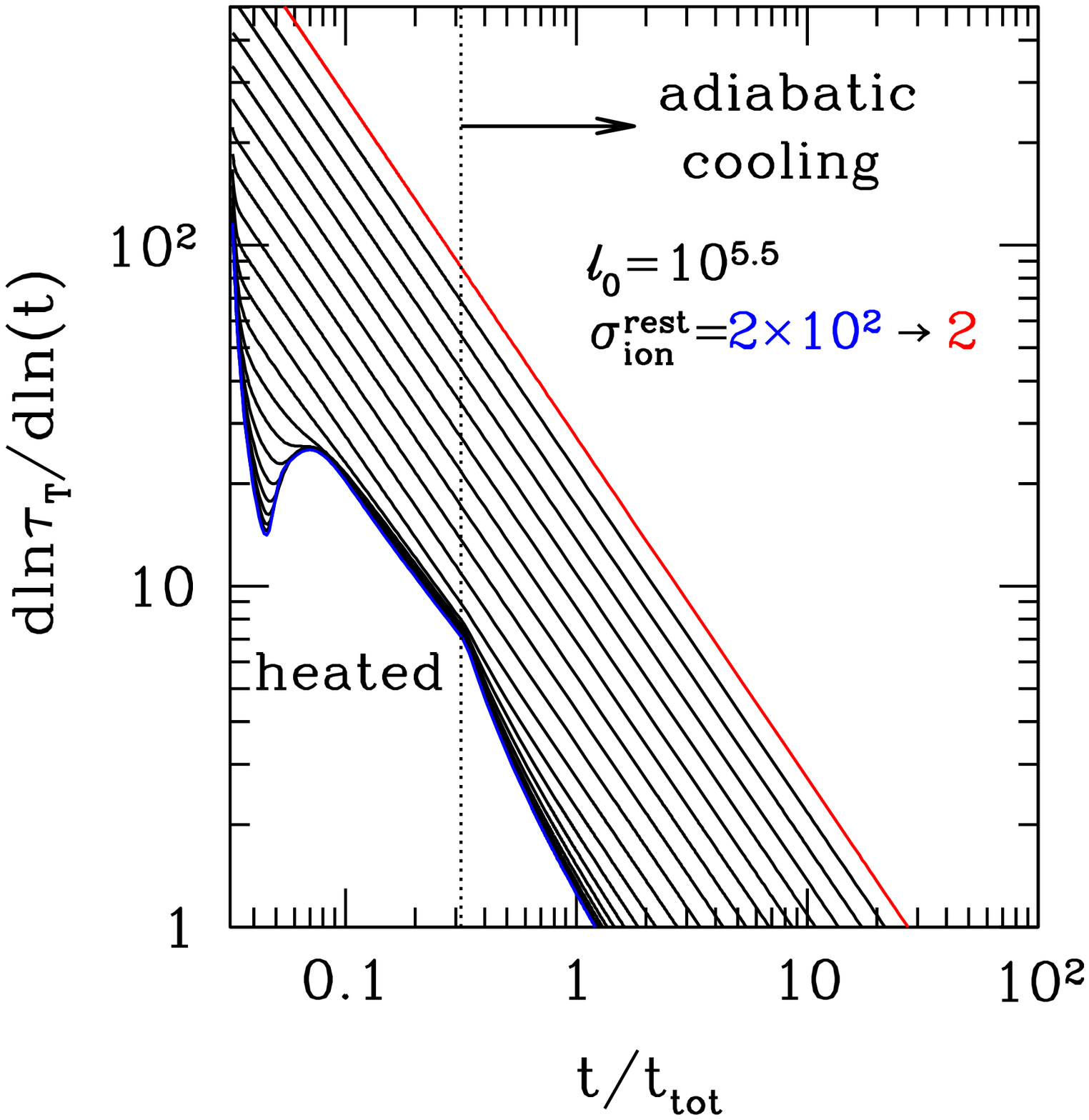}
\caption{Scattering depth in outflows of initial thermal compactness $10^{5.5}$, thermal energy density 
$f_{\rm th} = 0.1$ and various bulk-frame magnetizations, $\sigma_{\rm ion}^{\rm rest} = B^2/4\pi \rho_{\rm ion} c^2 = 2$
(top, red), up to $200$ (bottom, blue).  Flows dominated by the electron-ion component show
$\tau_T \propto t^{-1}$, whereas the pair-dominated flows follow a shallower profile after a 
transient dominated by the annihilation of an initial excess of pairs.} 
\label{fig:tauTb}
\vskip .1in
\end{figure}

\begin{figure} 
\epsscale{0.9}
\plotone{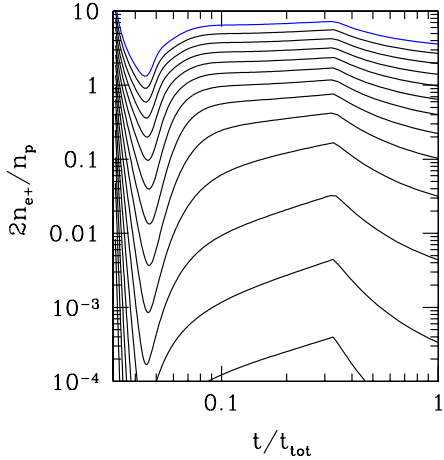}
\caption{Ratio of scattering optical depth to $e^+-e^-$ pairs and that due to the electron-ion
component, in the same sequence of outflows shown in Figure \ref{fig:tauTb}.}
\label{fig:npos}
\vskip .1in
\end{figure}

The optical depth as a function of time is shown in Figure \ref{fig:tauTb} for a bulk-frame 
magnetization 
\be\label{eq:magb}
\sigma_{\rm ion}^{\rm rest} \equiv {B^2\over 4\pi\rho_{\rm ion} c^2}
\ee
varying from $200$ down to $2$.  (Here $\rho_{\rm ion}$ and $B$ are the proper ion rest mass density and
magnetic flux density.)
At the higher values of $\sigma_{\rm ion}$, the scattering depth is dominated by the pairs, but 
there is a transition to a nearly pair-free gas as the magnetization is reduced (Figure \ref{fig:npos}).

\begin{figure} 
\epsscale{0.9}
\plotone{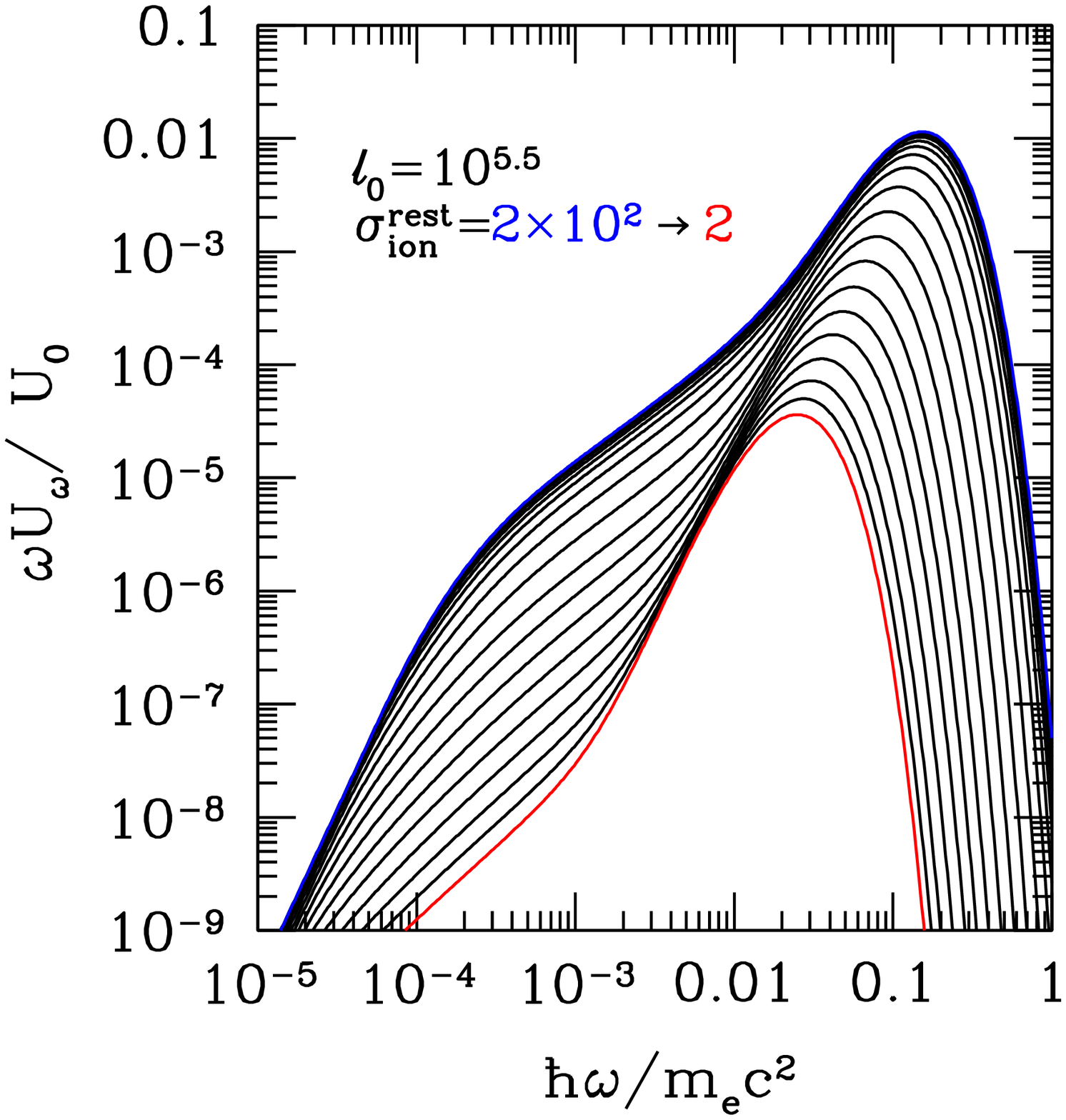}
\plotone{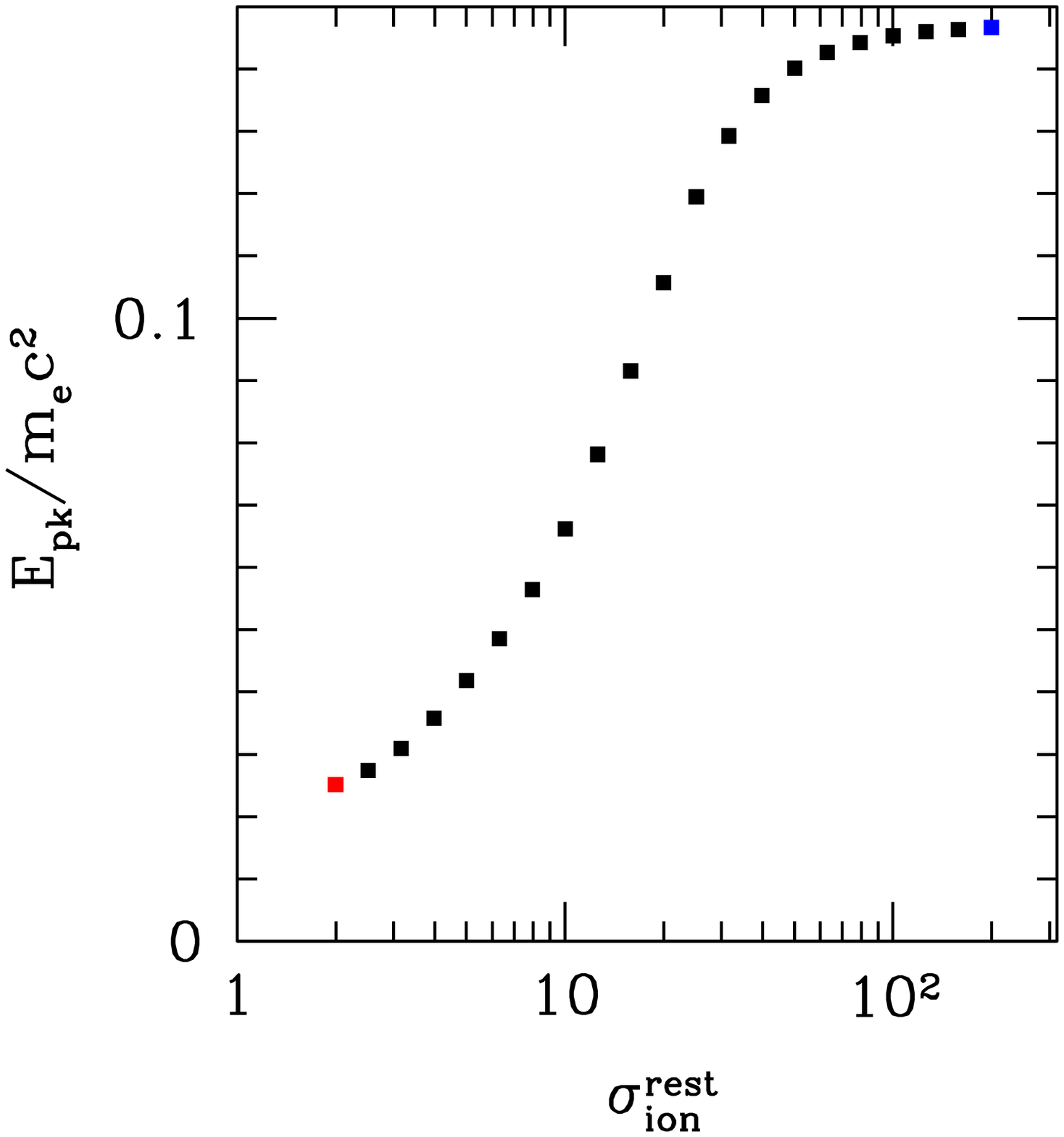}
\plotone{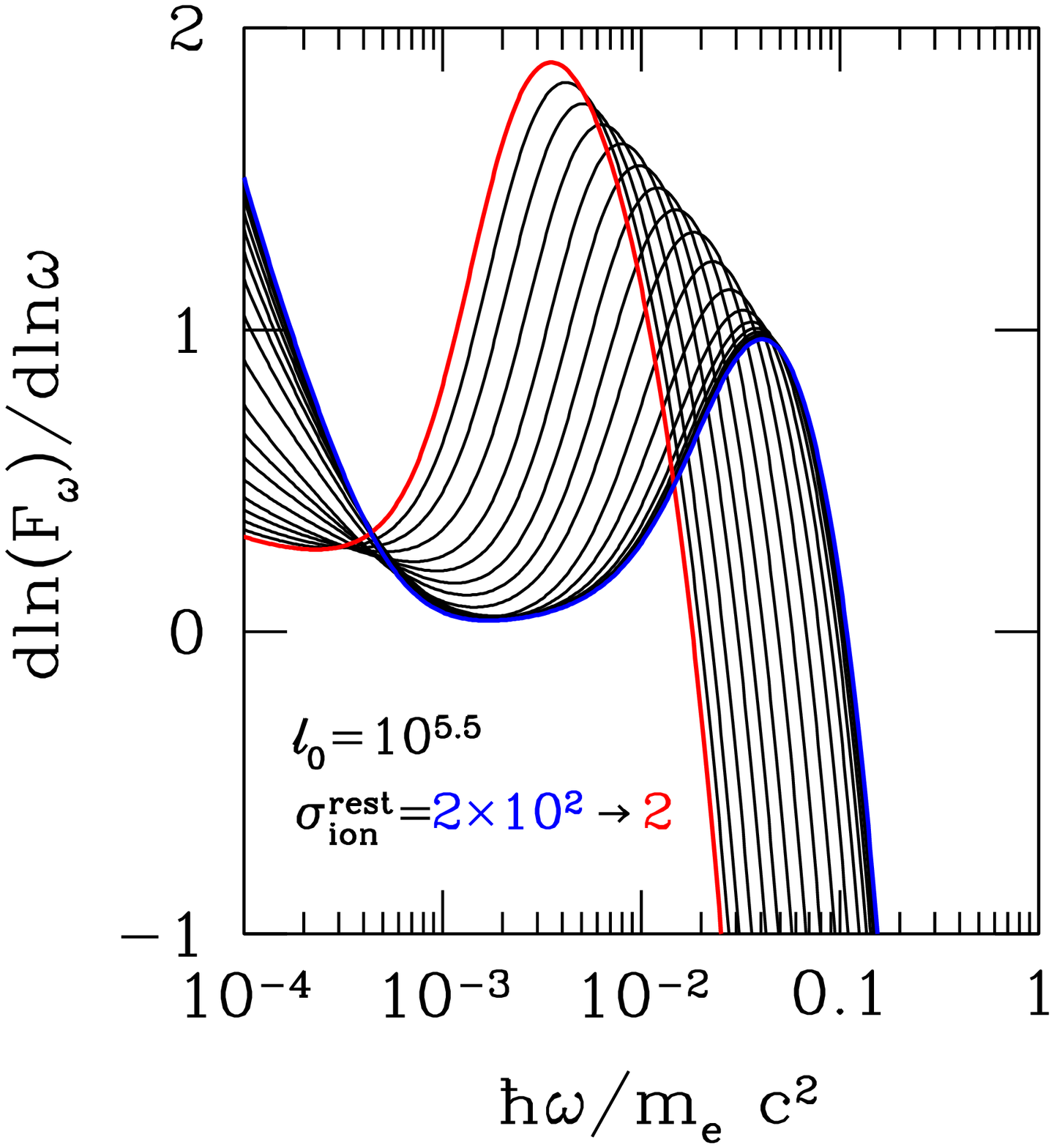}
\caption{Output spectrum of the same sequence shown in Figures \ref{fig:tauTb}, \ref{fig:npos}.
In more baryon-dominated flows, the spectral slope below the peak rises significantly,
and part of the reduction in $E_{\rm pk}$ is due to enhanced upscattering of soft photons
($y_C$ is larger).  The remaining reduction in $E_{\rm pk}$ is due to adiabatic cooling, 
and is therefore model-dependent.  Processing of this spectrum through the photosphere
of an accelerating MHD jet reduces the peak spectral index by $\sim 0.5$ (Section \ref{s:multscatt}).}
\label{fig:specb}
\vskip .1in
\end{figure}

The effect on the output spectrum, including the peak energy and the spectral slope below the peak,
is shown in Figure \ref{fig:specb}.  A reduction in $E_{\rm pk}$ without change in slope is
mainly caused by the additional expansion. But eventually, as the magnetization is decreased
further, the spectrum hardens below the peak.  This is due to the saturation of $y_C$, and
the increased soft photon flux toward the peak.

There is an interesting application to X-ray flashes here, which is discussed further in 
Section \ref{s:summary}.

\subsection{Low Emergent Peak Energy}\label{s:epeakbar}

\begin{figure} 
\epsscale{0.9}
\plotone{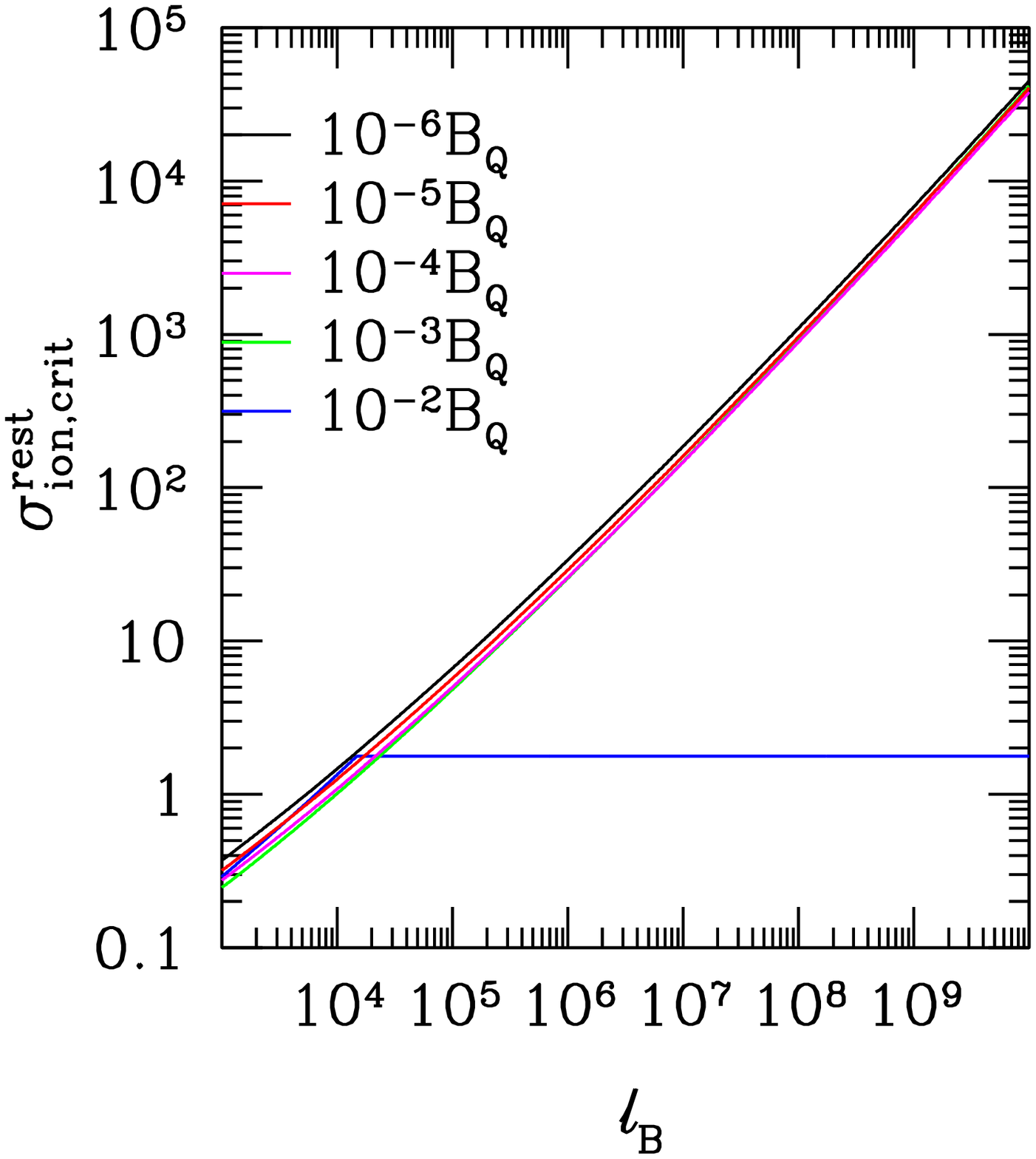}
\caption{Critical magnetization, defined in equation (\ref{eq:sigcr}), below which the electron-ion 
component dominates the scattering optical depth, and pairs largely freeze out.  Applies to the 
scaling solution for an expanding pair plasma derived in Section \ref{s:scaling}.}
\label{fig:sigmab_crit}
\vskip .1in
\end{figure}

The transition between pair-dominated and baryon-dominated outflows can also be considered using the
semi-analytic model of Section \ref{s:scaling}.  The critical baryonic magnetization is obtained by setting 
$Y_e\rho_{\rm ion}/m_p = n_{e^+}+n_{e^-}$, where $Y_e$ is the electron fraction:
\be\label{eq:sigcr}
\sigma_{\rm ion, crit}^{\rm rest} = {B^2 Y_e\over 4\pi (n_{e^+} + n_{e^-})m_pc^2}.
\ee
The result is shown in Figure \ref{fig:sigmab_crit}.

\begin{figure} 
\epsscale{0.9}
\plotone{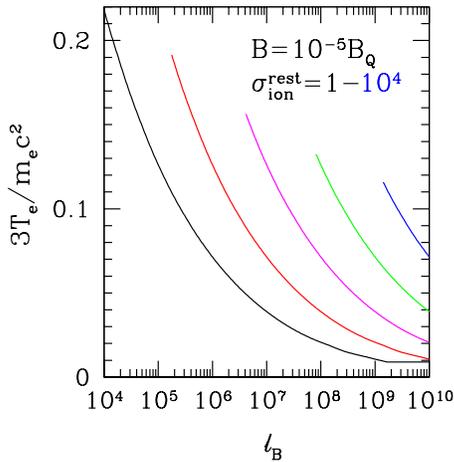}
\caption{Limiting temperature resulting from thermal cyclotron emission in a electron-ion dominated
plasma.   Magnetic compactness varies from $10^4$ down to $1$ (blue to black).
}
\label{fig:temp_baryon}
\end{figure}

During a phase of continuous heating, the analog to equation (\ref{eq:phcrbal}) for the temperature is
\be
m_*^2(T_e) \Te^3 = {\pi f_{\rm th}\over 36\alpha_{\rm em} \tau_T} = 
                  {\pi f_{\rm th}\sigma_{\rm ion}^{\rm rest}(m_p/m_e)\over 72\alpha_{\rm em} \ell_B}.
\ee
Here we have taken $f_\omega = \langle\hbar\omega\rangle/T_e = 3$ and made use of 
$n_\gamma \simeq {1\over 2}F_\gamma t$.  
The result is shown in Figure \ref{fig:temp_baryon}.  It is always larger than the
equivalent black-body temperature $T_{\rm bb} = (f_{\rm th} B^2/8\pi a_{\rm SB})^{1/4}$, except
for the flat part of the lowest-magnetization curve.  

In contrast with the pair-dominated plasma,  the optical depth drops only gradually after 
heating turns off in a baryon-dominated plasma.  Therefore the Wien temperature shown
in Figure \ref{fig:temp_baryon} may significantly exceed the emergent temperature -- as
is demonstrated by the numerical solutions of the preceding section.  

\begin{figure} 
\epsscale{0.9}
\plotone{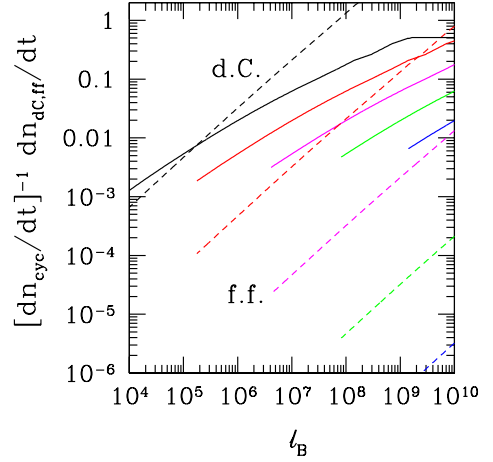}
\caption{Relative importance of double-Compton and free-free emission to the production of soft
photons, in an electron-ion dominated plasma.  Line colors correspond to those
in Figure \ref{fig:temp_baryon}.
}
\label{fig:coolchan_baryon}
\vskip .1in
\end{figure}

Finally, the soft-photon output through the double-Compton and free-free channels, related to thermal 
cyclo-synchrotron emission, is shown in Figure \ref{fig:coolchan_baryon}. 

\subsection{Neutron-rich Electromagnetic Outflows?}\label{s:neutron}

Gamma-ray burst outflows may contain a significant number of neutrons.  This is the case if
the nuclear composition evolves via weak interactions between nucleons and charged leptons
\citep{derishev99,bm00,rossi06}.
The radius at which neutrons and charged ions decouple depends strongly on the Lorentz factor
profile of the outflow, as well as on the magnetization (\ref{eq:magb}).  The neutron
loading of the outflow also will vary with angle:  a wind emanating from a 
torus orbiting a black hole should be neutron-rich.  We focus here on the jet core,
the source of the prompt gamma-ray emission.  

When the magnetization is very large, as considered here, two major changes occur.  First,
the dissipation associated with $n$-ion collisions becomes insignificant.  Second,
the $n/p$ ratio in a Blandford-Znajek jet may be very different (significantly lower) than
that in unmagnetized matter, because positron capture on neutrons is greatly accelerated 
by the large magnetic phase space factor.  We ignore this second complication here, and in 
order to explore the energetics assume comparable numbers of neutrons and protons in the outflow.  

Neutrons and ions (here idealized as protons) decouple where the optical depth for $n$-$p$ collisions (cross
section $\sigma_{pn} \sim 3\times 10^{-26}$ cm$^2$) is about unity \citep{rossi06,beloborodov10}.
In an outflow with large ion magnetization (\ref{eq:magb}), this occurs at a radius
\be
R_{pn} \sim {L_{\rm P\,iso}\sigma_{pn}\over 4\pi\sigma_{\rm ion} \Gamma^2 m_n c^3},
\ee
where $L_{\rm P\,iso}$ is the isotropic Poynting luminosity (\ref{eq:lpiso}).
Beyond this radius, the neutrons and ions develop a relative speed $\sim c$, and inelastic
collisions (with a cross section $\sim 0.1\sigma_{pn}$) create pions, and multiple
pairs by a cascade process.  In an unmagnetized fireball, one sets $L_{\rm P\,iso} \rightarrow L_{\rm m\,iso}$
(the matter kinetic luminosity) and $\sigma_{pn} \rightarrow \Gamma$.  Then one finds that the compactness at $R_{pn}$
takes a large value,
\be
\ell(R_{pn}) = {L_{\rm m\,iso}\sigma_T\over 4\pi m_e c^3 R_{pn}} \sim {\sigma_T\over\sigma_{pn}}
{m_n\over m_e} \sim 4\times 10^4,
\ee
independent of the details of the flow.

The spectral signature of this process has been calculated in detail by \cite{vurm11}
for a relatively weak magnetization, $\sigma_{\rm ion} \lesssim \Gamma$
($\sigma_{\rm ion}^{\rm rest} \lesssim 1$), and in the presence of 
a seed black-body radiation field.  Synchrotron cooling off the magnetic field
prevents the formation of a hard spectral tail when $\sigma_{\rm ion} \gtrsim 0.1\,\Gamma$,
but supplements the low-frequency spectrum by self-absorbed synchrotron emission
with a frequency-dependent photospheric radius.  (Essentially the same mechanism
was invoked by \cite{bk79} to explain flat-spectrum radio emission
from relativistic jets with power-law particle distributions.)

The size of this decoupling zone $R_{pn}$ depends sensitively on the acceleration
profile of the jet.  \cite{beloborodov10} considered an unmagnetized, neutron-loaded
fireball with Lorentz factor increasing linearly with radius from an engine of
size $R_s \sim 100$ km.   Then $\Gamma(R_{pn}) \sim 270\,L_{\rm P\,iso,51}^{1/4}$ and 
$R_{pn} \sim 2.7\times 10^9\,L_{\rm P\,iso,51}^{1/4}R_{s,7}^{3/4}$ cm.  

A strongly magnetized jet will have a very different radial flow profile.
A self-similar jet relaxes typically to $\Gamma \sim 1/\theta$ \citep{lyubarsky09},
although strong departures from this are possible in the non-self-similar density
profile of a collapsing stellar core.  Considering a magnetized outflow with 
$\Gamma \sim 1/\theta$ and $\sigma_{\rm ion} \gg \Gamma$, one finds instead
\be
R_{pn} \sim 6\times 10^{10}\,L_{\rm P\,iso,51}\sigma_{\rm ion,5}^{-1}\left({\Gamma\over 3}\right)^{-2}
\quad{\rm cm}.
\ee
Here we have normalized the magnetization to the level (\ref{eq:sigbreak}) that gives a pair-dominated
thermal photosphere.  Only a tiny fraction $\sim \Gamma/\sigma_{\rm ion} = 3\times 10^{-5}\sigma_{\rm ion,5}^{-1}
(\Gamma/3)$ of the outflow energy is thermalized by $n$-ion collisions.  

How sensitive is this conclusion to the assumed value of $\sigma_{\rm ion}$?  
A robust lower bound $\sigma_{\rm ion} \gtrsim \Gamma \sim 10^2$-$10^3$ is needed to create a GRB.  
Let us suppose that $\Gamma \sim \theta^{-1}$ out to a distance $R_{\rm exp}
\sim 10^{11}$ cm from the engine, followed by free expansion, $\Gamma \sim
\theta^{-1}(r/R_{\rm exp})$.  Then one finds $R_{pn} \sim 4\times 10^{11}
L_{P,51}^{1/3} (3\theta)^{2/3}R_{\rm exp,10}^{2/3}\sigma_{\rm ion,3}^{-1/3}$~cm, and $n$-ion collisions 
thermalize only a fraction 
\be
{\Gamma \over \sigma_{\rm ion}} \sim 4\times 10^{-3}\,{L_{P,51}^{1/3}(3\theta)^{2/3}\over R_{\rm exp,11}^{1/3}
\sigma_{\rm ion,3}^{4/3}}
\ee
of the outflow energy.  Only $\sim 10\%$ of this is converted to non-thermal pairs through the pion-creating
channel.

\begin{figure}
\epsscale{0.9} 
\plotone{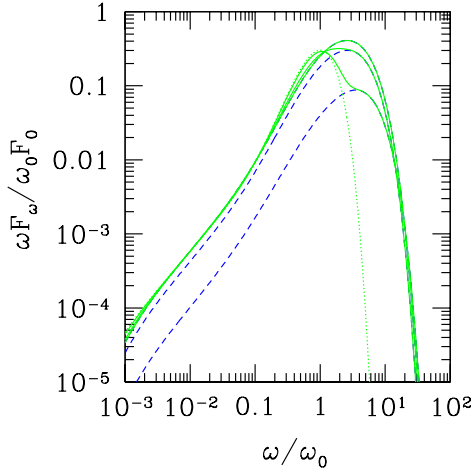}
\caption{Imprint of bulk Compton scattering on the output spectrum 
from the photospheric region of an accelerating, magnetized jet.  Dotted green line:  input spectrum from 
Section \ref{s:expand}, corresponding to initial (final) thermal compactness
$\ell_0 = 10^6$ ($\ell_{\rm f} = 10^4$).   
Dashed blue lines:  only the scattered spectrum without the partially transmitted input.
The solid curves show various values of the net radial optical depth  starting
from the input radius, which sits at $\sim 0.8$ times the jet breakout radius.
(See \citealt{russo13b} for futher details of the jet model.)}
\label{fig:jet_spec}
\vskip .2in
\end{figure}
\begin{figure}
\epsscale{0.9}
\plotone{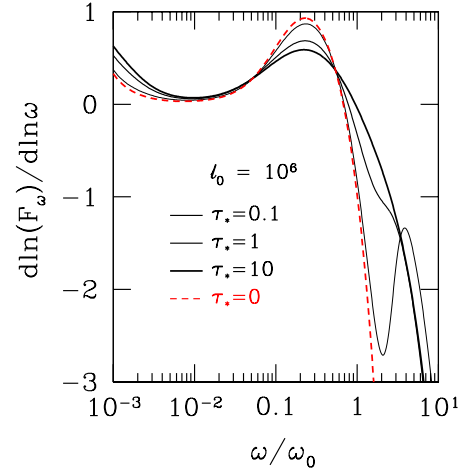}
\plotone{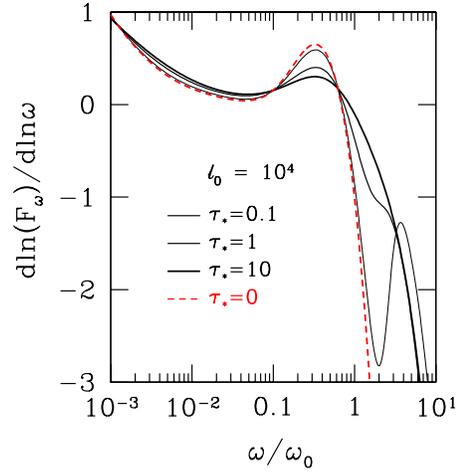}
\caption{Top panel: spectral slope $d\ln(F_\omega)/d\ln\omega$ corresponding to Figure
\ref{fig:jet_spec} (compactness $\ell_{\rm f} = 10^5$ at jet breakout), 
and the analogous result for $\ell_{\rm f} = 10^2$.  Black curves show
the effect of raising the radial scattering depth $\tau_*$ at breakout.  Effect
of delayed dissipation on the hard spectral tail is not included.}
\label{fig:jet_spec_slope}
\vskip .1in
\end{figure}

\section{Spectral Imprint of Bulk Compton Scattering During Jet Breakout}\label{s:multscatt}

As a jet breaks out of a confining medium, it experiences rapid acceleration
as magnetic flux surfaces diverge (e.g. \citealt{tchek10}).   This fast expansion
is also plausibly associated with a sudden drop in bulk heating, and a rapid
annihilation of electron-positron pairs.  In \cite{russo13a,russo13b} we calculated 
the imprint of bulk Compton scattering on the outgoing gamma-ray spectrum, assuming that
the input spectrum was GRB-like ($F_\omega =$ const at low frequency, with an
exponential cutoff above a frequency $\omega_0$).   The net result was that
the spectral peak was pushed higher in frequency, along with the low-frequency
flat spectrum

Here we replace this simplified input spectrum with the one obtained
in Section \ref{s:expand} for a continuously heated, and expanding, magnetized pair plasma.
We extract the spectrum after heating has stopped, and as the pairs are annihilating.
To facilitate comparison, the frequency is normalized to the peak $\omega_0$ of $\omega F_\omega$
in this input spectrum.  

The result, for an initial (final) thermal compactness $\ell_0 = 10^6$ ($\ell_{\rm f} = 10^4$),
is shown in Figure \ref{fig:jet_spec}.  The slope of $F_\omega$ is shown in Figure
\ref{fig:jet_spec_slope}, along with the result for the $\ell_{\rm 0} = 10^4$ plasma.
One observes in the slope an extended low-frequency plateau with $d\ln(F_\omega)/d\ln\omega = 0$,
and a localized bump that extends to $\sim 0.7 - 0.9$ at $\omega \sim \omega_0/3$.  This
peak in $d\ln(F_\omega)/d\ln\omega$ is significantly reduced in the scattered spectrum, 
by $\sim -0.5$, starting from an input optical depth $\sim 10$.  

\section{Focused Poynting-Dominated Jets}\label{s:poyntingjet}

Gamma-ray bursts emit such extreme fluxes of radiant energy -- isotropic energies
reaching at least $E_{\gamma\,\rm iso} \sim 10^{55}$ erg in the source rest frame \citep{abdo09} -- that 
purely hydrodynamic modes of energy transport are disfavored.  In principle, a jet could be 
accelerated within a de Laval nozzle \citep{br74} that forms in the envelope 
material:  for example, along the rotation axis in a collapsar, where the ram pressure of 
the infalling material is reduced.  

We first consider such a hydrodynamic jet, showing that for realistic pre-collapse mass profiles
it is inconsistent with the largest observed burst energies.  Poynting-dominated jets, by contrast, allow for a strong 
focusing of streamlines toward the jet axis by magnetic pressure gradients (\citealt{lyndenbell03},
and references therein). 

 We next work out the relation between isotropic jet luminosity
and opening angle in a steady, axisymmetric, and highly magnetized jet.  This relation, in 
combination with the results of Section \ref{s:spectrum}, is used in Section \ref{s:amati} to 
give a first-principles derivation of the Amati et al. boundary in the $E_{\rm pk}$-$E_{\gamma\,\rm iso}$ 
plane.  

\begin{figure}[h]
\plotone{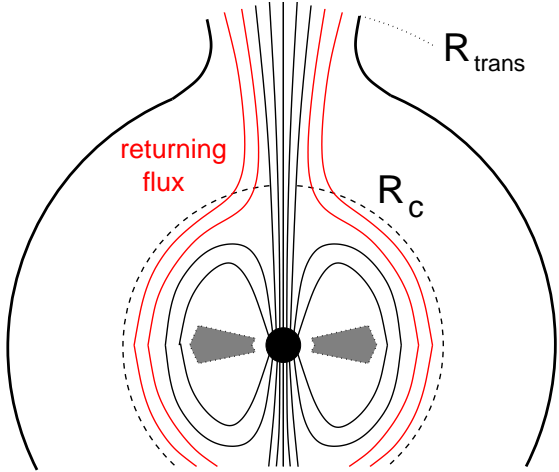}
\vskip .1in
\caption{A jet transmits energy from the ergosphere of a stellar-mass black hole (not to scale) that has formed
by gravitational collapse. A fraction $f_{\rm B,j}$ of the magnetic flux threading the hole
penetrates the surrounding envelope.  When this outgoing flux has constant sign, as depicted, it must return
through an annulus of shocked jet material (red lines).   The remaining flux is trapped in a relativistic bubble of
radius $\gtrsim R_c \sim 10^9$ cm, corresponding to the size of the pre-existing CO core in a collapsar,
or to the size of the neutron-rich debris cloud that surrounds the remnant of a binary neutron star
merger.  The connectivity of the trapped magnetic field lines is only schematic:  some may still connect
to the collapsed torus, and turbulent mixing with some stellar material is likely.  The envelope of shocked
jet material expands at most trans-relativistically.  It provides a confining sheath for the relativistic
jet core out to a radius $R_{\rm trans} \sim c t_{\rm col}$ which may exceed the original envelope radius.}
\label{fig:jet}
\vskip .2in
\end{figure}

\subsection{Peak Isotropic Luminosity of a Hydrodynamic Jet Flowing from a Confined Bubble}\label{s:lisomax}

\begin{figure}[h]
\epsscale{0.9}
\plotone{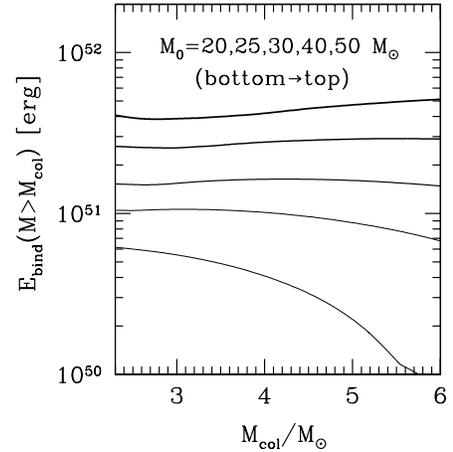}
\caption{Binding energy of core material lying outside a given collapsed mass $M_{\rm col}$, as a function of the
zero-age main sequence (ZAMS) mass $M_0$ of the progenitor.  $M_{\rm col}$ enters
into equation (\ref{eq:lptheta}) for the isotropic Poynting luminosity of a jet emerging from 
a trapped relativistic bubble in a massive stellar core; and equation (\ref{eq:amatijet}) normalizing
the spectral peak frequency of a heated jet that freely expands after
breakout from a Wolf-Rayet star.  Core profiles are obtained from the MESA star integrator \citep{paxton13}
in the simplest case of non-rotating single stars.  The mapping between ZAMS mass and pre-collapse mass
profile will vary as these other (uncertain) degrees of freedom are included; the goal here is to probe
a plausible range of pre-collapse profiles.}
\label{fig:binding}
\vskip .1in
\end{figure}

First consider a hydrodynamic jet emerging from a bubble of hot plasma injected into
the core of a massive star (Figure \ref{fig:jet}).  The net binding energy of the CO material is $E_{\rm bind} 
\sim 10^{51}$ erg outside a radius $R_c \sim (1-3)\times 10^9$ cm:  Figure \ref{fig:binding} shows the 
result for stellar models of various mass constructed using the MESA code \citep{paxton13}.

If the bubble material is sufficiently relativistic to
drive a GRB, its pressure is dominated by radiation, $P_{\rm rad,0} \sim E_{\rm bind}/4\pi R_c^3$.
The maximum energy flow out of such a static, confined bubble is \citep{br74} 
$F_{\rm rad}^* A^* \sim (8/3^{3/2})P_{\rm rad,0} A^* c$, where the jet cross section $A$ 
and energy flux $F_{\rm rad}$ are measured at the sonic radius $R^* \sim 2R_c$.  The radiation energy flux must decrease
from this point outward as the jet expands, because the radiation is still tied to high density of $e^\pm$ pairs.

We therefore obtain a strong upper bound on the isotropic jet luminosity,
\ba\label{eq:lmax}
L_{\rm iso} &<& 4\pi R_c^2 F_{\rm rad}^* = {8c\over 3^{3/2}R_c} E_{\rm bind}\nn
           &\sim& 4.6\times 10^{52}\,{E_{\rm bind,51}\over R_{c,9}}\quad{\rm erg~s^{-1}},
\ea
and on the jet energy
\ba\label{eq:emax}
E_{\rm P\,iso} \;&\sim&\; L_{\rm P\,iso} t_{\rm col}\nn
              \;&<&\; 7.3\times 10^{52}\,{E_{\rm bind,51}R_{c,9}^{1/2} \over (M_{\rm col}/3~M_\odot)^{1/2}}\,
\left[{t_{\rm col}\over 2t_{\rm ff}(R_c)}\right]^{1/2}\quad{\rm erg}.\nn
\ea
Here $M_{\rm col}$ the collapsed mass inside $R_c$, and $t_{\rm ff} \sim [R_c^3/2GM_{\rm col}]^{1/2} = 
1.1\,R_{c,9}^{3/2}(M_{\rm col}/3~M_\odot)^{-1/2}$~s the free-fall time.

\begin{figure}[h]
\epsscale{0.9}
\plotone{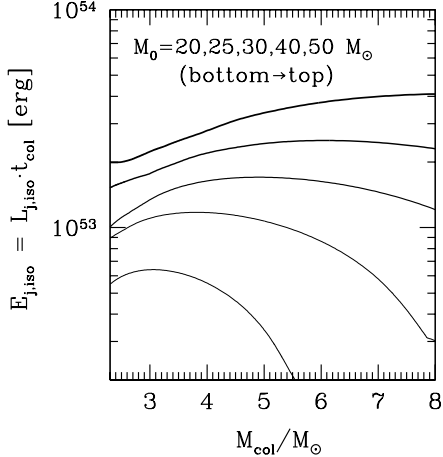}
\caption{Maximum isotropic jet energy (\ref{eq:emax}) carried by relativistic fluid flowing through
a nozzle out of a confined bubble of pre-collapse radius $R_c$ surrounding a collapsed core of mass 
$M_{\rm col}$.   These curves set a strong upper bound, because they do not allow for jet
expansion outside the nozzle. The bubble pressure $P_{\rm rad}$ is determined by setting the bubble energy 
$4\pi R_c^3 P_{\rm rad}$ equal to the binding energy of material outside the mass cut.}  
\label{fig:jetflux}
\vskip .1in
\end{figure}

\begin{figure}[h]
\epsscale{0.9}
\plotone{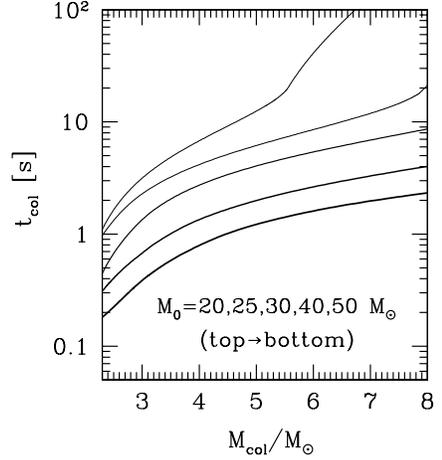}
\caption{Collapse time of material inside the mass cut $M_{\rm col}$, estimated to be twice
the free-fall time $(R_c^3/2GM_{\rm col})^{1/2}$  This sets a lower bound to the burst duration,
which is also influenced by continuing accretion through a collapsed torus.}
\label{fig:tcol}
\vskip .2in
\end{figure}

The results obtained from MESA stellar models are shown in Figures \ref{fig:jetflux} and \ref{fig:tcol}.
For each model, we vary the collapsed mass, taking the net binding energy of all material outside that mass cut.  
The collapse time $t_{\rm col}$, estimated to be twice the free-fall time from radius $R_c$, sets 
a lower bound to the burst duration (Figure \ref{fig:tcol}).  The observed isotropic jet energy 
$E_{\rm P\,iso}$, radiated over a duration $t_{\rm col} \sim 10$ s, must lie below the value shown.

The ram pressure of infalling core material does not allow a significant increase in
jet energy.  At a radius $r < R_c$, the confining pressure imparted to a jet of opening
angle $\theta_j$ is $P_{\rm ram,\perp} \sim \rho (\theta_j v_r)^2$, where $v_r$ is the net infall 
velocity.  This sets a limit
\be
4\pi r^2 (4P_{\rm ram,\perp}) c \;\sim\; \theta_j^2 {4c\over R_c} E_{\rm bind}\left({r\over R_c}\right)^{-1/2}
{t_{\rm ff}(R_c)\over t_{\rm col}}
\ee
to the isotropic-equivalent flow of enthalpy.

Greater confinement is afforded by a collapsed torus orbiting a black hole engine.
The hydrostatic pressure that accumulates parallel to the rotation axis greatly exceeds
the ram imparted by continuing infall.  But this confinement only extends over a limited
range in radius, pointing to an additional focusing mechanism at larger distances from the engine.


\subsection{Magnetized, Axially Symmetric, Relativistic Jet}

Even in a very focused jet, the outflowing material extends across a transverse dimension
$\theta_j R_c$ much larger than the light-cylinder radius $c/\Omega_f$ of the engine.\footnote{Here and
in the remainder of this Section, all variables refer to the inertial frame in which the jet is
propagating.  Quantities in the frame co-moving with the jet fluid are primed.}
Magnetic field lines tied to the engine rotate, in a steady,
axisymmetric jet, with a constant pattern angular velocity $\Omega_f$.  This is the angular velocity
of a material star, or $1/2$ the angular velocity of a black hole \citep{bz77}.
The toroidal magnetic field is then
\be
B_\phi = {v_\phi - \Omega_f r\sin\theta\over v_P} B_P \simeq -{\Omega_f r\theta\over c}B_r
\ee
in a jet with poloidal flux density $B_P$ and velocity $\vec v = \{v_P,v_\phi\}$.  We
focus on narrow, relativistic jets, within which $B_P \simeq B_r$, $v_P \simeq v_r \simeq c$ 
in spherical coordinates.  

The radial Poynting flux is
\be
S_r = {E_\theta B_\phi\over 4\pi} c \simeq {B_\phi^2\over 4\pi} c
\ee
and the isotropic Poynting luminosity
\be\label{eq:lpiso}
L_{\rm P\,iso} = 4\pi r^2 S_r = {1\over \theta^2} [B_r (r\theta)^2]^2 {\Omega_f^2 \over c}.
\ee

A simple relation between $L_{\rm P\,iso}$ and the core parameters
can easily be derived on the premise that some of the poloidal magnetic flux extending
from the engine is distributed in a broad fan and captured into a relativistic 
bubble within the core.  The remaining fraction $f_{\rm B,j}$ propagates out of the star behind the 
jet head.  Then
\be
L_{\rm P\,iso} = \left({f_{\rm B,j}\over \theta}\right)^2 B_H^2 R_H^4{\Omega_f^2\over c}.
\ee
Here $B_H$ is the poloidal flux density threading a black hole of radius $R_H$.  

As the collapse continues and magnetic flux builds up around the black hole, the total
luminosity $L_j$ in the two counter-propagating jets increases with time, until 
$L_j > E_{\rm bind}/t_{\rm col}$.  When threaded by a split-monopolar magnetic
field, the black hole releases $L_j \sim 2B_H^2 R_H^4\Omega_f^2/3 c$ \citep{bz77,tchek10}.
The isotropic energy of a single jet is, then,
\be\label{eq:lptheta}
E_{\rm P\,iso} = L_{\rm P\,iso} t_{\rm col} 
  \sim 1.5\times 10^{51}  \left({f_{\rm B,j}\over \theta}\right)^2 E_{\rm bind,51}\quad{\rm erg}.
\ee

Even though the isotropic energy flux must vary with angle within the jet, this expression has an
important feature:  $E_{\rm P\,iso}$ depends only on one local variable $\theta$ in addition to
the global parameters $f_{\rm B,j}$ and $E_{\rm bind}$.  This has interesting implications for 
GRB color-luminosity relations, which we address in Section \ref{s:amati}.

\subsection{Compactness and Co-moving Energy Density at Jet Breakout}\label{s:breakout}

The breakout of a relativistic jet from a Wolf-Rayet star must be accompanied by a broader, 
trans-relativistic cocoon (e.g. \citealt{rr02}, \citealt{lazzati09}).  This cocoon provides pressure that helps to confine 
the jet, at least out to a radius
\be
R_{\rm trans} \sim c t_{\rm col} \sim 3\times 10^{11} \left({t_{\rm col}\over 10~{\rm s}}\right)\quad {\rm cm},  
\ee
where $t_{\rm col}$ is the collapse time of the material powering the jet.

There is a downward gradient in Lorentz factor away from the relativistic jet and into the coccon.  As fresh
relativistic material continues to be injected from the engine, this gradient can be maintained on
an angular scale $\delta\theta \lesssim 1/\Gamma$, so that causal contact is maintained across the gradient.\footnote{Even
in a part of this heated boundary layer that is strongly depleted in baryons, we find optical depths large enough to suppress
the diffusion of photons across the layer (see Figure \ref{fig:tauT}).  The photons therefore do not see the gradient,
as envisaged by \cite{lundman13}.  The available shear kinetic energy can still couple to the photons through higher-frequency 
Kelvin-Helmholtz modes.} 

Material of an intermediate Lorentz factor $\Gamma_2$ provides confinement for a faster
core out to a distance $\sim 2(\Gamma_2)^2 R_{\rm trans}$.  Therefore complete deconfinement of a Lorentz factor
$\Gamma$ jet is delayed out to a radius 
\be\label{eq:rstar}
R_{\rm sheath}(\Gamma) \sim 2\Gamma^2 R_{\rm trans}.
\ee
Evidence for this type of extended structure is seen in the 2D hydrodynamic simulations of \cite{lazzati09}.

Even after taking into account this re-scaling, the radiation compactness remains very 
high at breakout.  The net isotropic luminosity, including contributions from
both thermal energy and toroidal magnetic field, is
\ba
L_{\rm iso} &=& \left[{4\over 3}\Gamma^2 U'_{\rm th} + {(\Gamma B_\phi')^2\over 4\pi}\right] 4\pi r^2 c\nn
            &=& {4\over 3}\Gamma^2 U'_{\rm th}\left(1 + {3\over 2f_{\rm th}}\right) 4\pi r^2 c.
\ea
Focusing on the thermal component, which is dominated by photons, and normalizing the energy density
to $m_ec^2/\bar{\lambda}_c^3$ in the co-moving frame, one finds
\be
U'_{\rm th} = 5.3\times 10^{-13}\,{L_{\rm \gamma\,iso,51}\over t_{\rm col,1}^2}\,
\left({\Gamma\over 3}\right)^{-5}\;{m_ec^2\over\bar{\lambda}_c^3}.
\ee
The radiation field is dilute in the sense that a thick pair gas cannot be maintained
in full thermodynamic equilibrium:
\be
T_{\rm bb}' = \left({U_{\rm th}'\over a_{\rm SB}}\right)^{1/4} = 0.48\,
{L_{\rm \gamma\,iso,51}^{1/4}\over t_{\rm col,1}^{1/2}}\left({\Gamma\over 3}\right)^{-5/4}\quad{\rm keV}.
\ee
In spite of this, the bulk-frame compactness is still large:
\be\label{eq:compbreak}
\ell_{\rm th}' = {\sigma_T U_{\rm th}' (r/\Gamma)\over m_e c^2} = 1.1\times 10^7\,  
                 {L_{\rm \gamma\,iso,51}\over t_{\rm col,1}}\left({\Gamma\over 3}\right)^{-3}.
\ee

\subsection{Minimal Magnetization}

\begin{figure}
\epsscale{0.9}
\plotone{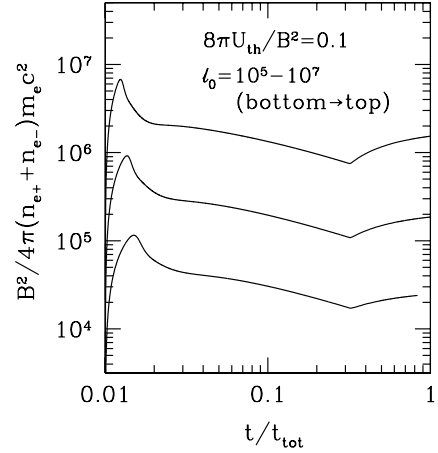}
\caption{Magnetization $\sigma_\pm^{\rm rest} = B^2/4\pi (n_{e^+} + n_{e^-})m_ec^2$ in the rest frame
of expanding outflows of initial thermal compactness $\ell_0 = 10^5$-$10^7$,
with heating turning off at $10^{-0.5}t_{\rm tot}$.  See also Figures \ref{fig:Te}-\ref{fig:fintemp}.
When the plasma is pair-dominated, the magnetization due to an electron-ion component is at
most a fraction $Y_e/1836$ of this.  A somewhat higher compactness, and therefore magnetization, is implied 
during jet breakout (equations (\ref{eq:compbreak}) and (\ref{eq:sigbreak})) than is provided by our
calculations.}
\label{fig:sigma}
\vskip .1in
\end{figure}

After heating turns off and the pair plasma reaches a scattering depth $\sim$ few, further 
annihilation freezes out.  We then obtain an estimate of the magnetization in the outflowing material.  
The magnetization as a function of time, up until freeze-out, is shown in Figure \ref{fig:sigma}.

By considering the evolution of a pure pair plasma, we are implicitly setting a lower bound on the 
magnetization imposed by the ion inertia.  Setting the number density of protons in the outflow
equation to $n_{e^-} + n_{e^+}$ gives the critical magnetization (\ref{eq:sigcr}).   Re-writing
this in terms of the breakout compactness (\ref{eq:compbreak}) and transforming to the frame of the engine
gives
\ba\label{eq:sigbreak}
\sigma_{\rm ion,crit} &=& \Gamma\sigma_{\rm ion,crit}^{\rm rest} = 
{2\Gamma\over f_{\rm th}}\left({m_e\over m_p}\right)\ell'_{\rm th} \nn
&=& 3.6\times 10^4 {L_{\gamma\,\rm iso,51}\over f_{\rm th} t_{\rm col,1}}
\left({\Gamma\over 3}\right)^{-2}.
\ea


\section{Implications for GRBs}\label{s:amati}

\subsection{Observed Relation between Spectral Peak and Isotropic Burst Energy}

Gamma-ray bursts with known redshifts can be labeled by the isotropic-equivalent bolometric energy
$E_{\gamma\,\rm iso}$ in the hard X-ray/gamma-ray band, and the photon energy $E_{\rm pk}$ where
the spectral energy flux $E^2 dN_\gamma/dE$ peaks.  The measured events generally sit above the Amati et al. 
line \citep{amati02},
\be\label{eq:amati}
E_{\rm pk} = 100~{\rm keV}\,\left({E_{\gamma\,\rm iso}\over 10^{52}~{\rm erg}}\right)^{1/2}.
\ee
For a recent re-analysis, which emphasizes this line as a boundary in the $E_{\rm pk}$-$E_{\gamma\rm\, iso}$ plane,
see \cite{heusaff13}. 

Because GRBs typically show extended hard tails of emission extending above $E_{\rm pk}$, bursts
with high $E_{\gamma\rm\, iso}$ but low $E_{\rm pk}$ would be detected if they existed \citep{piran96}.
Hence it appears that $E_{\rm pk}$ is buffered from below.  The simplest candidate mechanism
involves a thermal photon gas, which supplies Compton seeds.  

We have demonstrated that $E_{\rm pk}$ in a strongly magnetized pair gas
lies well above that encountered in a baryon-dominated outflow of the same compactness, due to the 
buffering of the Compton parameter at temperatures below $m_ec^2$.

In addition, there is only a modest adiabatic drop in temperature (a factor $\sim 0.5$) after heating 
ends, because the pairs rapidly annihilate.  The bulk-frame peak energy adjusts to $E_{\rm pk}' 
\simeq 0.1\,m_ec^2$ for a photospheric compactness $\gtrsim 10^4$ (see Figure \ref{fig:fintemp}).  
The observed peak energy is then
\be\label{eq:epkobs}
E_{\rm pk} \simeq {4\over 3}\Gamma\cdot 0.1 m_ec^2 = 70\,\Gamma\quad{\rm keV}.
\ee

We have also, in Section \ref{s:lisomax}, considered the binding energy of the massive CO cores which are believed
to be hosts for long GRBs.  This provides a rough upper envelope to the total energy released by a GRB jet,
if the accretion time through the torus surrounding the black hole is shorter than the collapse time.  For
the most massive progenitors, the collapse time approaches $\sim 10$ s at an enclosed mass of $4\,M_\odot$
(Figure \ref{fig:tcol}).  Longer $T_{90}$ burst durations may imply collapse from 
larger $R_c$ (but with a somewhat smaller $E_{\rm bind}$); or a long viscous time in a collapsed and centrifugally 
supported torus, which would require a rapidly-rotating progenitor.

Our goal here is to work out the minimum $E_{\rm pk}$ corresponding to a given $E_{\gamma\rm\, iso}$.
We only consider the GRB emission up to, and including, the peak, with the
implication that the bolometric gamma-ray energy would typically be at least $\sim 2$ times larger. 
This energy also depends on the efficiency of conversion of Poynting flux to
photons; we take a maximum value $E_{\gamma\,\rm iso}/E_{\rm P\,iso} \sim 1/2$ at breakout.  The core binding
energy reaches $E_{\rm bind} \sim 4\times 10^{51}$ erg at a progenitor ZAMS mass $40\,M_\odot$.  

Substituting these maximum parameter values into equation (\ref{eq:lptheta}) and inverting gives
\be\label{eq:amatijet}
E_{\rm pk,min} = 130\,E_{\gamma\,\rm iso,52}^{1/2}\,{\Gamma\theta\over f_{\rm B,j}}
\left({E_{\rm bind}\over 4\times 10^{51}~{\rm erg}}\right)^{-1/2} \quad{\rm keV}.
\ee
This lies close to equation (\ref{eq:amati}) if $\Gamma\theta \sim 1$.  Note that GRB jets with
\vskip .05in\noindent
i) lower radiative efficiency;
\vskip .05in\noindent
ii) lower escaping magnetic flux fraction $f_{\rm B,j}$; and/or
\vskip .05in\noindent
iii) originating from CO cores with lower binding energy, 
\vskip .05in\noindent
have higher $E_{\rm pk}$ for a given $E_{\gamma\rm\,iso}$ and sit {\it above} 
the Amati et al. line.  Bulk Compton scattering of the thermal emission during breakout also tends to raise
$E_{\rm pk}$.

One advantage of this relation is that it does not depend on the distance from the engine, because
$E_{\rm pk}'$ is directly related to the electron rest mass.  Related attempts based on a local
black-body approximation do not have this feature \citep{tmr07,lmmb13}.

\subsection{Implications for Magnetic Reconnection}\label{s:rec}

Magnetic reconnection is a promising source of variability and non-thermal emission in GRB outflows 
\citep{thompson94,spruit01,giannios07,zhang11,mckinney12}, as well as pulsar winds \citep{coroniti90,lyubarsky01} and
more dilute radio-emitting jets from black holes \citep{romanova92}.  

\subsubsection{Magnetic Field Geometry}

The geometry of the magnetic field depends on the type
of source.  A striped toroidal field geometry has been demonstrated in force-free
calculations of winds from rotating neutron stars with tilted magnetic dipoles \citep{spitkovsky06}.
This result may not, however, be relevant for GRBs: we argue in this
paper that the extreme baryon purity of GRB outflows points to the rapid formation of an event horizon
in the engine.  Calculations of jets from black hole magnetospheres which are fed by magnetic flux of
variable sign \citep{beckwith08} suggest that the flux threading the horizon rapidly reconnects
and maintains a uniform sign that reflects an average over the accretion history.    This uniform sign
of poloidal field then translates into a uniform sign of the wound-up toroidal field.  

The outgoing magnetic flux in the jet core -- which connects to the central black hole -- is surrounded
by an annulus of returning flux (see Figure \ref{fig:jet}).  The outgoing and returning magnetic fields
are separated by a cylindrical current sheet.  Reconnection could occur at this sheet, although it must be 
at least partly suppressed by strong radial velocity shear:  the return flux contains jet fluid that has 
shocked at the jet head and then fallen behind it as it escapes the star.  

In magnetic tower models
\citep{lyndenbell03,um06} the outgoing and returning flux both connect to a differentially rotating
object, and are both treated in the force-free approximation.  But in the collapsar context, the 
returning flux will have a significantly different magnetization, temperature, and velocity than the outgoing flux.  

Note also that, after the jet fluid escapes the star, and accelerates to $\Gamma \sim 10^2-10^3$, a diminishing fraction
$\sim (\Gamma\theta_j)^{-1}$ of the toroidal magnetic field sees this cylindrical current sheet.  While
reconnection at this sheet is a possible source of GRB variability, it is not on energetic grounds likely
to be the dominant source.

\begin{figure}
\epsscale{0.8} 
\plotone{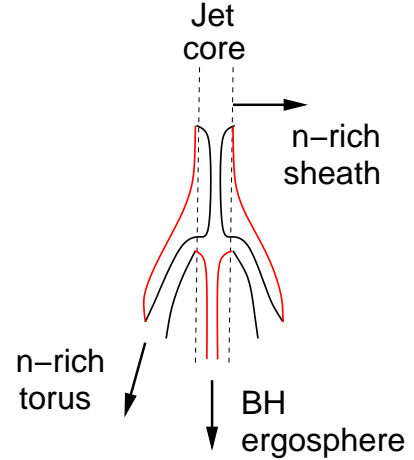}
\caption{Geometry of the poloidal magnetic field, when the sign of the magnetic field threading the black
hole varies stochastically in time.  The return flux (red lines) connects to a torus that is the source of a neutron-rich wind.  
After the flux threading the black hole flips in sign, the return flux in the neutron-rich annulus surrounding
the fast jet also flips, and can reconnect with previously ejected flux.}
\label{fig:jetflip}
\vskip .1in
\end{figure}

More complicated radial structure for the magnetic field is possible.
Radial reversals in the magnetic field on a lengthscale $\sim \pi c/\Omega_f$ are not expected
in a Blandford-Znajek jet.  More stochastic reversals, occurring on the timescale of the dynamo in
the orbiting neutron torus, are still possible \citep{thompson06}.  For example, a dynamo operating over
$\sim 100$ orbital periods ($t_{\rm dy} \sim 30$ ms) would generate flips in the magnetic field on
a characteristic lengthscale $\Delta r \sim ct_{\rm dy} \sim 10^9$ cm.  As long as the jet Lorentz factor remains modest,
domains of opposite toroidal field would come into causal contact at a radius $r > 2\Gamma^2 ct_{\rm dy} \sim 10^{10}$-$10^{11}$~cm.
 The relevant field geometry is a current sheet stretched out across the width of the jet, and therefore 
extending beyond the transverse causal scale $\sim r/\Gamma$.
The return flux in a neutron-rich annulus surrounding the fast jet core would also vary in sign
(Figure \ref{fig:jetflip}).

Finally, we note that reconnection is possible if the field becomes disorganized due to global current-driven instabilities
(e.g. \citealt{levinson13}).  A jet powered by such a hot bubble is subject to similar limitations
on isotropic jet energy to those analyzed in Section \ref{s:lisomax}, and therefore has difficulty supplying
the largest observed GRB output.

\subsubsection{Limitations to Reconnection Rate}

The rate of reconnection is potentially influenced by two physical effects which
we now discuss.  The first (which has received
less attention in the astrophysical literature because it requires more extreme conditions) involves
the trapping of radiation near the sheet.  Such an `opacity limit' to the rate of reconnection has
been considered in intermediate parts of GRB jets \citep{thompson94} as well as close to the engine, 
where the pair density is very high \citep{mckinney12}.  It is only relevant in situations where the 
current sheet is very extended (as it is in the co-moving frame of a rapidly expanding GRB outflow), and where
the rate of reconnection is not significantly limited by resistive diffusion.

In a relativistic plasma with Alfv\'en speed close to $c$, we may normalize the speed of inflow to a current 
sheet as $V_{\rm rec} = \varepsilon_{\rm rec} c$.  Fast reconnection corresponds to $\varepsilon_{\rm rec}
\sim 0.1$.  Consider an outflow with a scattering depth $\tau_T$ across
a causal distance $ct'\sim r/\Gamma$.  The corresponding depth across a layer of thickness 
$\Delta_{\rm rec} = V_{\rm rec}t'$ is $\tau_T(\Delta) \sim \varepsilon_{\rm rec}\tau_T$.  Requiring that
radiation generated by dissipation of reconnecting magnetic field be able to escape the layer on a timescale
$t'$, one obtains an upper bound on $\tau_T$:
\be
\tau_T(\Delta_{\rm rec}){\Delta_{\rm rec}\over c} < t'\quad\Rightarrow\quad \tau_T < \varepsilon_{\rm rec}^{-2}.
\ee
We encounter optical depths $\tau_T \lesssim 10^2$ in continuously heated and strongly magnetized pair plasmas
with equivalent black-body temperature $< 0.05\,m_ec^2$  (corresponding to distances from the engine larger
than $\sim 10^9$~cm).  Reconnection is not limited by the build-up of radiation pressure beyond such a distance from the engine, 
even if it is maximally efficient ($\varepsilon_{\rm rec} \sim 0.1$).  The back pressure of slowly cooling ions is
also irrelevant in this context.

Discussion of a second constraint on reconnection has been motivated by plasma experiment \citep{yamada10}, which shows
that X-point reconnection only occurs in electron-ion plasmas of size $L$ if the thickness of a Sweet-Parker current 
sheet is smaller than the ion plasma length, 
\be\label{eq:sp}
\delta_{\rm SP} \sim {L\over S^{1/2}} \sim \left({\eta L c^2 \over 4\pi V_A}\right)^{1/2} < {c\over \omega_{\rm Pi}}.
\ee
Here $\eta$ is the ohmic resistivity, $V_A$ is the Alfv\'en speed, and $\omega_{\rm Pi} = (4\pi n_e e^2/m_p)^{1/2}$ 
in a hydrogen plasma.  The Lundquist number is 
\be
S = {V_A L\over \eta c^2/4\pi}.
\ee
If the inequality (\ref{eq:sp}) is not satisfied, then experiments achieving
$S \sim 10^{2-3.5}$ find that reconnection is limited by plasma outflow through a narrow 
current sheet of thickness $\delta_{SP}$.  

This leads to the interesting suggestion of delayed reconnection in GRB outflows as they become effectively
collisionless \citep{mckinney12}.  When considering such implications, one key issue is whether experiment has achieved 
large enough $S$ to reflect the behavior of astrophysical systems.  If fast X-point reconnection were to take place, then 
the magnetic Reynolds number of the fluid motions would be 
\be
{\cal R}_m \sim {V_{\rm rec} L\over \eta c^2/4\pi} \sim \varepsilon_{\rm rec} S \sim 0.1 S.
\ee
The point here is that because $\delta_{\rm SP}$ depends on the size of the plasma, one can always satisfy the
inequality (\ref{eq:sp}) by considering a narrower slice closer to the current sheet.  However, if $S$ is not too
large, then a fluid description of the plasma flow toward these `sub-scale X-points' would break down, and a Sweet-Parker 
layer should persist.

Numerical experiment shows that X-points do form in weakly-magnetized pair plasmas, as long as the box size exceeds
$\sim 10^2 c/\omega_{\rm Pe}$ \citep{swisdak08}; and robustly in relativistic pair plasmas \citep{zenitani07}.  Fast reconnection occurs
in spite of the absence of a Hall term in the conductivity (which requires a mass asymmetry between positive and negative charges).

The constraint analogous to (\ref{eq:sp}) in a strongly magnetized pair plasma ($\omega_{\rm ce} = eB/m_ec \gg \omega_{\rm Pe} = 
(4\pi n_e e^2/m_e)^{1/2}$, implying $B^2/4\pi \gg n_e m_ec^2$) is, roughly
\be\label{eq:spb}
\delta_{\rm SP} < {c\over \omega_{\rm ce}}.
\ee
Note also that $V_A \sim c$ on large scales (where radiation and charges are coupled) as long as $B^2/8\pi \gg U_{\rm th}$;
whereas $V_A$ is very close to $c$ on small scales.  The resistivity in a magnetized plasma with significant radiation pressure
is mainly due to Compton drag on the drift motion of the current-carrying electrons \citep{uzdensky08},
\be
\eta \sim {(4/3)U_\gamma \sigma_T c\over 4\pi n_e e^2}.
\ee
Then we find
\ba
S &=& {27\pi\over 2} {\tau_T\over f_{\rm th} \alpha_{\rm em}^2 (B/B_Q)^2}\nn
  &\sim& 10^{17}\,\left({f_{\rm th}\over 0.1}\right)^{-1} \left({\tau_T\over 10^2}\right)\left({B/B_Q\over 10^{-4}}\right)^{-2},
\ea
corresponding to $\delta_{\rm SP} \sim 3\times 10^{-9} (r/\Gamma)$.

Given that multiple X-point reconnection does occur sufficiently close to a current sheet at large $S$,
the remaining key question involves the non-linear development of this structure.  If the magnetic field becomes
disorganized, as suggested by recent large-scale PIC simulations of electron-ion plasmas \citep{loureiro12}, then reconnection 
is sped up by the appearance of multiple Sweet-Parker layers \citep{lazarian99,kowal09}.
One is led to the picture of an expanding, turbulent current sheet outlined by \cite{thompson06}.  Turbulent reconnection
in a strongly magnetized GRB outflow is only limited by causality, that is, by the ability of any given patch of magnetic field
to see a current sheet, and by also by radiation pressure close to the engine.

\subsection{Charge Starvation of Plasma Currents}\label{s:cs}

A promising mechanism for particle heating involves the damping of MHD turbulence.
In the present context, where the plasma is pair dominated
and reaches a scattering depth $\sim 10-30$ during the heating phase, the Compton drag
time is longer than the wave period.  Strongly coupled waves will cascade to a high 
wave number first.

What happens near the inner scale of this cascade depends essentially on the particle density.  At 
a very low density, Alfv\'en waves become charge-starved \citep{tb98}, but otherwise they
Landau damp on the electrons and positrons.  Charge starvation is only
possible at a high wavenumber: a large-scale breakdown of MHD is inconsistent with any 
significant scattering depth through the entrained charges.

Most numerical experiments now find that the energy spectrum of magnetic fluctuations is
somewhat flatter than Kolmogorov \citep{maron01,boldyrev06}, with scaling $(\delta B)^2 
\sim k^{1-\alpha} = k^{-1/2}$.  However, we leave the index $\alpha$ open here, and consider
also the Kolmogorov scaling $\alpha = 5/3$.  The wave shear appears to adjust so that
collisions between waves are strongly coupled, and the conserved energy flux in wavenumber
space imposes the constraint
\be 
\omega (\delta B)^2 = k_\parallel V_A (\delta B)^2 = {\rm const}
\quad \Rightarrow\quad k_\parallel \sim k_\perp^{\alpha-1}.
\ee
We normalize $k_{\parallel,0} \sim \Gamma/r$ and $k_{\perp,0} \sim (\delta B_0/B)^{-1}\,\Gamma/r$ at 
the outer scale.\footnote{If colliding waves are roughly cylindrically symmetric but strongly 
sheared, $k_\perp \gg k_\parallel$, then the condition for this type of `critical balancing' is 
that the wave amplitude is $\sim k_\perp^{-1}$, corresponding to $(k_\perp/k_\parallel)(\delta B/B) \sim 1$
\citep{gs95}.  There is, however, some numerical evidence that cascading wavepackets become increasingly elongated
in the plane transverse to ${\bf B}$, which changes this condition \citep{boldyrev06}.}

To determine whether charge starvation or Landau damping cuts of this spectrum, consider
the current fluctuation 
\be
\delta J \sim {c\over 4\pi}k_\perp \delta B \sim {cB\over 4\pi r/\Gamma}
\left({k_\perp\over k_{\perp,0}}\right)^{(3-\alpha)/2}
\ee
Defining the charge starvation scale by
\be
\delta J \sim en_e c,
\ee
one finds 
\be\label{eq:kstarve}
k_\perp^{\rm starve} \sim {\Gamma\over r}\left({4\pi en_e r\over \Gamma B}\right)^{2/(3-\alpha)}
{B\over \delta B}_0.
\ee

When the magnetic energy density dominates the rest energy density of the light charges, as is
the case here, Landau damping occurs when the perpendicular wavenumber reaches the electron
skin depth, $k_\perp \sim \omega_{\rm Pe}/c$.\footnote{Alfv\'en waves with such strong perpendicular 
shearing develop large parallel electric fields, which has the effect of pushing the Alfv\'en 
speed significantly below $c$.}  To compare (\ref{eq:kstarve}) with the Landau-damping scale, 
we can simply evaluate
\ba
\left({k_\perp^{\rm starve} c\over \omega_{\rm Pe}}\right)^2 &=& 
{\bar{\lambda}_c\over r/\Gamma}\left({B\over B_{\rm Q}}\right)^{-4/(3-\alpha)} \times\nn
&&\left({3\tau_T\over 2\alpha_{\rm em}}\right)^{(1+\alpha)/(3-\alpha)} {B^2\over \delta B_0^2}.
\ea
In a jet of isotropic Poynting luminosity $L_{\rm P\,iso} = (\Gamma B r)^2 c$ and
Lorentz factor $\Gamma$, this gives at a radius $r$
\be\label{eq:kcs}
\left({k_\perp^{\rm starve} c\over \omega_{\rm Pe}}\right)^2 = 0.2\,
{(r_{12}\,\tau_{T,1})^{5/3}(\Gamma/3)^{11/3} \over L_{\rm P\,iso,51}^{4/3} (\delta B_0/B)^2}
\quad\quad \left(\alpha = {3\over 2}\right).
\ee
Charge starvation effects are therefore potentially quite important inside $\sim 10^{12}$ cm
from the engine.  Note, however, that the coefficient in (\ref{eq:kcs}) is $\sim 10^3$ times larger if the
wave spectrum is Kolmogorov.


\section{Discussion}\label{s:summary}

The repeatable spectral behavior of GRBs deserves a simple and robust explanation.
We focus here on the spectral peak and the low-frequency tail below it.  Although relativistic
beaming introduces complexities by blending together angle, frequency, and time, 
a simple model of a Planckian emitted by a relativistically boosted photosphere does
not come close to reproducing the typical low-frequency spectrum of a GRB:  the emergent photon
index is only slightly flatter than Rayleigh-Jeans, $\alpha = 0.4$ \citep{beloborodov10}.  

For that reason, it has long been suspected that the spectral peak and low-frequency tail offer essential
clues to the emission mechanism (and, thence, to the underlying mechanism of energy transport and particle 
heating).  Here we have demonstrated that by abandoning a longstanding assumption of a significant {\it baryonic} 
component for GRB outflows, but maintaining the strong Poynting flux that is needed to extract energy from the engine 
at the rates observed, these two essential features of GRBs fall easily into place.  Baryons are still need to provide
a confining medium, to limit the expansion of the jet to $\Gamma \sim 1/\theta$ while still confined,
and possibly to induce variability via hydrodynamic instabilities at the jet head.  

In the framework advanced here, the hard tail of a GRB must originate outside breakout, after the outflow has achieved 
higher Lorentz factors by a combination of radiation and magnetic stresses.  
Separating its origin from the remainder of the spectrum is partly motivated by
the inability of `one-box' models to avoid fine tuning (of the radius or compactness); and to avoid
introducing more complicated, non-thermal particle populations.   

Our results are independent of the details of the heating mechanism, as long as it is gradual.  Consideration
of the lateral structure of the jet, including the presence of slower material, suggests that breakout and full
jet acceleration may be delayed to $\sim 10^4-10^6$ times the gravitational radius of the engine.  That is,
breakout may even be pushed close to the transition between `jet' and `pancake' geometry.  In such a situation,
the amount of heating is sensitive to the degree of causal contact both across the the jet (e.g. the value of
$\Gamma\theta_j$), as well as along the jet axis.  The $m=1$ modes seen in the 3D jet$+$torus simulations of
\cite{mckinney09}, when translated this far out, are a plausible source of the mild heating we require.  It is
not clear to us that realistic hydromagnetic simulations have yet fully captured the heating effects of 
radial velocity shear across the jet.   Ideal kink modes remain an interesting possibility, although they
have only been considered so far in the `magnetic tower' approximation \citep{lyutikov03,giannios06b}.

Existing attempts to decompose the spectrum of a GRB into thermal and non-thermal components assume that
the low-frequency spectrum of the thermal component is Rayleigh-Jeans, and that the non-thermal components
extends above and below the peak \citep{ryde04,ryde09}.  Based on the present results, such a decomposition should be
repeated with a more general low-frequency index in the thermal component.  


Our main conclusions can be separated into the prompt emission mechanism, and the
physical properties of the engine and the outflow that it generates:

\subsection{Prompt Emission of Gamma-ray Bursts}

1. {\it GRB-like low-frequency spectrum.}  
A low-frequency photon index near -1 naturally arises in a {\it strongly magnetized}
($B^2/8\pi \gtrsim U_{\rm th}$) and nearly thermal pair plasma.  The resultant spectral state
does not depend significantly on the initial compactness over a range $\sim 10^5-10^7$.
Even flatter low-frequency spectra, which are sometimes seen in GRBs, are found at 
lower values of the compactness ($\lesssim 10^5$).

2. {\it Scaling behavior during expansion.}
When such a strongly magnetized plasma expands over more than a decade in radius, while 
being continuously heated, the spectrum remains relatively flat below the peak.  This
spectral state appears to be an attractor:  it does not depend significantly on the
initial low-frequency slope of the thermal seed.  At a very high initial compactness
$\sim 10^7$, the photon index reaches a maximum 0 over a narrow ($\sim$ factor few) 
frequency band below the peak.

3. {\it Rapid transition to transparency.}  
Rapid pair annihilation allows a pair-rich jet that experiences a sudden drop in
heating to become transparent, and feel a strong outward radiation pressure force.  This connects
the thermal and strongly magnetized plasma state analyzed here with the magnetic jet 
solutions of \cite{russo13a,russo13b}.  As long as heating turns off during breakout
and transverse expansion of the jet, then it simultaneously becomes transparent -- without
any fine tuning of the scattering opacity.

4. {\it Spectral flattening during jet breakout.}
Modest bulk Compton scattering during this breakout forces some further flattening of 
the spectrum near its peak:  the maximum value of the photon index drops by $\sim -0.5$.
For example, the photon index maintains an average value $\sim -0.8$ from $10^{-3}E_{\rm pk}$ 
to $E_{\rm pk}$ for a final radiation (magnetic) compactness $\sim 10^4$ ($10^5$).

5. {\it Magnetic energy reservoir.}
The hard gamma-ray tails seen in GRBs are independent evidence that the magnetic field
dominates the energy flux after breakout.  For example, in a burst with a high-frequency 
photon index $-2.5$, the flux above the peak is at least comparable to the flux at and 
below the peak, and it dominates in bursts with harder spectra.   The kinetic energy 
of the entrained pairs and ions is negligible at breakout, and so the magnetic field is identified
by default as the energy reservoir for the hard tail.  The nearly complete reconnection 
and thermalization of the magnetic field before breakout (e.g. \citealt{levinson13})  
is disfavored for the same reason.

6. {\it Mapping of observed spectral peak to breakout plasma conditions.}
Because the pair plasma experiences only weak adiabatic cooling near breakout, and
maintains a relatively low Lorentz factor, there is
a direct connection between the spectral peaks of GRBs and the electron rest
mass.  In a compact, thermal pair plasma, the bulk-frame spectral peak sits at $\sim 0.2\,
m_ec^2$ at the end of the heating phase.  During cooling and pair annihilation, it drops
by a factor of only $\sim 0.5$ due to adiabatic cooling.  The measured spectral peak
is therefore largely a measure of the bulk Lorentz factor at breakout:  $\Gamma \sim 3
(E_{\rm pk}/200~{\rm keV})$.  

7. {\it Free expansion phase.}
This mapping of the spectral peak back to conditions at jet breakout implies a significant
(but temporary) drop in the dissipation rate following breakout.  The bulk of the 
jet acceleration occurs during this intermediate phase.  An upswing in dissipation, 
required to explain the presence of the high-frequency spectral tail, is not addressed
in this paper.  Magnetic reconnection tends to freeze out during the
intermediate acceleration phase.  The collision between the bulk of the magnetized jet fluid,
and a forward shell that is swept up from the confining medium, is also delayed to a larger
radius \citep{thompson06}.

8. {\it Amati et al. boundary.} 
If $\Gamma \sim 1/\theta$ during the last stages of jet thermalization, then one
obtains the lower bound to $E_{\rm pk}(E_{\rm iso})$ obtained by Amati et al.
from a sample of BeppoSax bursts.  Such a relation between $\Gamma$ and $\theta$ has been
derived for self-similar, cold, magnetized jets \citep{lyubarsky09}.

In this derivation we have taken careful account of
the expected range of binding energies of CO cores of a range of masses.  The Amati et al.
boundary corresponds to the most massive cores,  to jets with the highest radiative
efficiency, and to situations in which a large fraction
of the magnetic flux threading the black hole engine escapes the star through the jet.
Outflows from less massive cores, with lower radiative efficiency, or lower fractions of
the engine output channeled through a relativistic jet, sit {\it above}
the boundary.  Bulk Comptonization during jet breakout also tends to raise the peak energy
\citep{russo13a,russo13b}.

9. {\it X-ray flashes from modest baryon loading.}
X-ray flashes have a similar duration to GRBs but are spectrally softer.  They naturally
arise from jets that have a high enough baryon loading that the electron-ion component
dominates the scattering opacity at the photosphere.  (The Poynting luminosity of such a jet
may still dominate the kinetic luminosity.) In this situation, the output spectrum
shows a more pronounced thermal peak, with a harder spectrum just below the peak.  The peak energy
$E_{\rm pk}$ drops due to more efficient photon creation, and also due to stronger
adiabatic cooling.  We observe similar effects in magnetized plasmas where
a modest fraction ($\sim 10^{-3}$-$10^{-2}$) of the dissipation is channeled through
relativistic particles.  GRB pulses with a low-frequency Rayleigh-Jeans
slope (e.g. \citealt{crider97,ryde04}) sometimes appear near the beginning of a burst, and could
represent an intermediate level of baryon loading or non-thermal pair creation.


10. {\it Absence of high-energy emission.}
The `no-high-energy' pulses, which are detected as subcomponents in many GRBs
\citep{pendleton97}, provide a possible exception to 5.  These may correspond to
components of GRB jets in which the magnetic field is significantly dissipated before breakout.

\subsection{Engine and Physical Properties of the Jet}

11. {\it Extreme magnetization.}
We find that the jet is very strongly magnetized: $\sigma_{\rm ion} > 10^5$ for breakout from a Wolf-Rayet
star at $\sim 10^{11}$ cm from the engine.  If the jet transitioned from an 
extended period of weak magnetization to stronger magnetization, as would be expected
if the engine were a rapidly rotating magnetar \citep{metzger07}, then one would observe
an extended, spectrally soft precursor.

12. {\it Horizon in the engine.}
 This strong magnetization is most easily generated by the horizon of a stellar-mass 
black hole.  In a collapsar or binary merger, the neutron-rich torus is a strong emitter
of neutrinos, whose collisions generate electron-positron pairs \citep{eichler89,zalamea11}.  The pressure of
this pair plasma in the jet funnel pushes any residual baryons down to the horizon.  

13. {\it Magnetic collimation required.}
We have calculated the maximum isotropic energy that could be carried by a hydrodynamic 
jet escaping through a nozzle from a trapped relativistic bubble inside a collapsing Wolf-Rayet core.
For progenitor masses in the range $20-50\,M_\odot$, this maximum energy is smaller 
than the largest observed $E_{\rm iso}$, thereby pointing to a role for magnetic collimation
in GRB jets.

14. {\it Gradual heating points to ideal hydromagnetic instabilities.}
We have constrained the heating mechanism in the magnetized jet.  The flattest low-frequency
spectra are obtained if heating is gradual enough that the pairs remain sub-relativistic.  Transient
and localized heating could create relativistic particles that generate a higher scattering depth
by a pair cascade.  In that case, we show that the low-frequency spectrum is harder than is typical
in GRBs.  The spectral peak is also reduced in frequency.  We infer that the heating mechanism
that forms the low-frequency spectrum is more consistent with the damping of large-scale hydromagnetic modes
in a jet, rather than with localized reconnection events. 

15. {\it Questions about collisional effects on magnetic reconnection.}
The pair plasma is dilute enough that the build-up of radiation pressure does not
limit magnetic reconnection at extended current sheets embedded in the outflow.  We
have also discussed the influence of a high Lundquist number on the development
of a tearing instability at a current sheet.  Existing reconnection experiments
may overestimate the importance of collisional effects in suppressing the formation of
X-points and slowing down the reconnection rate.  They should, therefore, be given limited
weight in constructing magnetic reconnection models of GRBs.

\acknowledgments
This work was supported by the NSERC of Canada.

\begin{appendix}

\section{A. Photon Emission and Absorption Processes}\label{s:appA}

A single electron or positron moving with speed $\beta c$ (component
$\beta_\parallel = \beta\cos\alpha$ parallel to the magnetic field ${\bf B}$) emits energy in cyclo-synchrotron
photons of frequency $\omega$ at the rate
\be
{dE_{\rm cyc}\over dt d(\cos\theta) d\omega} = {e^2\omega^2\over c} \sum_{n=1}^\infty 
   \delta\left[{n\omega_{ce}\over \gamma} - \omega(1-\mu\beta_\parallel)\right] 
\left[\left({\cos\theta-\beta_\parallel\over \sin\theta}\right)^2 J_n^2(z) + \beta_\perp^2{J_n'}^2(z)\right].
\ee
Here $\theta$ is the emission angle measured with respect to ${\bf B}$, $\omega_{\rm ce} = eB/m_ec$, $z \equiv 
\beta_\perp\gamma\sin\theta(\omega/\omega_{ce})$, and $\beta_\perp^2 = \beta^2-\beta_\parallel^2$.  
The integral over pitch angle can be eliminated using the delta function, giving
\ba\label{eq:dncycdt}
{dn_\gamma\over dt d\omega} 
&=& \int_{-1}^1{1\over 2}d\cos\alpha\int_{-1}^1 d(\cos\theta)
\int_0^\infty d\gamma {dn_e\over d\gamma}\, {1\over \hbar\omega}\cdot{dE_{\rm cyc}\over dt d(\cos\theta) d\omega}\nn
&=& {\alpha_{\rm em}\over 2}\sum_{n=1}^\infty \int_{-1}^1 d\cos\theta \int_0^\infty{d\gamma \over \beta}{dn_e\over d\gamma}
{\sqrt{\cos^2\theta-X^2}\over \cos^2\theta}\left[\left({\cos^2\theta - X\over\cos\theta\sin\theta}\right)^2 J_n^2(z)
+ \left(\beta^2 - {X^2\over\cos^2\theta}\right){J_n'}^2(z)\right],
\ea
where 
\be
X \equiv 1- {n\omega_{ce}\over\gamma\omega};\quad\quad z \equiv \beta\gamma\sin\theta{\omega\over \omega_{ce}}
\sqrt{1 - {X^2\over (\beta\cos\theta)^2}}.
\ee
The output spectrum is shown in Figure \ref{fig:cyclotron}.

\begin{figure}
\epsscale{0.5} 
\plotone{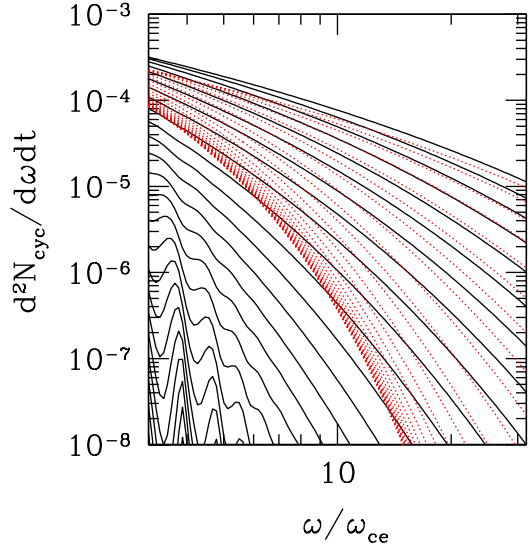}
\caption{Dotted red curves: relativistic synchrotron approximation (\ref{eq:synapp}) to the complete cyclo-synchrotron emissivity.
Temperature varies in increments $\Delta{\rm log}_{10}(T_e/m_ec^2) = 0.1$ from $T_e = 10^{-2.5} m_ec^2 = 1.6$ keV up to 
$T = m_ec^2 = 511$ keV (bottom to top).}
\label{fig:cychf}
\vskip .1in
\end{figure}

At high temperatures, we may compare equation (\ref{eq:dncycdt}) and Figure \ref{fig:cyclotron}
with the high-frequency synchrotron approximation, appropriately averaged over pitch angles:
\be
{d^2n_{\rm synch}\over d\omega dt} = \int_{-1}^1 {1\over 2}d\cos\alpha \int d\gamma {dn_e\over d\gamma}
{\alpha_{\rm em}\over \sqrt{3}\pi\gamma^2} {F_{\rm sync}(x)\over x},
\ee
where $x \equiv (2/3\gamma^2\sin\alpha)\omega/\omega_{\rm ce}$ and $F_{\rm sync}(x) \sim (\pi/2)^{1/2}x^{1/2}e^{-x}$.
This becomes
\be\label{eq:synapp}
{d^2n_{\rm synch}\over d\omega dt} = \int_0^\infty {d\gamma\over\gamma^2} {dn_e\over d\gamma} {\alpha_{\rm em}\over (6\pi\tilde\omega)^{1/2}}
\int_0^\infty d\beta {e^{-\widetilde\omega \cosh\beta}\over (\cosh\beta)^{3/2}},
\ee
where
$\tilde\omega \equiv (2/3\gamma^2)\omega/\omega_{\rm ce}$.  Equation (\ref{eq:synapp}) agrees to within a few
percent for $\Te \gtrsim 10^{-0.5}$ and $\omega/\omega_{\rm ce} > 10$.  A comparison between this approximation and
the full calculation is shown in Figure \ref{fig:cychf}.

The emission rate of bremsstrahlung photons is standard, and is given by 
\be
{d^2n_{\rm ff}\over d\omega dt} = \alpha_{\rm em}\bar{g}_{\rm ff} \left({8\over 3\pi\Te}\right)^{1/2} {\sigma_T c\over\omega}
\left[\left(n_{e-}+n_{e+}\right)n_p + 2^{3/2}n_{e-}n_{e+}\right]e^{-\hbar\omega/T_e},
\ee
where $\bar{g}_{\rm ff} \simeq \ln(2.2 m_ec^2/\hbar\omega)$.  Recall that collisions between $e^-$ and $e^+$,
and between $e^\pm$ and ions contribute to the bremsstrahlung emissivity, but not collisions between two electrons
or two positrons.

A similarly simple formula for double-Compton emission is available only at low frequencies \citep{thorne81,lightman81}:
\be
{d^2n_{\rm dC}\over d\omega dt} = {4\alpha_{\rm em}\over 3\pi} \left\langle\left({\hbar\omega\over m_ec^2}\right)^2\right\rangle
\,{\sigma_T n_e c\over\omega}{U_\gamma\over \langle \hbar\omega\rangle} f_{\rm rel}(\Te) \quad\quad(\hbar\omega \ll T_e),
\ee
with a correction factor for mildly relativistic temperatures \citep{svensson84},
\be
f(\Te) = (1 + 13.91\Te + 11.05\Te^2 + 19.92\Te^3)^{-1}.
\ee

Absorption through all channels is handled by Kirchoff's law,
\be
{d^2n\over d\omega dt} \longrightarrow {d^2n\over d\omega dt} \times \left[1 - {N(\omega)\over N_{\rm bb}(\omega)}\right],
\quad\quad N_{\rm bb}(\omega) = [\exp(\hbar\omega/T_e)-1]^{-1}.
\ee

\begin{figure}[h]
\epsscale{0.9}
\plottwo{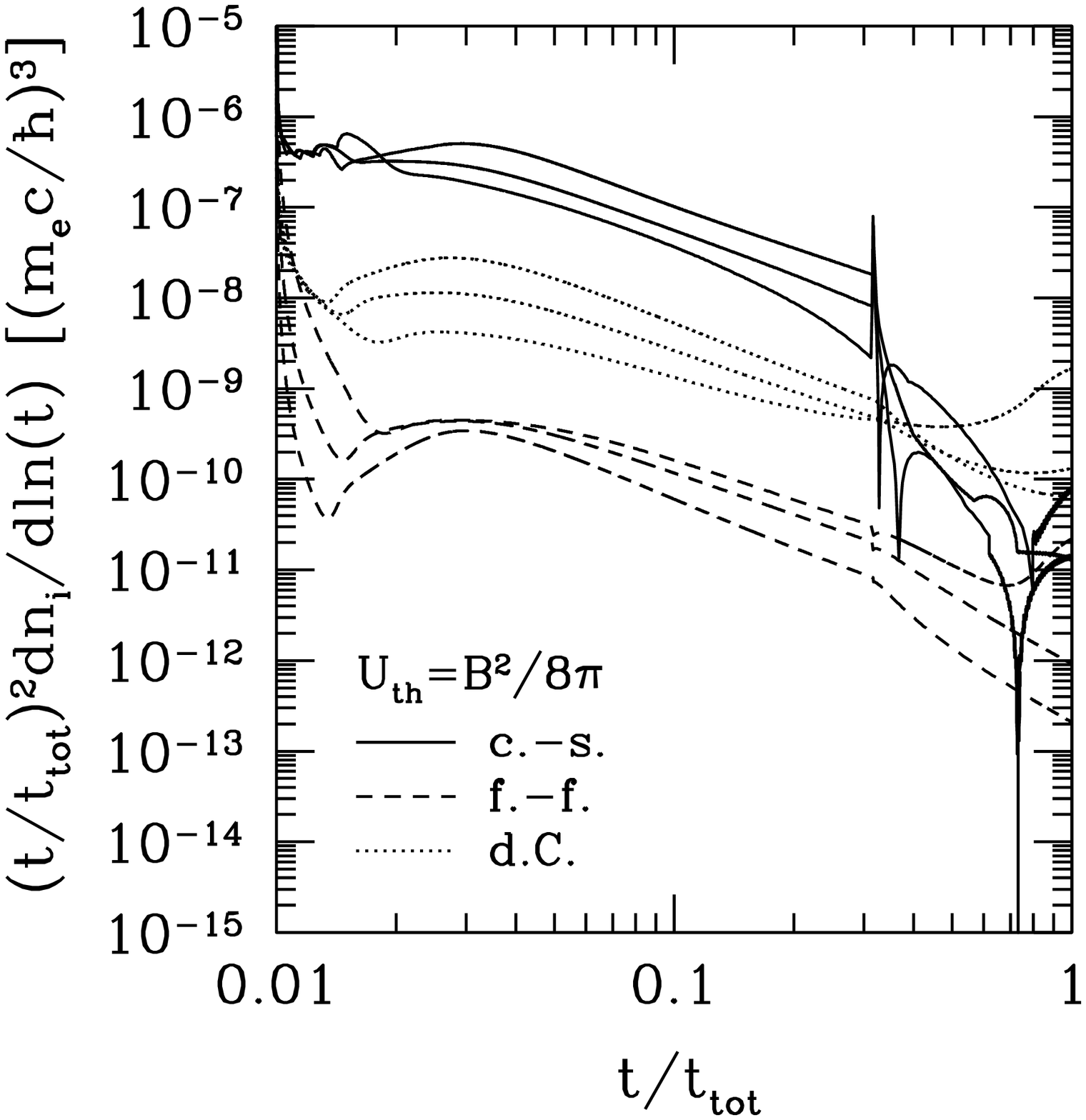}{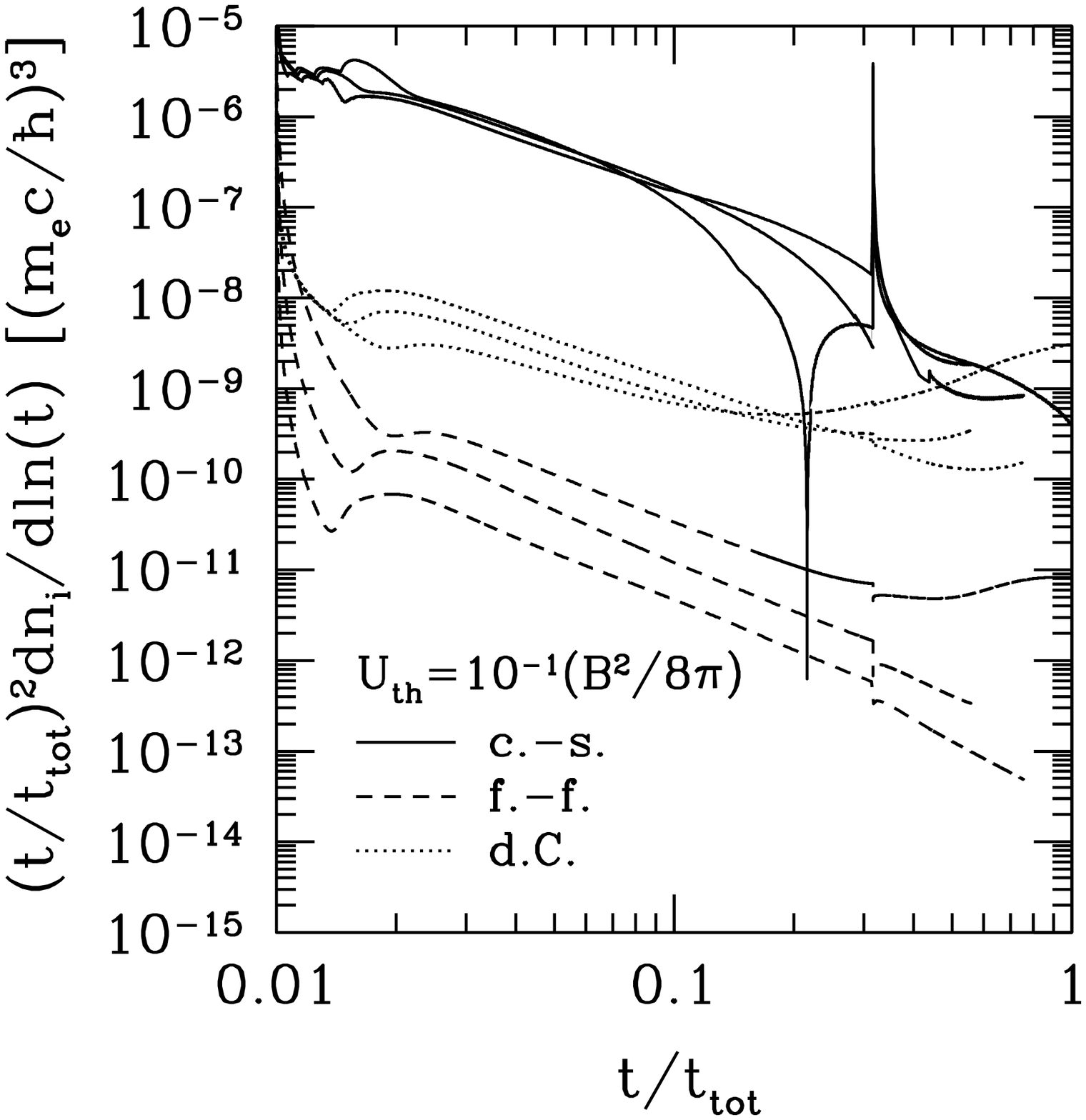}
\caption{Relative contribution of cyclo-synchrotron, free-free, and double-Compton soft photon emission
to the growth of photon density in an expanding pair plasma, as investigated in Section \ref{s:expand}.
The downward break in the curves occurs where heating is suddenly turned off, at time $10^{-0.5}t_{\rm tot}$.
Bremsstrahlung emission is briefly enhanced near the start of the calculation, as the pair density
relaxes to the equilibrium value.}
\label{fig:coolchan}
\vskip .1in
\end{figure}

The relative net contributions of these emission and absorption processes, integrated over frequency,
are shown in Figure \ref{fig:coolchan}, for an expanding plasma and two different values of the magnetization.  

\section{B. Pair Annihiliation and Creation}\label{s:appB}

Here we review the calculation of the creation of pairs by photon collisions,
$\omega + \omega' \rightarrow e^+ + e^-$ (see \citealt{svensson82} for a detailed treatment
in an astrophysical context).  A photon of frequency $\omega$ collides with another photon at a rate 
\be
\Gamma_{\gamma\gamma}(\omega) = 
{1\over 2}\int d\mu (1-\mu) \int^\infty_{\omega'_{\rm min}} d\omega'
\sigma_{\gamma\gamma}(\omega,\omega',\mu) {dn_\gamma\over d\omega'},
\ee
where the cross section 
\be
\sigma_{\gamma\gamma}(\omega,\omega',\mu) = 
{3\sigma_T\over 16}(1-\beta^2)\left[2\beta(\beta^2-2) + 
(3-\beta^4)\ln\left({1+\beta\over 1-\beta}\right)\right],
\ee
is defined in terms of the speed $\beta c$ of the resultant charged particles in the center-of-momentum frame,
\be
{m_ec^2\over \sqrt{1-\beta^2}} = E_{\rm cm}(\omega,\omega',\mu) = 
\left[{1\over 2}(\hbar\omega')(\hbar\omega) (1-\mu)\right]^{1/2}.
\ee
The low-energy threshold $\hbar\omega'_{\rm min}$ of the target photon that results in pair creation is obtained by setting 
$E_{\rm cm} \rightarrow m_ec^2$.  The net rate of pair creation per unit volume 
($n_e \equiv n_{e^+} + n_{e^-})$ is
\be\label{eq:dnedtgg}
{dn_e\over dt}\biggr|_{\gamma\gamma} = \int d\omega {dn_\gamma\over d\omega} \Gamma_{\gamma\gamma}(\omega).
\ee
Two charged particles are created in each photon collision, but the net collision rate is one-half
the integral on the right-hand side of (\ref{eq:dnedtgg}).

\end{appendix}



\end{document}